RESEARCH ARTICLE

# A Genomic Map of the Effects of Linked Selection in Drosophila

Eyal Elyashiv[1,2]\*, Shmuel Sattath[1], Tina T. Hu[3], Alon Strutsovsky[1], Graham McVicker[4], Peter Andolfatto[3], Graham Coop[5], Guy Sella[2]\*

1 Department of Ecology, Evolution, and Behavior, Hebrew University of Jerusalem, Jerusalem, Israel, 2 Department of Biological Sciences, Columbia University, New York, New York, United States of America, 3 Department of Ecology and Evolutionary Biology and the Lewis-Sigler Institute for Integrative Genomics, Princeton University, Princeton, New Jersey, United States of America, 4 The Laboratory of Genetics and The Integrative Biology Laboratory, Salk Institute for Biological Studies, La Jolla, California, United States of America, 5 Department of Evolution and Ecology, University of California, Davis, Davis, California, United States of America

\* eyalshiv@yahoo.com (EE); gs2747@columbia.edu (GS)

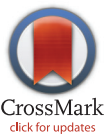

**Data Availability Statement:** All software files are available at http://github.com/sellalab/ LinkedSelectionMaps.

**Funding:** This work was funded by a Clore Foundation fellowship to EE, NIH grants GM107374 and GM83098 and NSF grant 1262645 to GC, NIH grant GM083228 to PA and GS, and Israel Science Foundation grant (no. 1492/10) to GS. The funders had no role in study design, data collection and analysis, decision to publish, or preparation of the manuscript.

## Abstract

Natural selection at one site shapes patterns of genetic variation at linked sites. Quantifying the effects of "linked selection" on levels of genetic diversity is key to making reliable inference about demography, building a null model in scans for targets of adaptation, and learning about the dynamics of natural selection. Here, we introduce the first method that jointly infers parameters of distinct modes of linked selection, notably background selection and selective sweeps, from genome-wide diversity data, functional annotations and genetic maps. The central idea is to calculate the probability that a neutral site is polymorphic given local annotations, substitution patterns, and recombination rates. Information is then combined across sites and samples using composite likelihood in order to estimate genome-wide parameters of distinct modes of selection. In addition to parameter estimation, this approach yields a map of the expected neutral diversity levels along the genome. To illustrate the utility of our approach, we apply it to genome-wide resequencing data from 125 lines in *Drosophila melanogaster* and reliably predict diversity levels at the 1Mb scale. Our results corroborate estimates of a high fraction of beneficial substitutions in proteins and untranslated regions (UTR). They allow us to distinguish between the contribution of sweeps and other modes of selection around amino acid substitutions and to uncover evidence for pervasive sweeps in untranslated regions (UTRs). Our inference further suggests a substantial effect of other modes of linked selection and of adaptation in particular. More generally, we demonstrate that linked selection has had a larger effect in reducing diversity levels and increasing their variance in *D. melanogaster* than previously appreciated.

## Author Summary

One of the major discoveries in modern population genetics is the profound effect that natural selection on one locus can have on genetic variation patterns at linked loci. Since the first evidence for linked selection was uncovered in *Drosophila melanogaster* over two decades ago, substantial effort has focused on quantifying the effects and on distinguishing the relative contributions of purifying and positive selection. We introduce an approach to jointly model the effects of positive and negative selection along the genome and infer selection parameters. To this end, we consider how closely linked each neutral site is to different types of annotations and substitutions. When we apply the inference method to genome-wide data from 125 *D. melanogaster* lines, our model explains most of the variance in diversity levels at the megabase scale and allows us to distinguish among the contribution of different modes of selection on proteins and UTRs. More generally, we provide a map of the effects of natural selection along the genome, and show that selection at linked sites has had an even more drastic effect on diversity patterns than previously appreciated. We also make a tool available to apply this approach in other species.

## Introduction

Selection at one site distorts patterns of polymorphism at linked neutral sites, acting as a local source of genetic drift. While the qualitative effects of "linked selection" are undisputed, quantifying them and understanding their source has been one of the central challenges in evolutionary genetics over the past two decades [1–17].

Indeed, characterizing the effects of linked selection is of central importance in many contexts. If linked selection introduces substantial heterogeneity in rates of coalescence along the genome, then obtaining accurate estimates of demographic parameters requires a genomic map of these effects [18,19]. Such maps would also serve as improved null models for other population genetic inferences, such as scans for recent targets of adaptation that rely on outlier approaches [20–22]. Moreover, an accurate characterization of the effects of linked selection carries extensive information about the selective pressures that shape genome evolution. Understanding how the effects vary among taxa would also inform long-standing questions about the determinants of levels of genetic diversity and genetic load within species [23,24,25, 26].

Patterns of genetic variation are informative about natural selection at linked sites because the effects of linked selection vary with the mode and parameters of selection. For instance, "classic" selective sweeps, in which a newly-arisen beneficial mutation is quickly driven to fixation, reduce genetic variation at nearby sites over a scale that depends on the strength of selection and rate of recombination [2,3]. Other modes of adaptation, including partial and soft sweeps, cause similar, although more subtle effects [27–31]. Background (purifying) selection against deleterious mutations also reduces diversity levels at linked sites over a scale that depends on the strength of selection and rate of recombination but to an extent that depends on the density of selected sites [5,8,9,32–34].

Until recently, evidence for the effects of linked selection was sought in the relationships between diversity patterns and factors that are expected to influence the strength and frequency of selection [13–15,17]. For example, both positive and negative linked selection should have a greater effect in regions with lower recombination rates, because, on average, a neutral site would be linked to more selected sites. Consistent with this expectation, diversity levels are positively correlated with rates of recombination in *Drosophila melanogaster* and several other species [4,35,36]. By a similar argument, linked purifying selection should be stronger in regions with a greater density of functional sites (e.g., coding regions) and the effects of sweeps should be greater in regions with more functional substitutions (e.g., non-synonymous substitutions). In accordance with these expectations, diversity levels decrease with the density of amino acid substitutions in Drosophila species [11,12] and in humans [37], and decrease with the density of coding and putatively functional non-coding regions in Drosophila [38], humans [18,35,37] and other species (e.g., [39,40] and cf. [17]).

Beyond providing compelling evidence for the importance of linked selection, these relationships can be used to estimate selection parameters [6,10–12]. These inferences, however, suffer from severe limitations. First, it is difficult to distinguish between the effects of different modes of linked selection, with two decades of effort focused on distinguishing the effects of classic selective sweeps from those of background selection [5,7,10,14,17,31]. Second, even when a specific mode of selection is assumed, some parameters remain poorly identifiable (e.g., the rate and strength of beneficial substitutions in sweep models [10,14]). These inferences also appear to be strongly affected by the genomic scale over which they are evaluated [14].

An alternative approach is to take advantage of spatial diversity patterns along the genome. Pioneering efforts in *D. melanogaster* used estimates of the genome-wide rate of deleterious mutations, genetic maps, and the spatial distribution of constrained genomic regions, to demonstrate that background selection could account for changes in diversity levels along chromosomes as well as for differences in diversity levels between X and autosomes ([41–43]). More recently, McVicker et al. [18] used ancestral diversity levels along the genome in order to build a map of the effects of background selection along the human genome. The central idea was to calculate the probability that a neutral site is polymorphic, given its genetic distance from conserved coding regions and the rate of deleterious mutation and distribution of selection effects at these regions; selection parameters were then estimated by maximizing the composite-likelihood for neutral polymorphisms along the genome. Although based on limited data, the map inferred by this approach provides an impressive fit to diversity patterns on the mega-base scale. However, the associated estimate of the deleterious mutation rate is unreasonably high, more than four-fold greater than estimates of the total spontaneous mutation rate [44–47], possibly reflecting the absorption of the effects of background selection from other, poorly annotated functional regions or the effects of positive selection [18].

Another recent approach to learn about selective sweeps relies on plots of the average levels of diversity as a function of distance from amino acid substitutions throughout the genome [48–50]. Assuming that some of the substitutions resulted from classic sweeps, we would expect a trough in diversity levels around substitutions, with the depth related to the fraction that were beneficial and the width (in units of genetic distance) reflecting the strength of selection. The rate and strength of classic sweeps can thus be inferred from the shape of the trough. Applying this methodology to data from *D. simulans*, Sattath et al. [48] found a trough in neutral diversity levels around amino acid substitutions that extended over ~15 kb, but not around synonymous substitutions (which served as a control). The collated plot approach has several limitations, however. First, application of the same approach to human data [49] suggests that background selection, which is concentrated in or near coding regions, may contribute to the troughs in diversity, and thus could bias estimates of positive selection parameters. Second, inferences based on collated diversity patterns account only for the average clustering of amino acid substitutions and not for their spatial distribution around every neutral site.

Here, we combine the advantages of these two recent approaches [18,48] in order to infer selection parameters and build a genomic map of the effects of linked selection, considering background selection and classic selective sweeps jointly. We model the effects of background selection using the annotations for linked sites, and those of classic sweeps by considering linked, putatively functional sites that experienced a substitution. The method is applicable to genome-wide polymorphism data, allowing for information to be combined across samples. As an illustration, we apply our method to genome-wide resequencing data from 125 lines of *Drosophila melanogaster* (from the DGRP [51]). We also make software available for the approach to be applied more broadly.

## Materials and Methods

### The model and inference method

We model the effects of background selection and classic sweeps on neutral heterozygosity (i.e., the probability of observing different alleles in a sample size of two), $\pi$, at an autosomal position $x$. In a coalescent framework, the model takes the form

$$\pi(x) = \frac{2u(x)}{2u(x) + 1/(2N_e B(x)) + S(x)}, \tag{1}$$

where $u(x)$ is the local mutation rate, $N_e$ is the effective population size without linked selection, $B(x)$ is the local (multiplicative) reduction in the effective population size due to background selection and $S(x)$ is the local coalescence rate caused by classic sweeps. This approximation can be arrived at by considering the probability that a mutation occurs (at a rate $2u(x)$ per generation) before our pair of lineages are forced to coalesce by either genetic drift ($1/2N_e B(x)$), which includes the effect of background selection, or by a selective sweep ($S(x)$). While we consider autosomes, the model can be extended to sex chromosomes with minor modifications.

The model for the effects of background selection, $B(x)$, follows Hudson & Kaplan [8] and Nordborg et al. [9] (Fig 1A). We assume a set of distinct annotations $i_B = 1, \ldots, I_B$ under purifying selection (e.g., exons, UTRs, introns and intergenic regions) and positions in the genome $A_B = \{a_B(i_B) | i_B = 1, \ldots, I_B\}$, where $a_B(i_B)$ denotes the set of genomic positions with annotation $i_B$. The selection parameters at these annotations are given by $\Theta_B = \{(u_d(i_B), f(t|i_B)) | i_B = 1, \ldots, I_B\}$, where $u_d$ is the rate of deleterious mutations and $f(t)$ is the distribution of selection coefficients in heterozygotes. The reduction in the effective population size is then

$$B(x|A_B, \Theta_B, R) = Exp\left(-\sum_{i_B} \sum_{y \in a_B(i_B)} \int \frac{u_d(i_B)}{t(1 + r(x,y)(1-t)/t)^2} f(t|i_B) dt\right), \tag{2}$$

where $R$ is the genetic map, $r(x, y)$ is the genetic distance between the focal position $x$ and positions $y$ (only positions on the same chromosome are considered). The integral reflects the effect that a site under purifying selection at position $y$ exerts on a neutral site at position $x$. This expression and its combination across sites should provide a good approximation to the effect of background selection so long as selection is sufficiently strong (i.e., when $2N_e t \gg 1$).

In turn, the model for the effect of selective sweeps follows from an approximation used by Barton [52] and Gillespie [53], among others (Fig 1A). Similarly to the model for background selection, we assume a set of distinct annotations $i_S = 1, \ldots, I_S$ subject to sweeps, but here we know the specific positions at which substitutions have occurred, $A_S = \{a_S(i_S) | i_S = 1, \ldots, I_S\}$, with $a_S(i_S)$ denoting the set of substitution positions with annotation $i_S$. The selection parameters at these annotations are $\Theta_S = \{(\alpha(i_S), g(s|i_S)) | i_S = 1, \ldots, I_S\}$, where $\alpha$ is the fraction of substitutions that are beneficial and $g(s)$ is the distribution of their additive selection coefficients. For autosomes, the expected rate of coalescent per generations at position $x$ due to sweeps is then approximated by

$$S(x|A_S, \Theta_S, R, \bar{N}_e, T) = \frac{1}{T} \sum_{i_S} \alpha(i_S) \sum_{y \in a(i_S)} \int Exp(-r(x,y)\tau(s, \bar{N}_e)) g(s|i_S) ds, \tag{3}$$

where $T$ is the length of the lineage (in generations) over which substitutions occurred, the positions of substitutions $y$ are summed over the chromosome with the focal site, $\bar{N}_e$ is the average effective population size and $\tau(s, N_e)$ is the expected time to fixation of a beneficial substitution

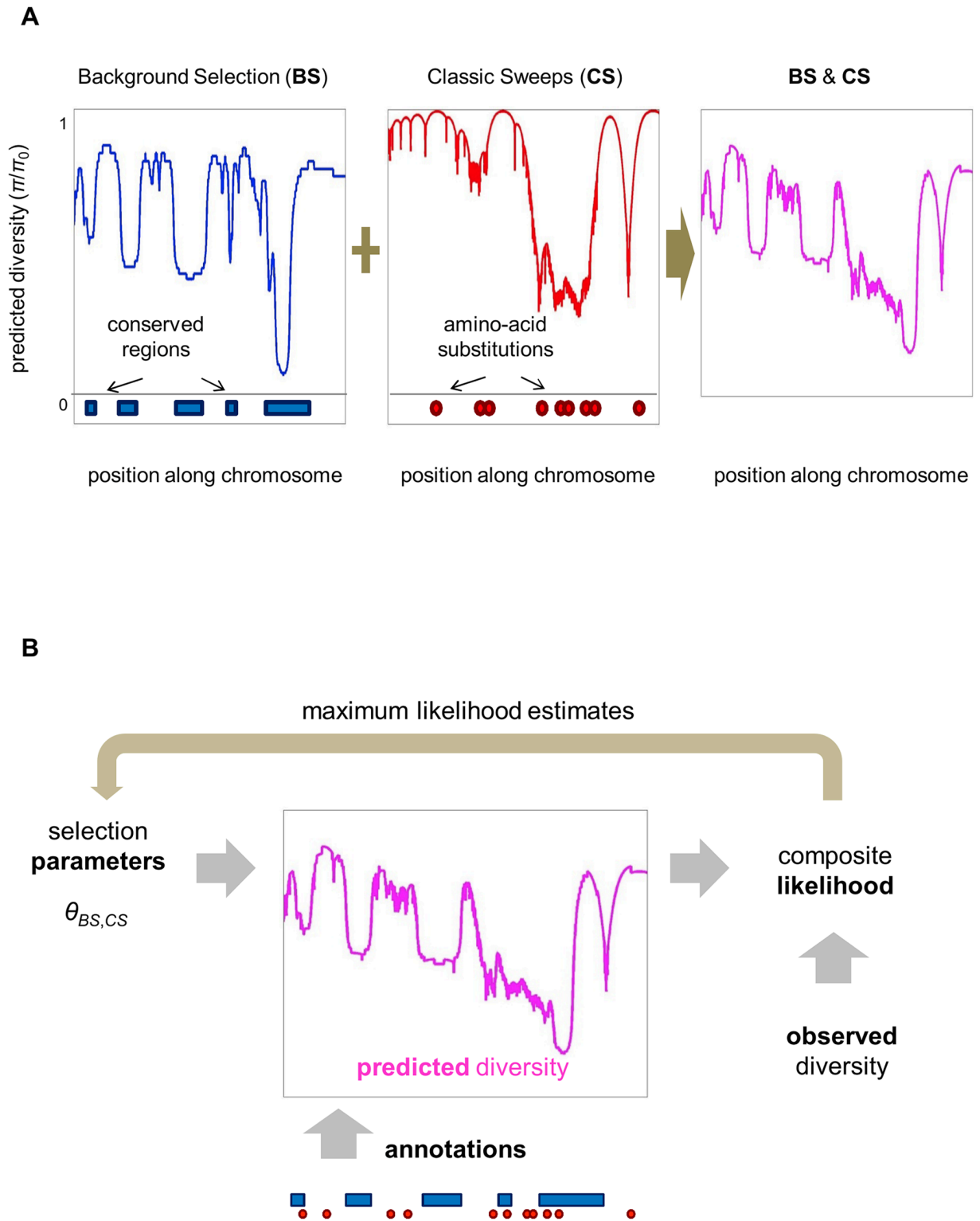

**Fig 1. Constructing a map of the effects of linked selection and inferring the underlying selection parameters.** (**A**) The expected neutral heterozygosity is estimated for each position in the genome, given the positions and selection parameters of different annotations. (**B**) To estimate selection parameters, their composite likelihood is maximized given the set of annotations and neutral polymorphism data throughout the genome.

doi:10.1371/journal.pgen.1006130.g001

with selection coefficient $s$ and given an effective population size $N_e$. We use the diffusion approximation for the fixation time

$$\tau(s, N_e) = \frac{2(\ln(4N_e s) + \gamma - (4N_e s)^{-1})}{s},$$ (4)

where $\gamma$ is the Euler constant (cf. [28]). This model relies on several simplifying assumptions and approximations. In particular, the term $1/T$ relies on an assumption of one substitution per site per lineage and neglects variation in the length of lineages across loci. In combining the effects over substitutions, we further assume that the timings of beneficial substitutions are independent and uniformly distributed along the lineage, and that they are infrequent enough such that we can ignore interference among them [54]. The exponent approximates the probability of coalescence of two samples due to a classic sweep with additive selection coefficient $s$ (where $2N_e s \gg 1$) in a panmictic population of constant effective size $\bar{N}_e$. (We consider the effects under more general sweep models later.) In principle, we should use the local $N_e$ incorporating the effects of background selection but given the logarithmic dependence of Eq (3) on $N_e$, we simply use the average.

To infer the selection parameters $\Theta_B$ and $\Theta_S$, we use a composite likelihood approach across sites and samples [55] (Fig 1B). We denote the positions of neutral sites by $X$ and the set of samples by $I$. We then summarize the observations by a set of indicator variables across sites and all pairs of samples $O = \{O_{i,j}(x) \mid x \in X, i \neq j \in I\}$, where $O_{i,j}(x) = 1$ indicates that samples $i$ and $j$ ($i \neq j$) differ at position $x$ and $O_{i,j}(x) = 0$ indicates that they are the same. In these terms the composite log-likelihood takes the form

$$LogL = \sum_{x \in X} \sum_{i \neq j \in I} \log(\Pr\{O_{i,j}(x) | \Theta_B, \Theta_S\}),$$

where

$$\Pr\{O_{i,j}(x) | \Theta_B, \Theta_S\} = \begin{cases} \pi(x | \Theta_B, \Theta_S) & O_{i,j}(x) = 1 \\ 1 - \pi(x | \Theta_B, \Theta_S) & O_{i,j}(x) = 0 \end{cases}.$$ (5)

Using composite likelihood circumvents the complications of considering linkage disequilibrium (LD) and the more complicated forms of coalescent models with larger sample sizes. Importantly, maximizing this composite likelihood should yield unbiased point estimates [56,57]. Beyond losing the information in LD patterns and in the site frequency spectrum, the main cost of this approach is the difficulty in assessing uncertainty in parameter estimates (as standard asymptotics do not apply). We therefore use other ways to assess the reliability of our inferences.

To make the composite likelihood calculations (i.e., the calculation of $\pi(x | \Theta_B, \Theta_S)$) feasible genome-wide, we discretize the distribution of selection coefficients on a fixed grid. Given a grid of negative and positive selection coefficients, $t_g$ and $s_k$, $g = 1, \ldots, G$ and $k = 1, \ldots, K$, the distribution of selection coefficients for each annotation becomes a set of weights on this grid, $w(t_g | i_B)$ and $w(s_k | i_S)$. (In principle, the grid could also be annotation-specific.) For background selection, these weights reflect the rate of deleterious mutations with a given selection coefficient and their sum should therefore be bound by the maximal deleterious mutation rate per site. For sweeps, the weights reflect the fraction of beneficial substitutions with a given selection coefficient and their sum should be bound by 1. In these terms, the effect of background selection takes the form

$$B(x | \Theta_B) = Exp\left(-\sum_{i_B} \sum_{g=1}^{G} w(t_g | i_B) b(x | t_g, i_B)\right),$$ (6)

where $Exp(-b(x|t_g, i_B))$ is the proportional reduction in the effective population size induced by having one deleterious mutation per generation per site with selection coefficient $t_g$ at all the positions in annotation $i_B$. By the same token, the effects of sweeps take the form

$$S(x|\Theta_S) = \frac{1}{T} \sum_{i_S} \sum_{k=1}^{K} w(s_k|i_S) s(x|s_k, i_S), \qquad (7)$$

where $\frac{1}{T} s(x|s_k, i_S)$ is the probability of coalescence per generation induced by sweeps in annotation $i_S$, if all the substitutions in this annotation are beneficial with selection coefficient $s_k$. Thus, by using a grid, we can calculate a lookup table of $b(x|t_g, i_B)$ and $s(x|s_k, i_S)$ once and then use it to calculate the likelihood for a given set of weights. Moreover, the interpretation of estimated distributions on a grid is arguably simpler than that of the continuous parametric distributions commonly used (e.g., gamma and exponential), for which densities associated with different selection coefficients are highly interdependent. In the Supplementary Material (S1B Text), we describe additional simplifications in the calculation of $b(x|t_g, i_B)$ and $s(x|s_k, i_S)$.

Other parameters are estimated as follows. Consider Eq (1) rewritten as

$$\pi(x) = \frac{\pi_0 \cdot (u(x)/\bar{u})}{\pi_0 \cdot (u(x)/\bar{u}) + 1/B(x) + S(x; \bar{N}_e, T)}, \qquad (8)$$

to clearly specify all the additional parameters required for inference. $\pi_0 \equiv 4N_e\bar{u}$ is (approximately) the average neutral heterozygosity, given the effective population size in the absence of linked selection ($\bar{u}$) and the average mutation rate per site ($\bar{u}$); $\pi_0$ is estimated through the likelihood maximization. The local variation in mutation rate $u(x)/\bar{u}$ is estimated by averaging substitution patterns at putatively neutral sites among closely related species in sliding windows, with a window size chosen to balance true variation in mutation rates and measurement error (see S1B Text). Finally, $\bar{N}_e$ is estimated based on the average genome-wide heterozygosity at putatively neutral sites, after dividing out by a direct estimate of the spontaneous mutation rate per site, and $T/2\bar{N}_e$ is estimated by $(\bar{K}/2)/\pi_0$, where $\bar{K}$ is the average number of substitutions per neutral site on the lineage.

The software package implementing the inference and construction of the map of the effects of linked selection is available online (http://github.com/sellalab/LinkedSelectionMaps). In the Supplementary Material (S1B Text), we describe the steps that were taken to check the proper convergence of the likelihood maximization.

## Application to data from Drosophila

We apply our method to population resequencing data from *Drosophila melanogaster*. The data analyses are briefly described here, with further details provided in S1A Text. As a proxy for neutral variation, we use synonymous polymorphism within *D. melanogaster*, based on resequencing data from the Drosophila Genetic Reference Panel (DGRP) [51] consisting of 162 inbred lines derived from the Raleigh, North Carolina population. The rate of synonymous divergence used to control for local variation in mutation rates is estimated using the aligned reference genomes of *D. simulans* and *D. yakuba* [58]. As potential targets of selection (annotations), we use coding regions, untranslated, transcribed regions (UTRs), long introns (>80bp) and intergenic regions, downloaded from FlyBase [59] (http://flybase.org, release 5.33), all of which have been inferred to be under extensive purifying selection in *D. melanogaster* [60–63], and which together cover ~98.5% of the euchromatic genome. Substitutions that occurred in these annotations on the *D. melanogaster* lineage since the common ancestor with *D. simulans* are inferred from a three-species alignment of reference genomes from *D. melanogaster*, *D.*

*simulans* and *D. yakuba* [58]. We do not include substitutions in intergenic regions, which are not included in the three-species alignment, and our treatment of missing data, e.g., due to gaps in the alignment, is detailed in S1B Text.

For the genetic map, we rely on estimates of the cM/Mb rates recently published by Comeron et al. [64]. Because our inferences are sensitive to errors in the genetic map in regions of low recombination, we exclude the distal 5% of chromosome arms (in which rates are known to be low in *D. melanogaster*) and regions with a sex-averaged recombination rate below 0.75cM/Mb.

We perform the inference under a variety of selection models. In the Results, we primarily compare the models incorporating classic sweeps, background selection or both, including all of the annotations listed above using a grid of selection coefficients which consists of five point masses on a log-linear scale, with $t$ and $s = 10^{-5.5}$, $10^{-4.5}$, $10^{-3.5}$, $10^{-2.5}$ and $10^{-1.5}$. Our maps of the effects of linked selection corresponding to the model incorporating both classic sweeps and background selection are available online (http://github.com/sellalab/LinkedSelectionMaps/melanogaster_maps). In the Supplementary Material we study the sensitivity of our results to: selection on synonymous mutations—using a subsets of synonymous differences (S1H Text), the recombination thresholds (S1H Text), the grid of selection coefficients (S1I Text), and to using subsets of annotations (S1I Text) and an upper bound on the deleterious mutation rate (S1E Text).

## Results

### Maps of the effects of linked selection along the genome

Our inference yields a map of the expected neutral diversity levels at every position along the genome. One way to evaluate these predictions is to compare them with observed diversity levels (Fig 2). A quantitative comparison at the 1Mb scale suggests that our map accounts for 71% of the variance ($R^2$) in diversity levels of non-overlapping autosomal windows. To address the concern that the high $R^2$ is the result of over-fitting, we perform a *leave-one-out cross-validation* (LOOCV) analyses [65] in which we divide the genome into non-overlapping 1Mb windows, using only data outside a window to make our predictions about diversity levels in it (S1C Text; Table S2 in S1 Text). This analysis shows that over-fitting has a negligible effect on our prediction, which is to be expected: while our model has many parameters (36), the data set is much larger (consisting of $1.7 \times 10^6$ codons, and levels of linkage disequilibrium are low).

In interpreting the fit, both model misspecification and the stochasticity inherent to the evolutionary process need to be considered. Importantly, even if our model provided an accurate description of the processes generating genetic diversity, we would not expect a perfect fit to the data because of the randomness of the processes being modeled. Notably, our model assumes that a substitution at a given annotation could have occurred with uniform probability at any time along the *D. melanogaster* lineage and that it had a certain probability of being beneficial with a given selection coefficient. Any evolutionary realization of the model would have that substitution occur at a particular time—more often than not, too far in the past to affect extant diversity patterns—and with a given selection coefficient, thus generating considerable variance in predicted diversity levels at linked sites. In addition, both genealogical and mutational processes are stochastic. Averaging over 1Mb windows partially reduces this stochasticity and in that regard, it is not surprising that our predictions become less precise when we use smaller windows (Fig 2B). However, even with 1Mb windows, we would still expect considerable variance in diversity levels around the expectation.

In addition, although we assume that the genetic maps and annotations are known, there is error in both. Imprecision of the genetic map and imperfect annotations (e.g., our clumping

**A**

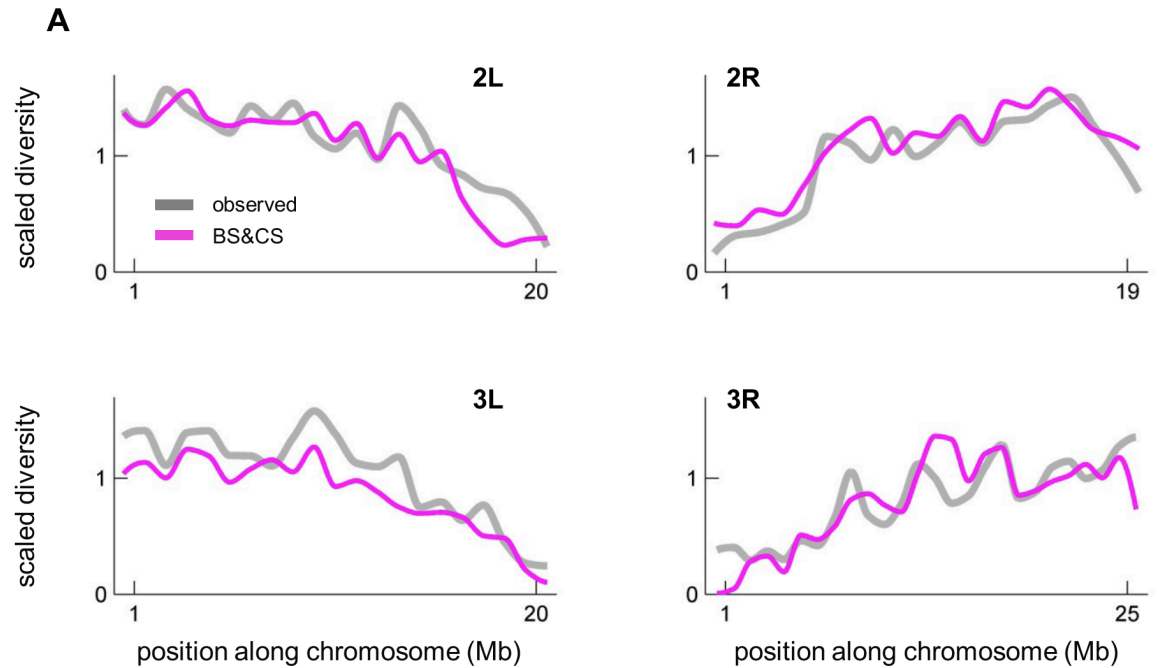

**B**

| | BS&CS | BS | CS |
|---|---|---|---|
| $R^2$ 1 Mb | 71% | 76% | 67% |
| $R^2$ 100 Kb | 44% | 42% | 39% |
| $R^2$ 10 Kb | 26% | 23% | 21% |
| $R^2$ 1 Kb | 20% | 18% | 16% |

**Fig 2. A comparison of observed and predicted scaled diversity levels along the major autosomes of *Drosophila melanogaster*.** Throughout, we refer to "scaled diversity" as synonymous heterozygosity divided by synonymous divergence, to control for variation in the mutation rate (as detailed in S1C Text); scaled diversity is shown relative to the genome average. (**A**) Observed and predicted scaled diversity over non-overlapping 1 Mb windows across chromosomal arms. (**B**) Summaries of the goodness of fit for models including background selection (BS), classic sweeps (CS) and both (BS & CS). $R^2$ is calculated for autosomes using non-overlapping windows of different sizes. Selection parameters are inferred using synonymous sites with recombination rate >0.75cM/Mb, while the predictions and corresponding summaries are calculated for sites with recombination rate >0.1cM/Mb.

doi:10.1371/journal.pgen.1006130.g002

together of all coding, UTR, intronic and intergenic substitutions and regions) decrease our predictive ability. As genetic maps and annotations become better, we should therefore expect our predictions to improve. Another class of assumptions relates to processes that we did not model, including changes in population size [61,66,67]. In spite of many potential factors contributing to noise in our predictions, the fit to data is very good.

In the Supplementary Materials (S1F Text) we compare our predictions to those based on a map of the effects of background selection generated using the methodology developed by Charlesworth [41] and recently extended by Charlesworth [42] and Comeron [43]. This approach differs from ours in several ways, most notably in being based on estimates of selection parameters from the literature, which themselves do not rely on the effects of linked selection on diversity patterns. While it performs impressively well at the 1Mb scale (though not as well as ours) the quality of the predictions becomes much worse than ours as the scale becomes smaller (Table S5 in S1 Text). (Note that Comeron [43] uses rank correlations to evaluate his predictions; the explained variance using rank correlations are much higher than the quantitative predictions we use here, which is why his result might appear comparable at first sight.)

Using $R^2$ values for window sizes varying from 1kb to 1Mb, we can ask which model(s) are best supported. We find that the one combining both background selection and classic sweeps almost always does better than the models with a single mode of selection (Fig 2). Our leave-one-out cross-validation analysis confirms that this finding is not the result of over-fitting in the combined model (Table S2 in S1 Text; see S1C Text for details). Thus, our combined model of the effects of linked selection captures much of the variation in diversity levels at the mega-base scale, and provides an improved null model in scans for targets of positive selection or for the purposes of demographic inference. Because using $R^2$ has its limitations, we use a variety of other statistical approaches to evaluate our inferences in the sections that follow.

## The effects of linked selection around different annotations

We can also use our analysis to learn about the effects of linked selection for different annotations. If a feature is enriched for targets of purifying or positive selection, then we expect to see a reduction in diversity levels around it due to linked selection. Collating diversity levels around all instances of a feature averages over confounding effects at specific genomic positions as well as over the inherent stochasticity in diversity levels, allowing us to isolate the selection effects [18,48–50].

We first consider how diversity levels vary with genetic distance from amino acid and synonymous substitutions (Fig 3). There is a trough in diversity around both, but the one around amino acid substitutions is substantially deeper (Fig 3A). Fig 3B compares the predicted diversity levels around amino acid substitutions based on Sattath et al. [48] and our inference. A rough quantitative comparison suggests that our method fits the data better than that of Sattath et al. ($R^2 = 62\%$ for our method compared to $R^2 = 56\%$ for Sattath et al.; see S1G Text for more details). Moreover, the new method also predicts more of the detailed variation in diversity levels, presumably because it accounts for the statistical properties of genome architecture, e.g., the density of coding regions at given genetic distances up or downstream of substitutions.

In principle, our approach should allow us to tease apart the contributions of classic sweeps and background selection to these diversity patterns (Fig 4). Comparing the predictions of each model alone is less informative for this purpose, because when only one is considered, it likely absorbs some of the effects of the other (see next section). In contrast, with the inference based

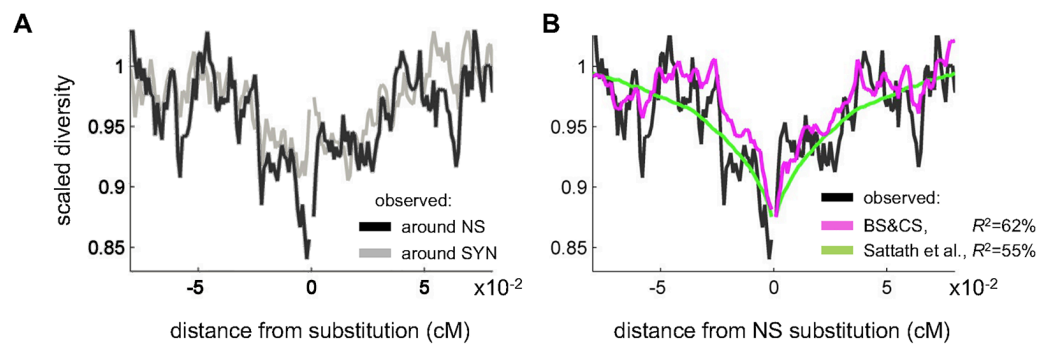

**Fig 3. Observed and predicted scaled diversity levels around amino acid substitutions.** (A) Comparison of scaled diversity levels around non-synonymous (NS) and synonymous (SYN) substitutions. (B) Comparison of predicted, scaled diversity levels based on our method and that of Sattath et al. (2011) [48].

doi:10.1371/journal.pgen.1006130.g003

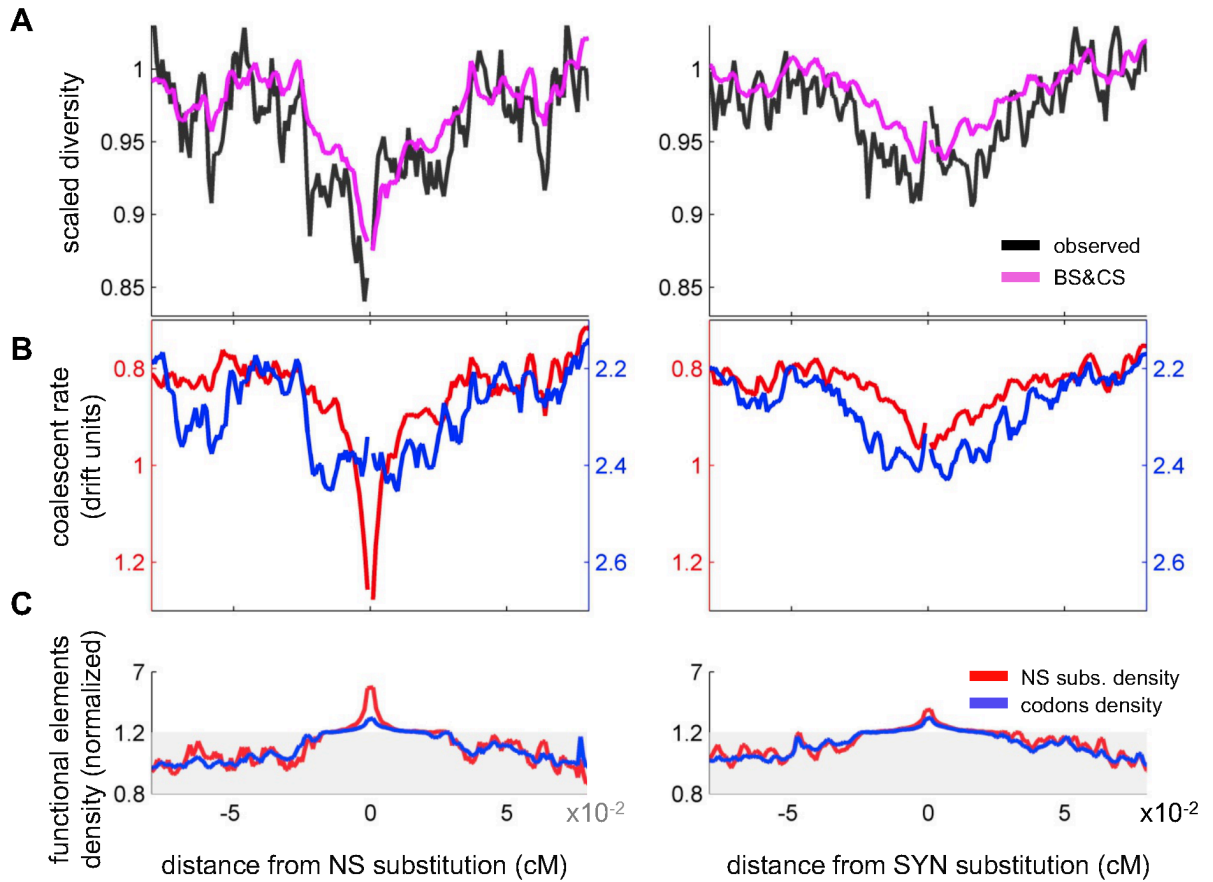

**Fig 4. The contribution of background selection and classic sweeps to scaled diversity levels around non-synonymous and synonymous substitutions.** (**A**) Observed and predicted scaled diversity levels around non-synonymous (left) and synonymous (right) substitutions. The predictions are based on the joint model for background selection and classic sweeps. (**B**) The contribution of background selection (blue) and classic sweeps (red) measured in terms of the coalescent rates that they induce. The rates are measured in units of $1/2N_e$, where $N_e$ is our estimate of the effective population size in the absence of linked selection. To make these graphs comparable to the scaled diversity levels in (A), with lower rates corresponding to higher scaled diversity levels, the direction of the y-axis is reversed. (**C**) The density of exonic sites (blue) and non-synonymous substitutions (red) as a function of distance from non-synonymous and synonymous substitutions. Densities are normalized by the average densities at distance >0.06cM; the shaded areas correspond to the use of a different linear scale.

doi:10.1371/journal.pgen.1006130.g004

on the combined model, the contribution of each mode should be identifiable from its specific functional forms and annotations. When we focus on the contribution of background selection (blue lines in Fig 4B), we see a reduction in diversity around both synonymous and non-synonymous substitutions because both types of substitutions occur in coding regions, in which background selection effects are strongest (e.g., [18,68]). Moreover, because the density of coding regions and other annotations (blue lines in Fig 4C and Fig S6 in S1 Text) is similar around the two kinds of substitutions, the shape and magnitude of the reductions in diversity are also similar (blue lines in Fig 4B). In contrast to background selection, the reduction around non-synonymous substitutions due to classic sweeps is much greater than for synonymous substitutions (red lines in Fig 4B). This results not only from the focal non-synonymous substitution but also (and primarily) from the greater density of non-synonymous substitutions near a focal non-synonymous substitution than around a synonymous one (red lines in Fig 4C). Whereas the clustering of non-synonymous substitutions around synonymous substitutions primarily reflects the greater density of coding sites, the clustering around non-synonymous substitutions (beyond the focal amino acid substitution) presumably reflects correlated evolution of nearby residues and other adaptive processes (e.g., [69]).

These findings illustrate that, at least as modeled, background selection and classic sweeps are identifiable. Intuitively, the information about classic sweeps at non-synonymous substitutions comes from the comparison of neutral diversity levels between sites near many non-synonymous substitutions versus near few, given a similar density of other annotations. After properly accounting for the effects of classic sweeps, information about the background selection pressure exerted by exons comes from contrasting the diversity levels among regions that vary in the density of codons but are otherwise similar. In practice, we do not learn about these processes in a stepwise fashion, as presented here, but instead maximize the probability of the data considering all of the annotations simultaneously.

We can therefore use these findings to revisit the enduring question of the relative contribution of background selection and classic sweeps to shaping diversity patterns (Fig 4). In particular, the negative correlation between diversity levels and the density of non-synonymous substitutions previously reported in Drosophila [11,12] likely reflects a substantial contribution of background selection in addition to positive selection. In contrast, the greater reduction in diversity levels at non-synonymous compared to synonymous substitutions in Drosophila is almost entirely the outcome of classic sweeps [48]. A caveat is that the parameter estimates obtained from the approach based on collated plots likely absorb some of the effects of background selection and thus overestimate the effects of linked selection due to sweeps (see next section and Tables S6 and S7 in S1 Text). More generally, in interpreting the results, an important consideration is the presence of other modes of selection that are not modeled explicitly, e.g., soft and partial sweeps. As we discuss at greater length below, our inferences about classic sweeps may reflect a mixture of different kinds of sweeps that result in substitutions while our inferences about background selection may reflect a contribution from other modes of linked selection, including sweeps that do not result in substitutions.

We can also consider how well the relationships between diversity levels and various genomic features are explained by models with a single mode of selection. As an illustration, Fig 5A shows that the background selection model does better than the model with classic sweeps at predicting diversity levels far from non-synonymous substitutions. Also visually apparent is that, in contrast to the background selection model, the classic sweeps model explains the

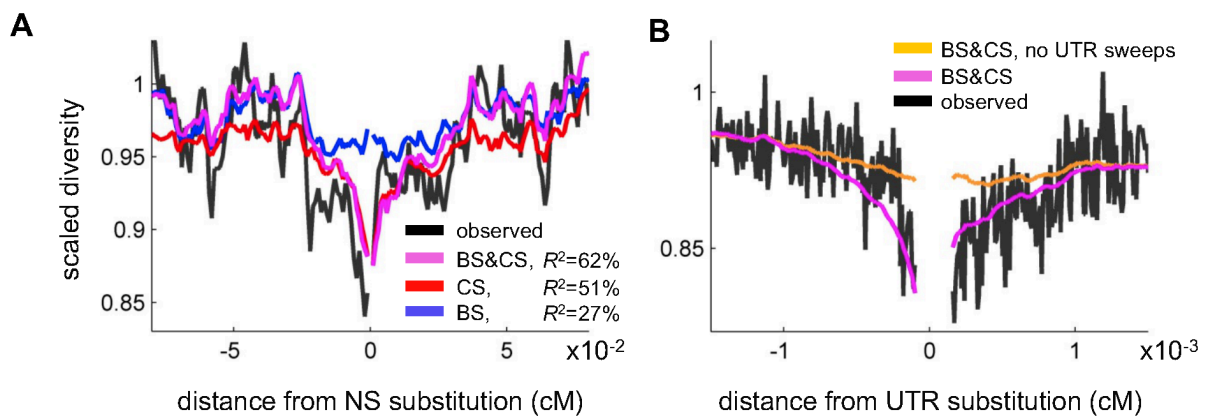

**Fig 5. Comparing alternative models around substitutions in proteins and UTRs.** (**A**) Comparison of predicted scaled diversity levels around non-synonymous substitutions based on models including background selection (BS), classic sweeps (CS) and both (BS & CS). (**B**) Comparison of predicted scaled diversity levels around substitutions in UTRs based on models with and without sweeps in UTRs.

doi:10.1371/journal.pgen.1006130.g005

narrow, deep trough close to non-synonymous substitutions. The combined model does well at predicting diversity levels both close to and far from non-synonymous and synonymous substitutions, again illustrating the need to consider both modes of linked selection in making inferences.

A similar approach can be used to examine the effects of selection acting on non-coding annotations. Notably, our inference suggests that a substantial fraction of substitutions at UTRs lead to classic sweeps (Table S11A in S1 Text and next section). To examine whether this feature of the model is required to explain the data, we look at average diversity levels as a function of genetic distance from substitutions in UTRs (Fig 5B). Our full model does much better at explaining these observations than a model without sweeps at UTRs. This provides the first evidence, to our knowledge, for sweeps at UTRs (or in any non-coding annotation) in Drosophila, and lends strong support to findings of pervasive adaptation in UTRs based on McDonald-Kreitman type approaches and genetic differentiation ($F_{ST}$) along clines [60,70].

## Estimates of sweep parameters

Our approach also provides estimates of selection parameters. We first consider those obtained for classic sweeps, for which the positions of potential targets of selection (i.e. substitutions) are known. For substitutions at non-synonymous sites and to a lesser extent in UTRs, the ability to localize substitutions and to measure diversity levels using nearby synonymous sites provides us with high spatial resolution about selection effects on diversity patterns.

If we exclude background selection from the model, the only notable difference is the addition of a probability mass of strong selection coefficients at amino acid substitutions (~0.3% of substitution with $s = 10^{-1.5}$), which affects diversity levels on a broad scale, in effect retracing large-scale variation in recombination rate and, to a lesser extent, coding density. When background selection is included in the model, this spatial effect becomes entirely associated with background selection (Fig S3 in S1 Text). This suggests that under a model of sweeps alone, the extra mass is absorbing some of the effects of other modes of selection that are not driven by substitutions.

In turn, under our combined models, the distribution of selection coefficient exhibits two dominant masses: ~4% of substitutions appear to have been strongly selected ($s \approx 10^{-3.5}$) and 35–45% of substitutions weakly so ($s$ between $10^{-5.5}$–$10^{-6}$; the ranges reported here and below correspond to grids of selection coefficients with 5 and 11 point masses; see S11 Text). The effects of both masses on diversity levels can be clearly seen in collated plots around substitutions (cf. Fig S8 in S1 Text) and accord with previous studies [48,71]. At UTRs, we find that 25–45% of substitutions are associated with weak to intermediate strength of selection ($s \approx 10^{-4.5}$–$10^{-5.5}$). While the effects of sweeps at UTRs are apparent in Fig 5B, our quantitative estimates are associated with greater uncertainty than those for non-synonymous substitutions because we have lower spatial resolution near substitutions at UTRs (see S1H Text). At long introns, we infer that none of the substitutions were driven by sweeps; this estimate, however, might also reflect low power in these regions, because we measure diversity levels at synonymous sites that are, on average, far from intronic substitutions (see S1I Text).

Intriguingly, our estimates of the fraction of beneficial substitutions in proteins and UTRs accord with those based on extensions of the McDonald-Kreitman test (i.e., between ~40–85% for amino acids and 30–60% from UTRs [12,38,60,72–74]), when previous estimates based on the effects of sweeps on polymorphism data were substantially lower [11,48]. A caveat is that this conclusion only holds when we include the contribution of weakly selected substitutions. Our inference about weakly selected substitutions is based on diversity patterns very close to substitutions (roughly equivalent to 50 bp on average) and at these distances, considerable uncertainty about the genetic map and limited polymorphism data preclude us from distinguishing between selection coefficients ranging between $10^{-5.5}$ and $10^{-6}$. Because selection coefficients at the lower end of this range could be nearly neutral, the substitutions could partially reflect the fixation of slightly deleterious mutations rather than beneficial ones and more generally compensatory evolution [75]. We note further that our approach is not necessarily expected to agree with McDonald-Kreitman based estimates, which reflect adaptive rates over different time scales (i.e., on the order of $N_e$ in our case [76], as opposed to the time scale of divergence). These reservations notwithstanding, our approach suggests that properly accounting for weakly selected substitutions leads to a convergence of estimates based on linked selection and McDonald-Kreitman based approaches, and provides, to our knowledge, the first corroboration of these elevated estimates.

With recent research highlighting the potential role of modes of adaptation other than classic sweeps, e.g., partial and soft sweeps [27–31,77–80], which we do not model explicitly, it is natural to ask how they might affect our inferences. To a first approximation, the effects of other kinds of sweeps on diversity levels around the selected site can be viewed as a superposition of the effects of classic sweeps with varying selection coefficients at different distances from the selected site (see [31,81] and S1D Text). This property implies that our parameter estimates for classic sweeps can be translated into rates and strengths of other types of sweeps.

As an example, consider our estimates that ~4% of amino acid substitutions were driven by selection coefficients of $s = 10^{-3.5}$ and ~35% by a selection coefficient of $10^{-5.5}$. An approximately similar effect on diversity levels along the genome could be explained by assuming that 39% of substitutions are caused by partial sweeps that are driven to a frequency of $x = 0.34$ with a selection coefficient of $s = 10^{-3.9}$, then to fixation with a selection coefficient of $s = 10^{-5.8}$ (see S1D Text). Similar parameter estimates could also be generated by mixtures of partial and full sweeps, described by the fraction of full and partial sweeps and associated selection coefficients and distributions of frequencies ($x$) for each kind of partial sweep. In S1D Text, we detail how other kinds of sweeps (soft, from multiple mutations or standing variation, or on recessive alleles) would be recorded by our approach and thus how the effects of mixtures of sweeps would translate into our parameter estimates.

In other words, in the presence of different kinds of sweeps, our parameter estimates reflect the effects of the mixture on diversity levels around substitutions. A given set of estimates designates a continuous class of mixtures and, in principle, one can write down equations for the parametric family of mixtures that would yield the same estimates. Further narrowing down the underlying mixtures, however, will require developing inferences that use other aspects of the data.

## Estimates of background selection parameters

Parameter estimates for purifying selection are fairly insensitive to the exclusion of classic sweeps from our model (e.g., Table S5 in S1 Text). When we do not impose an upper bound on the rate of deleterious mutations, we observe two main selection strengths, both of which are localized in exons and UTRs. The dominant one is extremely strong selection ($s = 10^{-1.5}$), which affects diversity over a spatial scale of ~4Mb (or ~7cM, the distance at which the diversity levels reach 90% of baseline levels). As noted previously, such selection coefficients lead diversity levels to follow large-scale variation in recombination rate and to a lesser extent coding density. In this regard, it is important to note that we have to rely on relatively crude annotations, rather than accounting for the fine-scale location of sites under purifying selection within each annotation. As a result, our inference is likely to capture an average effect over considerably larger spatial scales than is actually the case, thereby leading to somewhat inflated selection coefficients (akin to what is seen for classic sweeps when background selection is not considered).

The strong selection coefficient is also associated with unreasonably high estimates of the deleterious mutation rate, which far exceed direct estimates of the total mutation rate (by 4-9-fold in exons and UTRs; Table S12 in S1 Text) [82]. A plausible interpretation is that these high rates reflect the absorption of linked selection effects that evade direct capture by our inference. For example, they might absorb the effects of sweeps at introns (or intergenic regions) that evade our inference because of the crude annotation of substitutions in these regions. They might also absorb the effects of other modes of linked selection, which are not modeled explicitly. Notably, population genetic models of quantitative traits suggest that the response to changing selection pressures could involve many soft and partial sweeps that do not result in fixation [83,84] and therefore would not be included in our estimates for classic sweeps. The effects of such soft and partial sweeps on diversity levels can be similar to those of background selection [31,81,85,86]. Moreover, because we lack localized annotations for such sweeps (when they do not result in fixation), we would tend to associate them with stronger selection coefficients of background selection, whose effects on diversity are less localized. If this interpretation is correct, then our inference suggests that modes of linked selection other than classic sweeps and background selection have a substantial effect on diversity levels around coding regions.

We also find evidence for somewhat weaker purifying selection (centered around $s = 10^{-3.5}$) associated with a more realistic deleterious mutation rate (e.g. ~50–60% of the overall mutation rate in exons), but which may still reflect a contribution from other forms of linked selection. These values are in agreement with those obtained for exons by approaches that do not rely on the signatures of linked selection (cf. [42,43], and S1F Text). Purifying selection of this strength should affect diversity levels on spatial scale of ~40 kb (or 0.07cM, defined as above), a footprint that is visible in our analyses of diversity levels around synonymous and non-synonymous substitutions (blue lines in Fig 4B).

In the Supplementary Material (S1E and S1F Text), we present additional analyses that support this interpretation of background selection parameters, based on models in which we impose a biologically plausible upper bound on the deleterious mutation rate and use the modeling approach of Charlesworth [41,42].

## The impact of linked selection on diversity levels

We next examine the extent to which linked selection decreases the mean and increases the variance in diversity levels throughout the genome. The average reduction quantifies the effects of linked selection on the effective population size, a key parameter for many aspects of genome evolution [24,25]. The heterogeneity in diversity levels is of interest because it quantifies the deviation from the uniform neutral null model that is implicitly assumed in most, if not all, demographic inferences and scans for targets of adaptation.

We focus on the impact of linked selection in coding regions with recombination rates above 0.1cM/Mb, because our predictions become less reliable in regions with lower recombination rates (see S1H Text). To this end, we sort genomic positions according to their predicted levels of diversity (Fig 6A). For 1600 bins with equal amounts of data, the concordance between observed and predicted levels is extremely high (Spearman $\rho = 0.91$), indicating that the variation predicted by our model is real (and not due to over-fitting; Table S2 in S1 Text). Sorting based on our predictions, we find substantial variation in the observed diversity levels across bins (approximately five-fold difference between the upper and lower 2.5%; Fig 6B). Moreover, we see that the effects of linked selection are visible across all bins, rather than being restricted

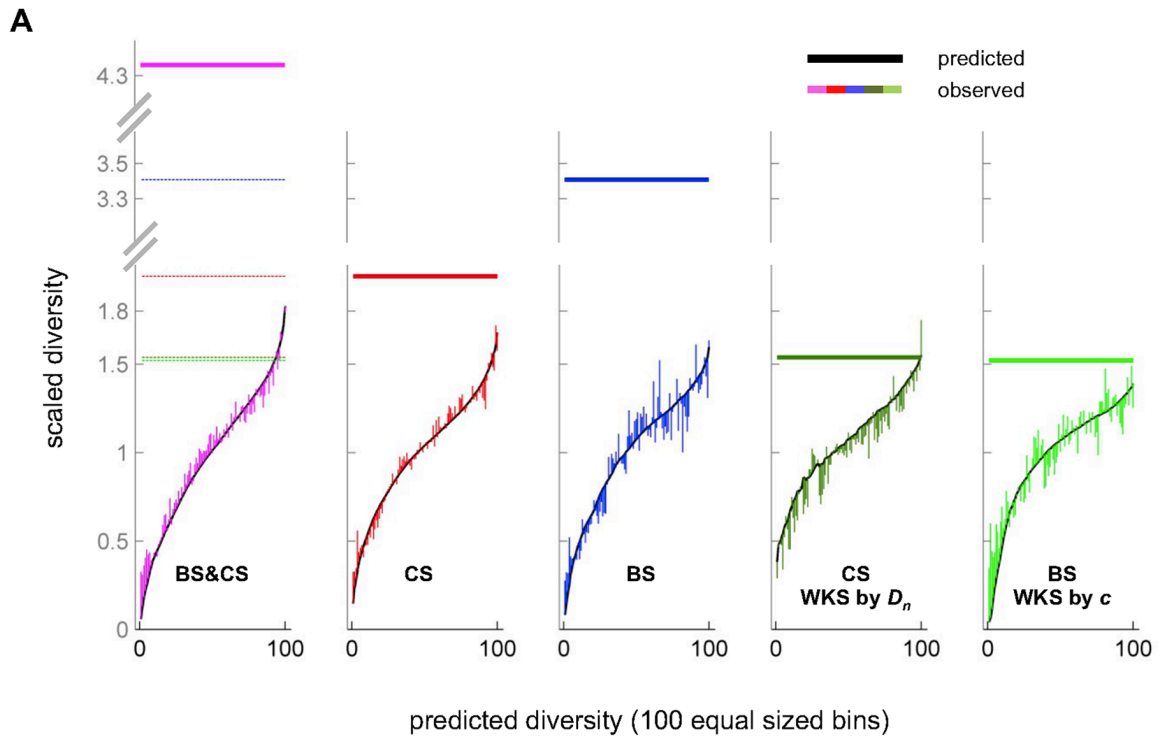

**predicted diversity (100 equal sized bins)**

| | BS&CS | CS | BS | CS<br>Wiehe&Stephan<br>(by NS divergence) | BS<br>Kim&Stephan<br>(by recombination rate) |
|---|---|---|---|---|---|
| diversity reduction,<br>mean | 77% | 50% | 71% | 35% | 36% |
| diversity reduction,<br>upper 2.5% tail | 60% | 18% | 57% | -5%[1] | 11% |
| observed diversity span,<br>upper vs lower 2.5%<br>tails of predicted effect | 5.3 | 5.1 | 4.4 | 4.1 | 3.2 |
| Spearman's $\rho$<br>(predicted vs. observed,<br>across 1600 bins) | 0.91 | 0.89 | 0.74 | 0.81 | 0.73 |
| $\sum_s \alpha(s) \cdot s$<br>compound parameter<br>of sweeps | $3.5 \times 10^{-5}$ | $1.5 \times 10^{-4}$ | - | $3.8 \times 10^{-5}$ | - |
| $U_{\text{del}}$<br>autosomal diploid rate of<br>deleterious mutations | 1.60 | - | 1.46 | - | 0.74 |

**Fig 6. The impact of linked selection on scaled diversity levels. (A)** Observed scaled diversity levels stratified by model predictions. Shown here are the results based on our method with both background selection and classic sweeps (pink), background selection alone (blue) and classic sweeps alone (red), as well as for the Wiehe and Stephan (1993) [6] method for classic sweeps based on the density of non-synonymous substitutions (dark green) and the Kim and Stephan (2000) [10] method for background selection based on recombination rates (light green). The stratification is described in the text. Predicted levels are shown in black, the observed deviations

from the predictions are shown as vertical lines, with the colors corresponding to different models, and the estimated scaled diversity levels in the absence of linked selection are shown as horizontal bars. **(B)** Summaries of the mean reduction and heterogeneity in scaled diversity levels based on the different methods and models. Also shown are estimates of compound selection parameters and the Spearman correlation between predicted and observed levels. (1) The negative value reflects the fact that the observed scaled diversity level is higher than the level predicted in the absence of linked selection.

doi:10.1371/journal.pgen.1006130.g006

to bins with lower expected diversity. In other words, almost no region in the genome is free from the effects of linked selection (with the exception of the correlation coefficient, none of these results are sensitive to the number of bins).

We quantify the average reduction due to linked selection as the ratio of the average observed diversity level, $\bar{\pi}$, to the predicted level without linked selection, $\pi_0$. Doing so indicates an average reduction of 77%-89% in neutral diversity levels genome-wide (excluding low-recombination regions for which the reduction should be greater). Strikingly, even in the upper 1%-tile, linked selection is predicted to have reduced diversity levels by ~60–80%. Given the uncertainty about the parameter estimates associated with strong purifying selection (S1E Text and Table S4 in S1 Text), our inferences about $\pi_0$ may not be robust, however. Indeed, imposing a plausible bound on the rate of deleterious mutations results in fits that are only marginally worse but dramatically affects our estimates of $\pi_0$ (reducing it from 4.4 fold times the observed mean to 2.8-fold, with 5 point masses; S1E Text and Table S4 in S1 Text). In brief, this follows from the fact that strong selection affects diversity levels on broad spatial scales, leaving little signal of localization, and thus similar observed diversity levels can result from different combinations of deleterious mutation rates and $\pi_0$ values. Unfortunately, we cannot observe $\pi_0$ directly. What we can say, based on our stratification, is that linked selection reduces average diversity levels by at least two-fold (Fig 6A).

Our estimates suggest much stronger effects of linked selection than do previous methods. Notably, when we apply previous methods based on the relationship between diversity levels and rates of recombination or functional divergence [6,10–12,26] (see S1G Text for details), we infer an average reduction in diversity levels that lies between 34–36%, with no reduction in the upper 1%-tile of predicted diversity levels (Fig 6B and Table S12 in S1 Text). Comparing the stratification of diversity levels by the various methods (Fig 6A and 6B) indicates these previous methods do worse at predicting diversity levels, span a smaller range of diversity levels and under-estimate the effects of linked selection; specifically, their predictions of $\pi_0$ are lower than the upper 1%-tile of observed diversity levels based on our stratification (Fig 6A and 6B). The reason is that by relying on a single genomic feature (e.g., recombination rate) and averaging over others (e.g., non-synonymous divergence), these methods overlook much of the variation in diversity levels caused by linked selection, causing their estimates to suffer from the equivalent of regression toward the mean (the same problem applies to their estimated selection parameters; see S1G Text). A similar "averaging out" effect takes place when we consider a model with background selection or classic sweeps alone (Fig 6A and 6B).

This line of argument implies that even with the combined model, we still underestimate the heterogeneity in diversity levels because of imperfect annotations. Notably, this would be the case if our inferences about background selection are likely absorbing substantial effects of other modes of linked selection but are unable to capture them in full, let alone to do so with high spatial resolution. Thus, the heterogeneity in diversity levels due to linked selection in the *Drosophila melanogaster* genome is likely to be even greater than we have inferred. Similar speculation about the average reduction in diversity levels is more difficult, given the uncertainty associated with our parameter estimates for background selection (Tables S4 and S10 in S1 Text). What we can say is that our lower bound based on stratification is likely to increase as annotations improve.

## Discussion

### The relative contribution of different modes of linked selection

Over two decades of research have aimed to quantify the relative contributions of classic sweeps and background selection in shaping diversity patterns. If these were the only modes of linked selection, then we would now have an answer. We have shown that the contributions of background selection and classic sweeps are identifiable using our inference and, with the stated caveats about the effects of partial annotations, we can quantify their relative contributions. Based on the combined model and using the genome-wide average rates of coalescence induced by each mode of selection as a measure of their relative contribution, our findings would suggest that background selection has a ~1.6–2.5-fold greater effect than classic selective sweeps (Table S3 in S1 Text; see S1C Text for details and other metrics).

The question is complicated, however, by the contribution of other modes of linked selection. Our results strongly suggest that inferences about background selection include a major contribution of other modes of linked selection, plausibly the result of sweeps that do not result in substitutions. In turn, our inferences for classic sweeps may reflect a combination of different kinds of sweeps. These results echo other theoretical and empirical results highlighting the importance of other modes of positive selection, notably of partial and soft sweeps [27–31,77–81].

The question about the relative contribution of different modes of linked selection can therefore be rephrased in terms of the contributions of background selection, classic sweeps and other modes of linked selection. If we assume that our combined model fully accounts for the reduction in diversity levels due to linked selection and that the effects of background selection are captured by our inferences excluding the strong selected mass, then 12% of the increase in coalescence rate due to linked selection is the result of background selection (estimates in this paragraph correspond to the model with 5 point masses). Further assuming that our inferences about classic sweeps can reflect any combination of classic and other kinds of sweeps resulting in fixation, and that the remaining effects are the outcome of other modes of linked selection, then we would conclude that roughly 0 to 29% of coalescent events are due to classic sweeps and the remaining 88 to 59%, respectively, are due to other modes of linked selection.

### Implications for Drosophila and other taxa

Despite unresolved questions about linked selection, the maps do well at predicting diversity levels at the 1Mb scale (Fig 2), the substantial stratification of diversity levels throughout the genome (Fig 6) and the diversity patterns around different annotations (Figs 3, 4 and 5). This predictive ability is explained in part by the effects of linked selection already well captured by our current approach, e.g., the effects of sweeps that result in substitutions. Also important, however, is the robustness of the inferred map of linked selection to model misspecification. For instance, our map performs well even though the effects of background selection may reflect a substantial contribution of other modes of linked selection and despite an averaging effect owing to the imprecise annotations. Moreover, at this scale, the performance is fairly insensitive to variations of the model (e.g., imposing a bound on the deleterious mutation rate), suggesting that these features play a relatively minor role. Thus, while the spatial resolution of maps of linked selection in Drosophila (and other taxa) is expected to improve with better genetic maps, annotations and models, we can already do quite well. One implication is that our approach already generates substantially improved null models for population genetic inferences about demography and scans of selection.

The reliability of our inferences about selection critically depends on well-localized annotations and on the distance between these annotations and the putatively neutral sites used to measure diversity levels. For these reasons, we obtain reliable estimates for sweeps resulting in substitutions at exons and UTRs and distinguish their contribution from other forms of linked selection, but cannot achieve similarly reliable estimates for other modes and annotations. It follows that in applications to other species, we would expect the reliability of estimates to depend both on the quality of annotations and on genome architecture. Human data may be particularly well suited, as there are higher-resolution annotations as well as phylogeny-based information about conservation in both coding and non-coding regions. In addition, properties of the genome architecture, notably the lower density of selected regions [87], may help to distinguish effects of different annotations and modes of linked selection.

In both Drosophila and humans, one area that will need further work is the inclusion of other modes of selection. In that regard, it is interesting to note that our results mirror similar finding in humans: inferences about background selection in McVicker et al. [18] also led to too large a rate of deleterious mutation and work done since suggests that classic sweeps contribute little to the effects of linked selection on genetic variation [49,77,78]. Taken together with other empirical evidence and modeling [27–30,77,79,80,83], these results strongly suggest that other modes of linked selection and of adaptation in particular play a central role in both Drosophila and humans.

It might be difficult to distinguish between different kinds of sweeps based on their footprints around substitutions, especially given the many additional parameters for each if they act in concert (S1D Text). Additional footprints of selection are likely to be needed. Notably, there is likely to be important information about alternative modes of sweeps in diversity levels and patterns of linkage disequilibrium around amino acid polymorphisms [22,80,88].

Another pertinent extension will be to incorporate more realistic demographic assumptions. Like many other methods aimed at quantifying the genome-wide effects of linked selection to date [10,12,18], our model implicitly assumes a panmictic population of constant size. While we focus on a single population, and hence our assumption of random-mating is appropriate, our assumption of a constant size is likely invalid [66,67,89,90]. However, our inference method should be fairly insensitive to changes in the population size, because demographic history should affect different genomic regions similarly, regardless of annotations or other aspects of genomic architecture. Since our method learns about modes of selection and their parameters by contrasting diversity patterns among regions with different properties, it should implicitly control for much of the effects of demography. Having said that, drastic changes in population size could change the efficacy of selection and thus influence our estimates of the distribution of selection coefficients. In addition, regions with different effective population sizes due to linked selection could differ in their transient responses to demographic changes, potentially affecting our inferences. Accounting for these effects is difficult, however. Moreover, existing demographic inferences for North American *D. melanogaster* are confounded by the pervasive effects of linked selection. The methods developed here offer a way forward in inferring demography in the presence of linked selection as our map of linked selection could be factored into such analyses.

While these extensions will be important, our current application to Drosophila already reveals that the effects of linked selection are greater than previously assumed, by taking into account spatial features of genome architecture that were previously averaged out. Even excluding low recombination regions, our results suggest high heterogeneity in expected diversity levels due to linked selection (Fig 6B) and an overall reduction in diversity levels of at least two-fold. Applying our approach to other taxa will reveal whether linked selection is having a similarly large effect in other species, and is an important contributor to the apparent disconnect between census and effective population sizes [2,23–26].

## Supporting Information

**S1 Text. Supporting Online Materials.**
(DOCX)

## Acknowledgments

We thank Yosi Rinott, David Murphy and Guy Amster for helpful discussions and Molly Przeworski for many helpful discussions and comments on the manuscript. We also thank Nick Barton and two anonymous reviewers for many helpful comments on the manuscript.

## Author Contributions

**Conceived and designed the experiments:** EE GS.

**Analyzed the data:** EE.

**Contributed analysis tools:** EE GM SS.

**Wrote the paper:** EE GS.

Prepared the data: EE TTH AS. Wrote Supplementary Section D: GC. Provided input on the analysis and manuscript: PA GC.

## References

1. Hill WG, Robertson A (1966) The effect of linkage on limits to artificial selection. Genet Res 8: 269–294. PMID: 5980116

2. Maynard Smith JM, Haigh J (1974) The hitch-hiking effect of a favourable gene. Genet Res 23: 23–35. PMID: 4407212

3. Kaplan NL, Hudson RR, Langley CH (1989) The "hitchhiking effect" revisited. Genetics 123: 887–899. PMID: 2612899

4. Begun DJ, Aquadro CF (1992) Levels of naturally occurring DNA polymorphism correlate with recombination rates in *D. melanogaster*. Nature 356: 519–520. PMID: 1560824

5. Charlesworth B, Morgan MT, Charlesworth D (1993) The effect of deleterious mutations on neutral molecular variation. Genetics 134: 1289–1303. PMID: 8375663

6. Wiehe TH, Stephan W (1993) Analysis of a genetic hitchhiking model, and its application to DNA polymorphism data from *Drosophila melanogaster*. Mol Biol Evol 10: 842–854. PMID: 8355603

7. Hudson RR (1994) How can the low levels of DNA sequence variation in regions of the Drosophila genome with low recombination rates be explained? Proc Natl Acad Sci USA 91: 6815–6818. PMID: 8041702

8. Hudson RR, Kaplan NL (1995) Deleterious background selection with recombination. Genetics 141: 1605–1617. PMID: 8601498

9. Nordborg M, Charlesworth B, Charlesworth D (1996) The effect of recombination on background selection. Genet Res 67: 159–174. PMID: 8801188

10. Kim Y, Stephan W (2000) Joint effects of genetic hitchhiking and background selection on neutral variation. Genetics 155: 1415–1427. PMID: 10880499

11. Macpherson JM, Sella G, Davis JC, Petrov DA (2007) Genomewide spatial correspondence between nonsynonymous divergence and neutral polymorphism reveals extensive adaptation in Drosophila. Genetics 177: 2083–2099. PMID: 18073425

12. Andolfatto P (2007) Hitchhiking effects of recurrent beneficial amino acid substitutions in the *Drosophila melanogaster* genome. Genome Res 17: 1755–1762. PMID: 17989248

13. Wright SI, Andolfatto P (2008) The Impact of Natural Selection on the Genome: Emerging Patterns in Drosophila and Arabidopsis. Annu Rev Ecol Evol S 39: 193–213.

14. Sella G, Petrov DA, Przeworski M, Andolfatto P (2009) Pervasive natural selection in the Drosophila genome? PLoS Genet 5: e1000495. doi: 10.1371/journal.pgen.1000495 PMID: 19503600

15. Stephan W (2010) Genetic hitchhiking versus background selection: the controversy and its implications. Phil Trans R Soc B 365: 1245–1253. PMID: 20308100.

16. Charlesworth B (2013) Background Selection 20 Years on. J Hered 104: 161–171. doi: 10.1093/jhered/ess136 PMID: 23303522

17. Cutter AD, Payseur BA (2013) Genomic signatures of selection at linked sites: unifying the disparity among species. Nat Rev Genet 14: 262–274. doi: 10.1038/nrg3425 PMID: 23478346

18. McVicker G, Gordon D, Davis C, Green P (2009) Widespread Genomic Signatures of Natural Selection in Hominid Evolution. PLoS Genet 5: e1000471. doi: 10.1371/journal.pgen.1000471 PMID: 19424416

19. Li H, Durbin R (2011) Inference of human population history from individual whole-genome sequences. Nature 475: 493–496. doi: 10.1038/nature10231 PMID: 21753753

20. Sabeti PC, Reich DE, Higgins JM, Levine HZ, Richter DJ, Schaffner SF, Gabriel SB, Platko JV, Patterson NJ, McDonald GJ, Ackerman HC, Campbell SJ, Altshuler D, Cooper R, Kwiatkowski D, Ward R, Lander ES (2002) Detecting recent positive selection in the human genome from haplotype structure. Nature 419: 832–837. PMID: 12397357

21. Nielsen R, Williamson S, Kim Y, Hubisz MJ, Clark AG, Bustamante C (2005) Genomic scans for selective sweeps using SNP data. Genome Res 15: 1566–1575. PMID: 16251466

22. Voight BF, Kudaravalli S, Wen X, Pritchard JK (2006) A map of recent positive selection in the human genome. PLoS Biol 4: e72. PMID: 16494531

23. Lewontin RC (1974) The genetic basis of evolutionary change. New York: Columbia University Press. xiii, 346 p. p.

24. Lynch M (2007) The origins of genome architecture. Sunderland, MA: Sinauer Associates. xvi, 494 p. p.

25. Leffler EM, Bullaughey K, Matute DR, Meyer WK, Segurel L, Venkat A, Andolfatto P, Przeworski M (2012) Revisiting an old riddle: what determines genetic diversity levels within species? PLoS Biol 10: e1001388. doi: 10.1371/journal.pbio.1001388 PMID: 22984349

26. Corbett-Detig RB, Hartl DL, Sackton TB (2015) Natural selection constrains neutral diversity across a wide range of species. PLoS Biol 13: e1002112. doi: 10.1371/journal.pbio.1002112 PMID: 25859758

27. Przeworski M, Coop G, Wall JD (2005) The signature of positive selection on standing genetic variation. Evolution 59: 2312–2323. PMID: 16396172

28. Hermisson J, Pennings PS (2005) Soft sweeps: molecular population genetics of adaptation from standing genetic variation. Genetics 169: 2335–2352. PMID: 15716498

29. Pennings PS, Hermisson J (2006) Soft sweeps II—molecular population genetics of adaptation from recurrent mutation or migration. Mol Biol Evol 23: 1076–1084. PMID: 16520336

30. Pennings PS, Hermisson J (2006) Soft sweeps III: the signature of positive selection from recurrent mutation. PLoS Genet 2: e186. PMID: 17173482

31. Coop G, Ralph P (2012) Patterns of Neutral Diversity Under General Models of Selective Sweeps. Genetics 192: 205–224. doi: 10.1534/genetics.112.141861 PMID: 22714413

32. McVean GA, Charlesworth B (2000) The effects of Hill-Robertson interference between weakly selected mutations on patterns of molecular evolution and variation. Genetics 155: 929–944. PMID: 10835411

33. Gordo I, Navarro A, Charlesworth B (2002) Muller's ratchet and the pattern of variation at a neutral locus. Genetics 161: 835–848. PMID: 12072478

34. Good BH, Walczak AM, Neher RA, Desai MM (2014) Genetic diversity in the interference selection limit. PLoS Genet 10: e1004222. doi: 10.1371/journal.pgen.1004222 PMID: 24675740

35. Payseur BA, Nachman MW (2002) Gene density and human nucleotide polymorphism. Mol Biol Evol 19: 336–340. PMID: 11861892

36. Wright SI, Foxe JP, DeRose-Wilson L, Kawabe A, Looseley M, Gaut BS, Charlesworth D (2006) Testing for effects of recombination rate on nucleotide diversity in natural populations of Arabidopsis lyrata. Genetics 174: 1421–1430. PMID: 16951057

37. Cai JJ, Macpherson JM, Sella G, Petrov DA (2009) Pervasive Hitchhiking at Coding and Regulatory Sites in Humans. PLoS Genet 5: e1000336. doi: 10.1371/journal.pgen.1000336 PMID: 19148272

38. Begun DJ, Holloway AK, Stevens K, Hillier LW, Poh YP, Hahn MW, Nista PM, Jones CD, Kern AD, Dewey CN, Pachter L, Myers E, Langley CH (2007) Population genomics: whole-genome analysis of polymorphism and divergence in Drosophila simulans. PLoS Biol 5: e310. PMID: 17988176

39. Cutter AD, Payseur BA (2003) Selection at linked sites in the partial selfer Caenorhabditis elegans. Mol Biol Evol 20: 665–673. PMID: 12679551

40.  Nordborg M, Hu TT, Ishino Y, Jhaveri J, Toomajian C, Zheng H, Bakker E, Calabrese P, Gladstone J, Goyal R, Jakobsson M, Kim S, Morozov Y, Padhukasahasram B, Plagnol V, Rosenberg NA, Shah C, Wall JD, Wang J, Zhao K, Kalbfleisch T, Schulz V, Kreitman M, Bergelson J (2005) The pattern of polymorphism in *Arabidopsis thaliana*. PLoS Biol 3: e196. PMID: 15907155

41.  Charlesworth B (1996) Background selection and patterns of genetic diversity in *Drosophila melanogaster*. Genet Res 68: 131–149. PMID: 8940902

42.  Charlesworth B (2012) The Role of Background Selection in Shaping Patterns of Molecular Evolution and Variation: Evidence from Variability on the Drosophila X Chromosome. Genetics 191: 233–246. doi: 10.1534/genetics.111.138073 PMID: 22377629

43.  Comeron JM (2014) Background Selection as Baseline for Nucleotide Variation across the Drosophila Genome. PLoS Genet 10: e1004434. doi: 10.1371/journal.pgen.1004434 PMID: 24968283

44.  Nachman MW, Crowell SL (2000) Estimate of the mutation rate per nucleotide in humans. Genetics 156: 297–304. PMID: 10978293

45.  Kondrashov AS (2003) Direct estimates of human per nucleotide mutation rates at 20 loci causing Mendelian diseases. Human mutation 21: 12–27. PMID: 12497628

46.  Roach JC, Glusman G, Smit AF, Huff CD, Hubley R, Shannon PT, Rowen L, Pant KP, Goodman N, Bamshad M, Shendure J, Drmanac R, Jorde LB, Hood L, Galas DJ (2010) Analysis of genetic inheritance in a family quartet by whole-genome sequencing. Science 328: 636–639. doi: 10.1126/science.1186802 PMID: 20220176

47.  Kong A, Frigge ML, Masson G, Besenbacher S, Sulem P, Magnusson G, Gudjonsson SA, Sigurdsson A, Jonasdottir A, Jonasdottir A, Wong WS, Sigurdsson G, Walters GB, Steinberg S, Helgason H, Thorleifsson G, Gudbjartsson DF, Helgason A, Magnusson OT, Thorsteinsdottir U, Stefansson K (2012) Rate of de novo mutations and the importance of father's age to disease risk. Nature 488: 471–475. doi: 10.1038/nature11396 PMID: 22914163

48.  Sattath S, Elyashiv E, Kolodny O, Rinott Y, Sella G (2011) Pervasive Adaptive Protein Evolution Apparent in Diversity Patterns around Amino Acid Substitutions in *Drosophila simulans*. PLoS Genet 7: e1001302. doi: 10.1371/journal.pgen.1001302 PMID: 21347283

49.  Hernandez RD, Kelley JL, Elyashiv E, Melton SC, Auton A, McVean G, Sella G, Przeworski M, Project G (2011) Classic Selective Sweeps Were Rare in Recent Human Evolution. Science 331: 920–924. doi: 10.1126/science.1198878 PMID: 21330547

50.  Williamson R, Josephs EB, Platts AE, Hazzouri KM, Haudry A, Blanchette M, Wright SI (2014) Evidence for widespread positive and negative selection in coding and conserved noncoding regions of Capsella grandiflora. PLoS Genet 10: e1004622. doi: 10.1371/journal.pgen.1004622 PMID: 25255320

51.  Mackay TF, Richards S, Stone EA, Barbadilla A, Ayroles JF, Zhu D, Casillas S, Han Y, Magwire MM, Cridland JM, Richardson MF, Anholt RR, Barron M, Bess C, Blankenburg KP, Carbone MA, Castellano D, Chaboub L, Duncan L, Harris Z, Javaid M, Jayaseelan JC, Jhangiani SN, Jordan KW, Lara F, Lawrence F, Lee SL, Librado P, Linheiro RS, Lyman RF, Mackey AJ, Munidasa M, Muzny DM, Nazareth L, Newsham I, Perales L, Pu LL, Qu C, Ramia M, Reid JG, Rollmann SM, Rozas J, Saada N, Turlapati L, Worley KC, Wu YQ, Yamamoto A, Zhu Y, Bergman CM, Thornton KR, Mittelman D, Gibbs RA (2012) The *Drosophila melanogaster* Genetic Reference Panel. Nature 482: 173–178. doi: 10.1038/nature10811 PMID: 22318601

52.  Barton NH (1998) The effect of hitch-hiking on neutral genealogies. Genet Res 72: 123–133.

53.  Gillespie JH (2000) Genetic drift in an infinite population. The pseudohitchhiking model. Genetics 155: 909–919. PMID: 10835409

54.  Kim Y, Stephan W (2003) Selective sweeps in the presence of interference among partially linked loci. Genetics 164: 389–398. PMID: 12750349

55.  Hudson RR (2001) Two-locus sampling distributions and their application. Genetics 159: 1805–1817. PMID: 11779816

56.  Fearnhead P (2003) Consistency of estimators of the population-scaled recombination rate. Theor Popul Biol 64: 67–79. PMID: 12804872

57.  Wiuf C (2006) Consistency of estimators of population scaled parameters using composite likelihood. J Math Biol 53: 821–841. PMID: 16960689

58.  Hu TT, Eisen MB, Thornton KR, Andolfatto P (2013) A second-generation assembly of the *Drosophila simulans* genome provides new insights into patterns of lineage-specific divergence. Genome Res 23: 89–98. doi: 10.1101/gr.141689.112 PMID: 22936249

59.  St Pierre SE, Ponting L, Stefancsik R, McQuilton P, FlyBase C (2014) FlyBase 102—advanced approaches to interrogating FlyBase. Nucleic acids res 42: D780–788. doi: 10.1093/nar/gkt1092 PMID: 24234449

**60.** Andolfatto P (2005) Adaptive evolution of non-coding DNA in Drosophila. Nature 437: 1149–1152. PMID: 16237443

**61.** Haddrill PR, Charlesworth B, Halligan DL, Andolfatto P (2005) Patterns of intron sequence evolution in Drosophila are dependent upon length and GC content. Genome Biol 6: R67. PMID: 16086849

**62.** Halligan DL, Keightley PD (2006) Ubiquitous selective constraints in the Drosophila genome revealed by a genome-wide interspecies comparison. Genome Res 16: 875–884. PMID: 16751341

**63.** Casillas S, Barbadilla A, Bergman CM (2007) Purifying selection maintains highly conserved noncoding sequences in Drosophila. Mol Biol Evol 24: 2222–2234. PMID: 17646256

**64.** Comeron JM, Ratnappan R, Bailin S (2012) The Many Landscapes of Recombination in *Drosophila melanogaster*. PLoS Genet 8: e1002905. doi: 10.1371/journal.pgen.1002905 PMID: 23071443

**65.** Arlot S, Celisse A (2010) A survey of cross-validation procedures for model selection. Statist Surv 4: 40–79.

**66.** Thornton K, Andolfatto P (2006) Approximate Bayesian inference reveals evidence for a recent, severe bottleneck in a Netherlands population of *Drosophila melanogaster*. Genetics 172: 1607–1619. PMID: 16299396

**67.** Li H, Stephan W (2006) Inferring the demographic history and rate of adaptive substitution in Drosophila. PLoS Genet 2: e166. PMID: 17040129

**68.** Loewe L, Charlesworth B (2007) Background selection in single genes may explain patterns of codon bias. Genetics 175: 1381–1393. PMID: 17194784

**69.** Callahan B, Neher RA, Bachtrog D, Andolfatto P, Shraiman BI (2011) Correlated Evolution of Nearby Residues in Drosophilid Proteins. PLoS Genet 7: e1001315. doi: 10.1371/journal.pgen.1001315 PMID: 21383965

**70.** Kolaczkowski B, Kern AD, Holloway AK, Begun DJ (2011) Genomic differentiation between temperate and tropical Australian populations of *Drosophila melanogaster*. Genetics 187: 245–260. doi: 10.1534/genetics.110.123059 PMID: 21059887

**71.** Lee YCG, Langley CH, Begun DJ (2014) Differential Strengths of Positive Selection Revealed by Hitch-hiking Effects at Small Physical Scales in *Drosophila melanogaster*. Mol Biol and Evol 31: 804–816. doi: 10.1093/molbev/mst270 PMID: 24361994

**72.** Smith NG, Eyre-Walker A (2002) Adaptive protein evolution in Drosophila. Nature 415: 1022–1024. PMID: 11875568

**73.** Fay JC, Wyckoff GJ, Wu CI (2002) Testing the neutral theory of molecular evolution with genomic data from Drosophila. Nature 415: 1024–1026. PMID: 11875569

**74.** Wilson DJ, Hernandez RD, Andolfatto P, Przeworski M (2011) A population genetics-phylogenetics approach to inferring natural selection in coding sequences. PLoS Genet 7: e1002395. doi: 10.1371/journal.pgen.1002395 PMID: 22144911

**75.** Maruyama T (1974) The age of a rare mutant gene in a large population. Am J Hum Genet 26: 669–673. PMID: 4440678

**76.** Przeworski M (2002) The signature of positive selection at randomly chosen loci. Genetics 160: 1179–1189. PMID: 11901132

**77.** Coop G, Pickrell JK, Novembre J, Kudaravalli S, Li J, Absher D, Myers RM, Cavalli-Sforza LL, Feldman MW, Pritchard JK (2009) The role of geography in human adaptation. PLoS Genet 5: e1000500. doi: 10.1371/journal.pgen.1000500 PMID: 19503611

**78.** Pritchard JK, Pickrell JK, Coop G (2010) The genetics of human adaptation: hard sweeps, soft sweeps, and polygenic adaptation. Curr Biol 20: R208–215. doi: 10.1016/j.cub.2009.11.055 PMID: 20178769

**79.** Berg JJ, Coop G (2014) A Population Genetic Signal of Polygenic Adaptation. PLoS Genet 10: e1004412. doi: 10.1371/journal.pgen.1004412 PMID: 25102153

**80.** Garud NR, Messer PW, Buzbas EO, Petrov DA (2015) Recent Selective Sweeps in North American *Drosophila melanogaster* Show Signatures of Soft Sweeps. PLoS Genet 11: e1005004. doi: 10.1371/journal.pgen.1005004 PMID: 25706129

**81.** Berg JJ, Coop G (2015) A Coalescent Model for a Sweep of a Unique Standing Variant. Genetics 201: 707–725. doi: 10.1534/genetics.115.178962 PMID: 26311475

**82.** Haag-Liautard C, Dorris M, Maside X, Macaskill S, Halligan DL, Houle D, Charlesworth B, Keightley PD (2007) Direct estimation of per nucleotide and genomic deleterious mutation rates in Drosophila. Nature 445: 82–85. PMID: 17203060

**83.** de Vladar HP, Barton NH (2011) The statistical mechanics of a polygenic character under stabilizing selection, mutation and drift. J R Soc, Interface 8: 720–739. doi: 10.1098/rsif.2010.0438 PMID: 21084341

84. de Vladar HP, Barton N (2014) Stability and Response of Polygenic Traits to Stabilizing Selection and Mutation. Genetics 197: 749–767. doi: 10.1534/genetics.113.159111 PMID: 24709633

85. Santiago E, Caballero A (1998) Effective size and polymorphism of linked neutral loci in populations under directional selection. Genetics 149: 2105–2117. PMID: 9691062

86. Robertson A (1961) Inbreeding in Artificial Selection Programmes. Gene Res 2: 189–194.

87. Green P, Ewing B (2013) Comment on "Evidence of abundant purifying selection in humans for recently acquired regulatory functions". Science 340: 682. PMID: 23661742

88. Kim Y, Nielsen R (2004) Linkage disequilibrium as a signature of selective sweeps. Genetics 167: 1513–1524. PMID: 15280259

89. Campo D, Lehmann K, Fjeldsted C, Souaiaia T, Kao J, Nuzhdin SV (2013) Whole-genome sequencing of two North American Drosophila melanogaster populations reveals genetic differentiation and positive selection. Mol Ecol 22: 5084–5097. doi: 10.1111/mec.12468 PMID: 24102956

90. Duchen P, Zivkovic D, Hutter S, Stephan W, Laurent S (2013) Demographic Inference Reveals African and European Admixture in the North American Drosophila melanogaster Population. Genetics 193: 291–301. doi: 10.1534/genetics.112.145912 PMID: 23150605

# A genomic map of the effects of linked selection in Drosophila

Eyal Elyashiv[1,2,*], Shmuel Sattath[1], Tina T. Hu[3], Alon Strustovsky[1], Graham McVicker[4], Peter Andolfatto[3], Graham Coop[5] and Guy Sella[2,*]

1 Department of Ecology, Evolution and Behavior, Hebrew University of Jerusalem, Jerusalem, Israel

2 Department of Biological Sciences, Columbia University, New York, New York, USA

3 Department of Ecology and Evolutionary Biology and the Lewis-Sigler Institute for Integrative Genomics, Princeton University, Princeton, New Jersey, USA

4 Department of Genetics, Stanford University, California, USA

5 Department of Evolution and Ecology, University of California, Davis, California, USA

[*] Corresponding authors: eyalshiv@yahoo.com and gs2747@columbia.edu.

# Table of Contents

# A. Drosophila data set

**Overview**. Our approach relies on data from neutral sites to estimate the effects of linked selection on diversity levels. As a proxy for neutral sites, we use synonymous variation. Synonymous diversity is measured using re-sequencing data from the *Drosophila* Genetic Reference Panel (DGRP)[1]. The number of synonymous substitutions per codon, used to control for local variation in mutation rates, is estimated from the aligned reference genomes of *D. melanogaster*, *D. simulans* and *D. yakuba* [2]. Among possible choices, synonymous variation reflects a good compromise between the amount of data—since coding regions composing ~20% of the *D. melanogaster* euchromatic genome [3]—and the attempt to minimize the effects of direct selection on the sites [4-8]. In Section H we consider the robustness of our results when instead considering subsets of synonymous polymorphisms that should be even less affected by selection (such as synonymous codon bias).

Our inference further relies on knowledge of the locations and annotations of sites under negative (purifying) and positive selection—the sources of diversity reduction at linked neutral sites. As potential targets of selection, we consider coding regions; untranslated, transcribed regions (UTRs); and long introns and intergenic regions, all of which have been inferred to be under widespread purifying and positive selection in *D. melanogaster* [9-11]. Together, these annotations, which we downloaded from FlyBase [12], cover 98.5% of the euchromatic genome. We consider substitutions along the *D. melanogaster* lineage from the common ancestor with *D. simulans* in any of these annotations (with the exception of synonymous sites, and intergenic regions where we do not have data) to be putative targets of sweeps. We infer the substitutions from the three species alignment of reference genomes from *D. melanogaster*, *D. simulans* and *D. yakuba* [2]. More detail for each of these steps is provided below.

**Genetic maps**. We use the genetic map for *D. melanogaster* estimated by Comeron et al. [13], which is based on ~6000 female meioses, providing estimates at 100 kb resolution (http://bioweb.biology.uiowa.edu/labs/comeron/recombination/).

We note that while the genetic map inferred by Chan et al. [14] from patterns of linkage disequilibrium provides higher resolution, we cannot use it for our purposes. Chan et al. infer the population scaled recombination rate $\rho = 4N_e c$, where $N_e$ is the local effective population size. Because linked selection causes the effective population size to vary along the genome (and our inference suggests that this variation is considerable; cf. Fig 6 in the main text), the genetic map that they estimate confounds the effects of linked selection with variation in recombination rates, and thus hard to interpret in our inference framework.

Because linked selection affects diversity levels over a scale proportional to $1/c$, where $c$ is the recombination rate per bp, our estimates are sensitive to errors in the recombination rate in regions of low recombination. Moreover, our modeling assumption that interference among selective sweeps is negligible is more likely to be violated in these regions. We therefore exclude neutral polymorphism from regions with a sex-averaged recombination rate below 0.75 cM/Mb as well as centromeric and telomeric regions (i.e., 5% at either end of each arm in physical distance), which have low recombination rates in *D. melanogaster*. We do, however, use the information within these regions (i.e., the positions of annotations and substitutions within them and the number of synonymous substitutions per codon used to control for variation in mutation rates) to generate our predictions of diversity levels. In Section H, we show that our inference is largely robust to the choice of recombination rate threshold.

***Annotations.*** The genomic positions for exonic, intronic and untranslated, transcribed regions (UTRs) are from FlyBase [12], for release 5.33 of the *D. melanogaster* genome ([ftp://ftp.flybase.net/genomes/Drosophila_melanogaster/dmel_r5.33_FB2011_01/](ftp://ftp.flybase.net/genomes/Drosophila_melanogaster/dmel_r5.33_FB2011_01/)). The positions for UTRs are used as is. From the set of intronic regions, we remove short introns (<80 bp), because they appear to be under weak or no selection in *D. melanogaster* [10,15,16]; the remaining positions comprise our "long intron" annotation. In the exonic class, we include only the longest transcript for each gene and, in the rare cases in which genes overlap, we include only the one for which the maximal transcript is longer. All regions between neighboring transcripts are classified as intergenic. Together, the four labels cover 98.5% of the euchromatic autosomal genome, with 18.3% in the exonic class (consisting of 11,447 transcripts), 5.6% in UTRs, 39.2% in long introns, and 35.4% in intergenic regions.

Because we use synonymous polymorphism and divergence to estimate the effects of linked selection on neutral diversity levels, it is especially important to minimize the erroneous identification of codons. To do so, we restrict ourselves to the experimentally validated transcripts in the Gold set of the *Drosophila* Gene Collection (DGC) [17], consisting of 9,358 autosomal longest transcripts and amounting to 3,831,195 codons. Also, we exclude the first (start) and last (stop) codons from each transcript and codons that are split between exons.

***Substitutions and divergence rates.*** To identify substitutions in annotations that are possible targets of selection, we rely on the Hu et al. [2] ([http://genomics.princeton.edu/AndolfattoLab/w501_genome.html](http://genomics.princeton.edu/AndolfattoLab/w501_genome.html)) multiple sequence genic alignments of *D. yakuba*, *D. melanogaster* and *D. simulans* and on their reconstruction of the most recent common ancestor of *D. melanogaster* and *D. simulans*. Specifically, substitutions along the *D. melanogaster* lineage in transcribed regions (i.e., exons, UTRs and long introns) are inferred from sequence differences between the *D. melanogaster* reference genome and its reconstructed ancestor with *D. simulans* (thus neglecting multiple hits). The three species alignment covers only 70% of exons (with 64,205 non-synonymous substitutions, corresponding to 0.0092 per bp), 45% of UTRs (with 153,765 substitutions, corresponding to 0.025 per bp) and 43% of long introns (with 456,401 corresponding to 0.032 per bp). We describe how we handle missing data in Section B.

Because the inferred substitutions are based on a single reference genome from each species, some of them are in fact polymorphic in *D. melanogaster* (~18%), likely resulting in moderate underestimates of the proportion of beneficial substitutions. We use this definition rather than also considering polymorphism data because, under neutrality, conditioning on a site not being polymorphic distorts diversity levels nearby, introducing potential artifacts into our inferences.

We rely on the same multiple species alignment in order to estimate local number of synonymous substitutions per codon and correct for variation in mutation rates along the genome. For this purpose, we use only codons in the Gold set with the aforementioned filters. We estimate the number of synonymous substitutions in non-overlapping window using CODEML (for details see Section B). Specifically, we use the estimate of the number of synonymous substitutions between the common ancestor of *D. simulans* and *D. yakuba* (for which there is an average synonymous divergence of 0.146 synonymous substitutions per codon, or ~0.2 per synonymous site) to minimize statistical dependencies with our polymorphism measurements.

***Polymorphism data.*** We measure synonymous polymorphism using data from the *Drosophila Genetic Reference Panel* (DGRP) [1] ([http://dgrp.gnets.ncsu.edu/](http://dgrp.gnets.ncsu.edu/)). In brief, flies were collected from a farmers market in Raleigh, North Carolina, and inbred for 20 generations to generate 162 lines that are mostly isogenic (the average fraction of heterozygous exonic sites per line is 0.4%, with a maximum of 3.3%). The lines were sequenced using Illumina at an average coverage of ~20X and polymorphisms were called using the Joint Genotyper for Inbred Lines (JGIL)[18].

Some of the lines in the DRGP are closely related [14,19]. Given that we assume a random-mating population, we removed a subset of the closely related lines, as follows. First, we calculate the average pairwise differences per site among all pairs of lines. Based on the median of the distribution, we estimate that a threshold of <0.28% per bp corresponding to first cousins (i.e., 7/8 of the median distance across all pairs). We then apply a sequential algorithm in which, at each step, we remove one of the lines from the most closely related pair and recalculate the distribution, until none of the remaining pairs are below the threshold. After this process, 125 lines remain.

The set of codons used to measure pairwise synonymous differences are further filtered as follows. For codons that are heterozygous in a given line, we randomly sample one of the alleles and use it throughout our pairwise comparisons. We further exclude codons that have more than two alleles in our sample or for which the two alleles differ at more than one position (only 6% of polymorphic codons violated these conditions). Lastly, we retain only codons for which we have divergence data (reducing the number of codons by 52%).

After applying the filters, we have polymorphism data at 1,775,362 codons, ~45% of those that met our recombination threshold. At these codons, we have an average sample size of 124 lines, and fewer than 0.5% with sample size <100 lines. Average synonymous heterozygosity per codon is ~0.6% (which is ~0.8% per synonymous site).

### B. The inference procedure

***Estimating the local mutation rate.*** We use synonymous divergence data to estimate local mutation rates. First, we divide each chromosome into non-overlapping windows of 1780 bp for estimation (see below for the justification of window size). For each window that contains divergence data for at least 50 codons, we estimate the relative mutation rate at its mid-point by

$$\frac{\hat{u}}{\bar{u}} = \frac{\hat{K}_S}{\overline{K}_S},$$

where $\hat{u}$ is our estimate, $\bar{u}$ is the average mutation rate across windows, $\hat{K}_S$ is an estimate of the number of synonymous substitutions per codon in the window and $\overline{K}_S$ is the number of substitutions per codon, averaged across windows. We estimate $\hat{K}_S$ using the three species alignment from Hu et al. [2] for the codons in the window and apply CODEML [20] to obtain the number of substitutions between *D. yakuba* and *D. simulans* (with CODEML parameters: runmode = 0; seqtype = 1; CodonFreq = 2; clock = 0; model = 1; NSsites = 0; icode = 0; fix_kappa = 0; kappa = 1.6; fix_omega = 0; ncatG = 1; getSE = 0; RateAncestor = 2; Small_Diff = $3 \times 10^{-7}$; cleandata = 0; method = 0). Requiring 50 codons or more per window amounts to a relative sampling error of less than ~4%. To obtain point estimates at every genomic position, we use linear interpolation between the two closest flanking estimates.

The choice of window size should minimize sampling error while not masking true variation in mutation rates, as systematic changes in mutation rate near selected annotations (e.g., due to differences in base composition) could potentially bias our estimates of selection parameters. We therefore use a number of approaches to assess the effects of window size. First, we examine how the maximum composite likelihood and $R^2$ between predicted and observed diversity levels vary with window size (Fig S1A and S1B). These estimates can be viewed as measuring our ability to predict diversity levels along the genome and so we reason that a window size that best captures the true variation in mutation rates would lead to better predictions. The likelihood estimates indicate that window sizes between 1000-3000 bp provide a good balance between true variation and sampling error, and estimates of $R^2$ are also maximized within this range.

Second, we examine how the choice of window size affects estimates of positive selection parameters, i.e., the fractions of substitutions with a given selection coefficient. As can be seen (Fig S1C), using large window sizes (above our 1780 bp grid point) leads to lower estimates of the fraction of beneficial substitutions with intermediate selection coefficients (e.g., $s=10^{-3.5}$). This can be understood as follows. If functional substitutions tend to occur in regions with higher mutation rates, then we would also expect synonymous diversity and the number of substitutions per codon to be elevated in their vicinity. If we use too large a window size, we will fail to capture heterogeneity in mutation rates on spatial scales smaller than the chosen size, such that after dividing out the number of synonymous substitutions per codon, scaled diversity levels near substitutions at these smaller scales would appear to be greater than they should be. The result would be an underestimation of the effects of sweeps on these spatial scales, or more precisely, an underestimation of the fraction of sweeps with weak selection coefficients. Our window size of 1780 bp is within the range suggested by the goodness-of-fit statistics (Fig S1A) but is also sufficiently small such that our estimates for the fraction of beneficial substitutions corresponding to different selection coefficients appears to be stable. We note that there may well be mutation rate heterogeneity below this scale, which may lead us to slightly underestimate the proportion of substitutions associated with weak sweeps.

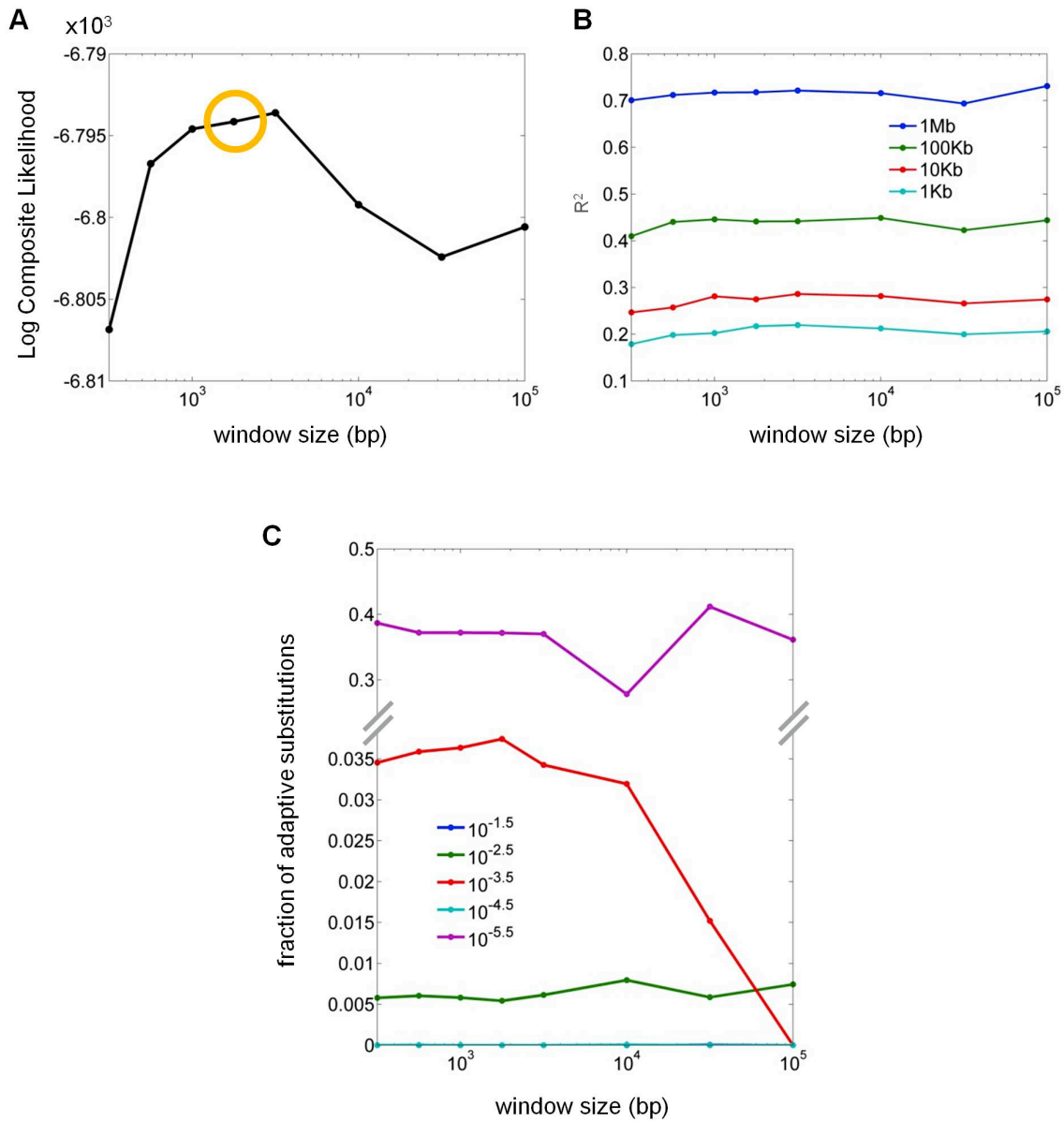

**Fig S1. Choosing a window size in which to estimate the mutation rate.** (**A**) The maximum likelihood as a function of window size. (**B**) $R^2$ as a function of the choice of window size, over several scales. (**C**) Estimated fractions of beneficial amino-acid substitutions with a given selection coefficient as a function of window size, for different selection coefficients. In all cases, we use the model combining background selection and classic sweeps. The circles denote the window size.

***Handling missing data***. Due to incomplete alignments and the application of quality filters, the positions of putatively selected substitutions cover only a subset of the genome. Yet substitutions in regions of missing data are likely to impact neutral diversity levels and their absence could affect parameter estimates and predictions. One possibility is to ignore this problem. Another is to add substitutions in regions with missing data [21]. To choose among these strategies, we compare the maximum likelihood values and $R^2$ statistics of diversity along the genome under the two approaches (relying on the same reasoning that we use in choosing the window size to estimate mutation rates). To incorporate missing substitutions, we randomly pick their positions in regions with missing data, with a number of substitutions per codon chosen based on the average over regions with the same annotation within 200 kb. The results suggest that inferences incorporating missing substitutions fit the data slightly better (Table S1).

***Calculating the effects of linked selection***. As detailed in the Results, we calculate the expected diversity levels along the genome given a set of selection parameters in two steps. First, we evaluate the functions $b(x|\ t_g, i_B)$ and $s(x|\ s_k, i_S)$ at each position $x$ in the genome, for each selection coefficient on the predefined grid ($g=1, \ldots, G$ and $k=1,\ldots, K$) and selected annotations ($i_B=1,\ldots,I_B$ and $i_S=1,\ldots,I_S$). These functions are pre-calculated and then used to obtain the expected diversity levels given a set of selection parameters (as specified in Equations (6) and (7)). Here, we specify how we calculate the functions $b(x|\ t_g, i_B)$ and $s(x|\ s_k, i_S)$.

## A. Goodness-of-fit measures.

| Model | | Background selection and classic sweeps with added "missing substitutions" | Background selection and classic sweeps without "missing substitutions" |
|---|---|---|---|
| $\Delta CL$ | | $3.9 \times 10^{-4}$ | $3.8 \times 10^{-4}$ |
| $R^2$ | 1 Mb | 0.71 | 0.70 |
| | 100 kb | 0.44 | 0.44 |
| | 10 kb | 0.26 | 0.25 |
| | 1 kb | 0.20 | 0.20 |

## B. Parameter estimates.

| Model | Background selection and classic sweeps with added "missing substitutions" | | | | Background selection and classic sweeps without "missing substitutions" | | | |
|---|---|---|---|---|---|---|---|---|
| $k_{B\&S}\,(1-\bar{\pi}/\pi_0)$ | 73% | | | | 74% | | | |
| Annotation | Exons | UTRs | Introns | Intergenic | Exons | UTRs | Introns | Intergenic |
| Background selection parameters | | | | | | | | |
| $u(t=10^{-1.5})\,/\,\mu$ | 377% | 577% | 19% | - | 273% | 695% | 55% | 2% |
| $u(t=10^{-2.5})\,/\,\mu$ | 2% | 2% | - | - | 1% | 4% | - | - |
| $u(t=10^{-3.5})\,/\,\mu$ | 56% | - | - | - | 75% | - | - | - |
| $u(t=10^{-4.5})\,/\,\mu$ | 2% | 23% | - | - | - | 6% | 1% | - |
| $u(t=10^{-5.5})\,/\,\mu$ | - | 2% | - | - | - | 1% | - | - |
| Classic sweeps parameters | | | | | | | | |
| $\alpha(s=10^{-1.5})$ | - | - | - | - | - | - | - | - |
| $\alpha(s=10^{-2.5})$ | 0.6% | - | - | - | 1.1% | - | - | - |
| $\alpha(s=10^{-3.5})$ | 3.5% | - | - | - | 2.0% | - | - | - |
| $\alpha(s=10^{-4.5})$ | - | 5.1% | - | - | 3.8% | 6.8% | - | - |
| $\alpha(s=10^{-5.5})$ | 36.3% | 42.1% | - | - | 36.5% | 34.5% | - | - |

1  **Table S1. Results with and without adding missing substitutions.** Goodness-of-fit measures (**A**)
2  and parameter estimates (**B**) obtained using the joint model with background selection and classic
3  sweeps. In (**A**), $\Delta CL$ is the difference between the model's likelihood and the likelihood of a neutral
4  model. $R^2$ is measured in non-overlapping windows of 1Mb, 100kb, 10kb and 1kb. In (**B**):
5  $k_{B\&S}\left(1-\bar{\pi}/\pi_0\right)$ is the average estimated reduction in diversity; the fraction of deleterious
6  mutations with a given selection coefficient is measured relative to our estimate for the overall
7  mutation rate $\mu=6.8\times10^{-9}$ per bp (see Section E); $\alpha$ denotes the fraction of substitutions driven by
8  sweeps with a given selection coefficient.

Our calculation of $b(x|\ t_g,\ i_B)$ follows McVicker et al. (2009) [22] and relies on their code. In brief, they consider a grid of selected segment lengths and a grid of genetic distances from the segment, and calculate $b$ values over these grids using Equation (2) [23]. For a grid of positions along the genome, they then calculate the $b$ value by summing over the effects of conserved segments, where the effect of a given segment is calculated based on bi-linear interpolation, using the grids over segment lengths and distances from the previous step. The spacing of grid positions along the genome is determined based on the first two derivatives of previous points, in order to ensure that changes in $b$ values between consecutive points are not too large. Finally, $b$ values between grid points are calculated based on linear interpolation. See McVicker et al. (2009) for further details.

The calculation of $s(x|\ s_k,\ i_S)$ is similar in spirit but simpler, because it depends on the discrete positions of substitutions and we have a closed form description of the effects of a sweep as a function of the selection coefficient and genetic distance (Equation 3). For a given selection coefficient $s_k$, we calculate $s(x|\ s_k,\ i_S)$ for a grid of positions along the genome, where grid points are spaced according to the maximum of $3.3 \times 10^{-6}$ and $1 \cdot s_k$ cM. This spacing is chosen because $3.3 \times 10^{-6}$ cM corresponds to the average genetic length of a single codon and the effect of a classic sweep with selection coefficient $s_k$ extends over a distance of $10 \cdot s_k$ cM. At each grid point, we sum over the effects of substitutions within a genetic distance of $100 \cdot s_k$ cM given by Equation 3. Classic sweeps beyond this distance have a negligible effect on diversity levels [24]. The value of $s(x|\ s_k,\ i_S)$ between grid points is calculated by linear extrapolation between values at flanking grid points.

***Likelihood maximization***. We maximize the likelihood function using a combination of optimization algorithms. We begin with each of the weights corresponding to the grid of selection coefficients set to a small positive value, amounting to a negligible deviation from neutrality, and with $\pi_0$ set to the observed genome-wide average. We then run three consecutive programs from the Matlab Optimization Toolbox: Active-Set, Interior-Point and Sequential-Quadratic-Programming [25]. We use the estimates obtained from the previous program as initial conditions for the next one, unless the run fails (either because the likelihood does not improve or because the exit flags reported by the output indicated an unsuccessful run). We consider the maximization successful if at least one of the programs completed successfully and, if this is the case, we use the parameter estimates obtained by the last successful one.

This algorithm was developed to assure that our parameter estimates are close to the maximum likelihood estimate (MLE) despite of the high dimensionality of the parameter space (with up to 78 parameters for the most complex model). We check the reliability of the algorithm in three sets of analyses:

**1. Choosing the sequence of optimization algorithms.** While using a single-algorithm or a different sequence of algorithms does not always lead to a successful maximization, the chosen sequence is successful for all the models described in the text. Moreover, in almost all cases all three algorithms complete successfully. Specifically, we confirm that this is the best of the six possible orderings of algorithms for the three main models: (i) background selection with four annotations, classic sweeps with three annotations, and a grid of five selection coefficients for each mode and annotation, (ii) the same for background selection alone, and (iii) the same for classic sweeps alone.

**2. More complex models lead to greater likelihood.** One indication that the algorithm converges to the MLE is that, for nested models, the maximum likelihood is always greater (or equal) for the more complex model. We confirm that this is the case using the following set of nested models: (i) the full model of background selection with four annotations and classic sweeps with three annotations, with 11 point masses; (ii) the same model with 5 point masses; (iii) the models for background selection and sweeps alone with 5 point masses; and (iv) the models with background selection and sweeps alone with 5 point masses and a single (exonic) annotation.

**3. Convergence to the same MLE from different initial conditions.** Another indication that the algorithm converges to the MLE is that it arrives at very similar parameter estimates based on a variety of initial conditions. While an exhaustive examination of initial conditions is infeasible, we run the inference using a series of initial conditions that should sample very different parts of the parameter space. We do so for the most complex model, including background selection with four annotations, classic sweeps with three annotations and a grid of 11 point masses for each annotation. The initial conditions that we examine include (i) no selection; (ii) strong selection (leading to a substantial reduction in diversity due to linked selection) with equal weights on selection coefficients for all masses and all annotations; and (iii) estimates inferred for simpler models, including a similar model but with 5 point masses and a similar model with background selection and sweeps alone and 5 point masses. From almost all initial conditions, the algorithm converges to very similar parameter estimates, as well as likelihoods and goodness of fit measures (i.e., $R^2$ in windows of 1 kb to 1 Mb; see Section C). The sole exception is when we begin with no selection, i.e. with null values for all weights, in which the maximization leads to a sub-optimal solution, presumably because the algorithms encounter difficulty starting the search from initial conditions on the boundary of the parameter space. When we modify the weights slightly to positive but tiny values that do not cause any appreciable effect of linked selection, the maximization converges to the same estimates obtained using the other initial conditions. Because this form of 'no selection' condition is the most conservative with respect to the effects of linked selection, we choose it as the standard initial condition for the maximization procedure.

## *C. Statistical analyses*

*Details about summaries and figures*.

**Comparing predicted and observed diversity levels in windows (Fig 2 in the main text)**. Each of the major autosomes (2 and 3) is divided into non-overlapping windows of 1 Mb, 100 kb, 10 kb or 1 kb. Observed and predicted average, scaled heterozygosity are then calculated for each window with data at more than 50 codons. Specifically, we calculate the observed levels by dividing the average synonymous heterozygosity per codon by the average number of synonymous substitutions per codon (see Section B). Predicted average levels are calculated using our predictions at the same codons. We calculate the $R^2$ for the model in which observed scaled diversity levels equal their predicted values (y=x) across windows, where the weight for each window equals the number of codons in it.

**Average diversity levels around substitutions (Fig 3, 4 and 5 in the main text)**. We divide the genetic distance between the focal substitution and a maximal distance of 0.11 cM into $10^{-6}$ cM bins. For each bin, we calculate the observed scaled diversity level (as detailed above) using all the codons that are found at the specified distance from a substitution. Similarly, we obtain the levels predicted based on our method by averaging the predictions over the same set of codons. Predicted levels at each distance based on the Sattath et al. method are calculated directly from their coalescent model.

**Observed diversity levels as a function of predicted levels (Fig 6 in the main text)**. For each model, we order codons according to the predicted diversity level and then divide them into bins with equal amounts of data. For the Wiehe-Kim-Stephan models, predicted levels are determined either by the local recombination rate or by the local density of non-synonymous substitutions; these are calculated in 0.03 cM non-overlapping windows along the genome and linearly interpolated between window midpoints. Observed and predicted diversity levels are then calculated for each bin, as described above. While we conduct this analysis using 25, 100, 400 and 1600 bins, we only show the graphs for 100 bins and the correlations for 1600 bins.

*Leave-one-out cross-validation analysis*. Because we use the same data to infer parameters and evaluate our predictions, over-fitting might inflate our goodness-of-fit estimates. We use a leave-one-out cross-validation (LOOCV) analysis [26] to assess the extent of this problem. Specifically: i) we divide the genome into non-overlapping windows of 1 Mb; ii) dropping each

1  window in turn, we infer selection parameters using the rest of the genome, excluding codons in the

2  window and at distance < 0.5 Mb from it (to avoid correlations between diversity levels at the

3  edges); iii) we use the inferred selection parameters to predict diversity levels within that window;

4  iv) by combining predictions over all windows, we derive a genome-wide map of predicted

5  diversity levels, where the prediction at any site does not rely on polymorphism data at that site or

6  anywhere in its vicinity. In Table S2, we compare our predictions on different spatial scales based

7  on the entire dataset and on LOOCV. We do so for both the recombination threshold of 0.75 cM/Mb

8  used in the inference and for the recombination threshold of 0.1 cM/Mb used in the main text

9  (where polymorphism data for regions with recombination rate < 0.75 cM/Mb are not used in the

10  inference). The results with LOOCV and the entire dataset are very similar, showing that over-fitting

11  has little effect on our results.

| | BS & CS (> 0.10 cM/Mb ) | | BS & CS (> 0.75 cM/Mb ) | |
|---|---|---|---|---|
| | LOOCV | Full | LOOCV | Full |
| **1 Mb** | 70% | 71% | 59% | 60% |
| **100 Mb** | 43% | 44% | 28% | 29% |
| **10 Kb** | 27% | 26% | 17% | 18% |
| **1 Kb** | 21% | 20% | 13% | 14% |

12  **Table S2. Leave-one-out cross-validation (LOOCV) analysis.**

13  ***Quantifying the relative contribution of background selection and sweeps.*** In the

14  Discussion, we consider the relative effects of different modes of linked selection and specifically

15  those of background selection and classic sweeps. One way to quantify these contributions relies on

16  our model for expected heterozygosity. Notably, based on Equation 1 and measuring time in units

17  of $2N_e$ generations, the coalescence rates due to genetic drift equals 1 throughout the autosomes,

18  the increase in rate at autosomal position $x$ due to background selection is $B^{-1}(x) - 1$ and the

19  increase due to classic sweeps is $2N_e S(x)$. The genome-wide average of these rates $r_B = \overline{B^{-1}} - 1$

20  and $r_S = 2N_e \overline{S}$ therefore provide a natural additive measure for the effect of each mode, with the

1   ratios $1 : r_B : r_S$ quantifying the relative contribution of drift, background selection and classic

2   sweeps and $r_B + r_S$ quantifying the total contribution of linked selection.

3   Although these measures are natural in some respects, they also have some limitations. For

4   instance, in the Discussion, we break up the contribution of background selection due to strong and

5   moderate selection. We cannot do the same with the above measures because the effects of

6   background selection due to different selection coefficients combine multiplicatively (we could,

7   however, do so for sweeps). Also, quantifying the relative contributions of background selection

8   and classic sweeps in terms of average coalescence rates could be misleading because the averages

9   might be dominated by regions with high coalescence rates (e.g., with low recombination) in which

10  diversity levels would be low even if only one mode of linked selection were present.

11  We therefore consider a second measure that quantifies the effects on diversity levels more

12  directly. For that purpose, we rely on our parameter estimates to build maps predicting diversity

13  levels due to classic sweeps or background selection alone, considering the parameter estimates

14  derived for the joint model; by the same token, we can also build maps corresponding to

15  background selection due to strong or moderate selection coefficients. Based on these maps, we

16  calculate the average relative reduction in heterozygosity if only background selection were

17  present, $k_B = 1 - \dfrac{\overline{\pi}_B}{\pi_0}$, or only classic sweeps, $k_S = 1 - \dfrac{\overline{\pi}_S}{\pi_0}$, or both, $k_{B\&S} = 1 - \dfrac{\overline{\pi}_{B\&S}}{\pi_0}$. Then we

18  quantify the relative reduction due to background selection by $\dfrac{k_B}{k_B + k_S}$ and due to sweeps by

19  $\dfrac{k_S}{k_B + k_S}$. Similar definitions are used for subsets of selection coefficients associated with

20  background selection. These measures are used in the Results section, and in Table S3 we provide

21  both kinds of measures for several models.

| Model | BS & CS | BS | CS | BS & CS 11 masses | BS & CS $u_{del}$ constrained | BS $u_{del}$ constrained | BS & CS excluding UTR sweeps | BS Charlesworth |
|---|---|---|---|---|---|---|---|---|
| **Diversity reduction measures** | | | | | | | | |
| $k_{B\&S}\ (1-\bar{\pi}/\pi_0)$ | 73% | 65% | 43% | 86% | 59% | 49% | 74% | 37% |
| $k_B$ | 67% | 66% | - | 81% | 49% | 49% | 69% | 38% |
| $k_S$ | 41% | - | 43% | 69% | 32% | - | 37% | - |
| $k_B/(k_B+k_S)$ | 62% | 100% | - | 54% | 61% | 100% | 65% | 100% |
| $k_S/(k_B+k_S)$ | 38% | - | 100% | 46% | 39% | - | 35% | - |
| **Coalescent rate measures** | | | | | | | | |
| $r_B+r_S$ | 3.18 | 2.13 | 0.87 | 7.59 | 1.66 | 1.07 | 3.28 | 5.23 |
| $r_B$ | 2.26 | 2.13 | - | 4.72 | 1.08 | 1.07 | 2.51 | 5.23 |
| $r_S$ | 0.92 | - | 0.87 | 2.87 | 0.58 | - | 0.77 | - |
| $r_B/(r_B+r_S)$ | 71% | 100% | - | 62% | 65% | 100% | 77% | 100% |
| $r_S/(r_B+r_S)$ | 29% | - | 100% | 38% | 35% | - | 23% | - |

1 **Table S3. The relative contribution of background selection and classic sweeps.** All genome-
2 wide average coalescence rates are shown in units corresponding to genetic drift in the absence of
3 linked selection.

### D. Interpreting inferences about classic sweeps

Our inference is predicated on a model of classic sweeps on semi-dominant alleles, which takes the form of an exponential decay of coalescence rates around beneficial substitutions, where the rate of decay depends on the selection coefficient. In the likelihood calculation, we model coalescence rates around observed substitutions as a superposition of such exponentials, with weights that follow from the distribution of selection coefficients. Other modes of linked positive selection have similar effects on the expected coalescence rates around substitutions. Below, we discuss how this similarity allows us to interpret our estimates.

***Partial sweeps.*** We rely on the work of Coop and Ralph [27] to illustrate the equivalence between the expected effect of partial sweeps, in which a new mutation rises quickly to intermediate frequency but fixes less rapidly [28,29], and of a mixture of classic sweeps (Fig S2). Consider a model with two types of classic sweeps, one driven by stronger selection than the other, and assume that strong selection drives a fraction $\alpha_1$ of substitutions to fixation over $t_1$ generations and weaker selection a fraction $\alpha_2$ to fixation over $t_2$ generations ($t_2 >> t_1$). Next consider a model in which a fraction $\alpha$ of substitutions are driven to fixation by a single kind of partial sweeps, where a new mutation is driven to frequency $x$ over $t_P$ generations and then to fixation over $t_F$ generations, and $t_F >> t_P$, reflecting much stronger selection in the initial stage. At neutral sites close to a substitution ($r << 1/t_F$), the expected coalescent rate is $\alpha \cdot Exp(-r(t_P + t_F))$, and at sites farther away ($r >> 1/t_F$), the expected rate is $\alpha \cdot x^2 \cdot Exp(-r \cdot t_P)$ [27]. It follows that the two models would generate similar expected diversity levels around substitutions if their parameters satisfy the requirements that $\alpha_1 = \alpha \cdot x^2$ and $\alpha_2 = \alpha \cdot (1 - x^2)$, as well as $t_1 = t_P$ and $t_2 = t_F + t_P$. More generally, there is a continuous spectrum of mixtures of partial and classic sweeps that would generate similar diversity patterns. Moreover, this equivalence can be generalized to partial sweeps with more than two different selected phases.

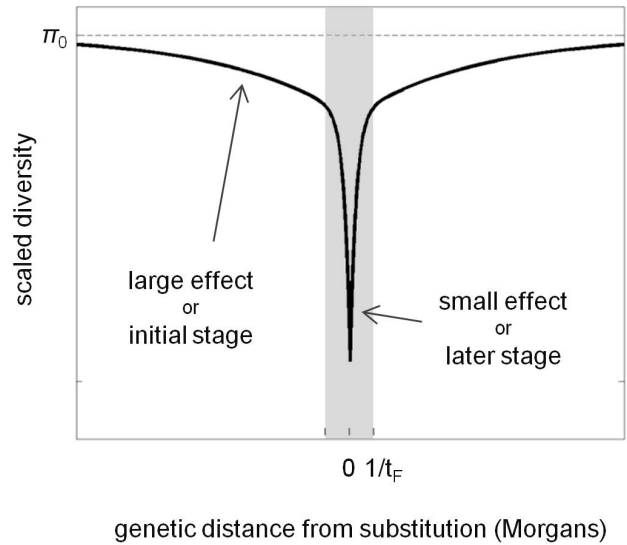

genetic distance from substitution (Morgans)

**Fig S2. Similarities between signatures of classic and partial sweeps**. The figure illustrates the similarity between expected diversity levels under a mixture of classic sweeps driven by either weak or strong selection and a single kind of partial sweeps initially driven by strong selection to intermediate frequency and then to fixation by weak selection. The strong classic sweeps or the initial stage of the partial sweeps govern diversity levels far from substitutions while the weak classic sweeps or the final stage of the partial sweeps govern them close to the substitution.

This equivalence suggests how to interpret our parameter estimates in the presence of a mixture of classic and partial sweeps. Namely, the total estimate of the fraction of substitutions driven by positive selection remains the same, but in this case, the fraction associated with a specific selection coefficient reflects both the fraction of beneficial substitutions driven by such selection sometime during their trajectory and the part of the trajectory in which it acts. More precisely, this estimate corresponds to either the fraction of full trajectories with the same selection coefficient, i.e. classic sweeps, or to a greater fraction of partial trajectories with equivalent effects on diversity.

***Soft sweeps (multiple mutations)***. We can make a similar argument for soft sweeps, where multiple beneficial alleles at the same locus sweep to intermediate frequencies in response to the same selection pressure [30]. During such a soft sweep, none of the beneficial alleles fix and thus no substitution occurs. Over a longer period, however, one of these alleles will fix due to drift or

1   selection. If it is due to drift then there will be sufficient time for recombination to reduce linkage

2   disequilibrium around the selected site to background levels, such that this fixation period will not

3   leave a signature of linked selection close to the selected site. If, instead, the beneficial alleles have

4   similar, but not identical, selection coefficients then the most favorable allele will likely fix, but will

5   do so under much weaker selection (depending on its selective advantage over the other beneficial

6   alleles), resulting in a process akin to the slower phase of the partial sweep considered above.

7   The resulting effect on diversity levels can be approximated as follows. Assume that the initial soft

8   sweep takes $t_S$ generations and that the subsequent fixation of the neutral or favored allele takes $t_F$

9   generations, where $t_F \gg t_S$. Pennings and Hermisson [31] showed that, with a population-scaled

10  beneficial mutation rate at the selected locus $\theta_B$, the probability that two lineages, sampled

11  immediately after the initial soft sweep, coalesce during the sweep is $(1/(1+\theta_B))Exp(-r \cdot t_S)$.

12  Following the same logic as above (i.e., that of Coop and Ralph [27]), after fixation, farther from the

13  selected site ($r \gg 1/t_F$), the expected rate of coalescence would be $(1/(1+\theta_B))Exp(-r \cdot t_S)$. In turn, the

14  expected rate of coalescence closer to the selected site ($r \ll 1/t_F$) depends on whether fixation

15  occurs by drift or selection. If it is due to drift then the expected rate of coalescence would be the

16  same as at a farther distance, i.e., $(1/(1+\theta_B))Exp(-r \cdot t_S)$, but if it is due to selection then lineages

17  would be affected by the entirety of the sweep and the expected rate of coalescence would be

18  $Exp(-r(t_S+t_F))$ (where we assume that $1/(1+\theta_B) \gg Exp(-r \cdot t_F)$).

19  These approximations suggest how our estimates of sweep parameters could be interpreted in the

20  presence of soft sweeps. A fraction $\alpha$ of substitutions due to soft sweeps with a neutral fixation

21  phase would appear in our estimates as $\alpha/(1+\theta_B)$ sweeps driven by the selection coefficient

22  characterizing the initial soft sweep phase. Thus, having multiple beneficial mutations would cause

23  the fraction that we infer to be an underestimate of the fraction of beneficial substitutions, while

24  our estimate of the selection coefficient would be unbiased. If, instead, the fixation phase is also

25  driven by selection, then we would estimate that a fraction of (approximately) $\alpha$ substitutions were

26  driven by weak selection corresponding to a beneficial sojourn time of $t_S+t_F$ and (approximately) a

27  fraction $\alpha/(1+\theta_B)$ were driven by strong selection corresponding to a beneficial sojourn time of $t_S$.

28  This case is similar to the case of partial sweeps outlined above, with $\alpha/(1+\theta_B)$ replacing $\alpha \cdot x^2$.

29  ***Sweeps from standing variation***. Another kind of sweep occurs when the selected allele that

30  fixes was common when selection began [32-35]. Specifically, consider an allele that was present in the population at frequency $f$ ($1/2N_e \ll f \ll 1$) at the onset of selection, either having drifted to that frequency or having been balanced close to that frequency for some (but not a very long) time. Selection then begins to favor the allele (in a semi-dominant manner) and it sweeps to fixation over $t_S$ generations. For a pair of lineages at genetic distance $r$ from the selected allele, the probability of coalescence due to the sweep is then well approximated by $1/(1+4N_e r \cdot f(1-f))Exp(-r \cdot t_S)$ [36]. Thus, far from the selected site, where the distance is defined by the initial frequency, i.e., $4N_e r \cdot f \gg 1$, the sweep will have little effect on diversity levels. Close to the selected site, i.e., when $4N_e r \cdot f \ll 1$, the probability of coalescence can be further approximated by $Exp(-r(t_S+4N_e f(1-f)))$.

The interpretation of our estimates in the presence of sweeps from standing variation follows. If the initial frequencies of beneficial alleles are large then the sweeps will not be captured by our inference because they have negligible effect on diversity levels. In other words, substitutions driven by such sweeps will be estimated as neutral. For substitutions driven by sweeps with sufficiently small initial frequencies, we will estimate their fraction correctly but the selection coefficients will be downwardly biased, corresponding to a beneficial sojourn time of approximately $t_S+4N_e f(1-f)$. This sojourn time is in fact roughly the coalescent time of two selected alleles in the population but it is greater than the classic sweep equivalent with the same selection coefficient. Overall, for substitutions caused by sweeps from standing variation, we will underestimate the selection coefficients and their fraction.

***Recessive and other classic sweeps.*** We could also have classic sweeps caused by beneficial alleles that are recessive, or more generally, not semi-dominant [37,38]. In that case, a recessive sweep would leave a signature that is similar to a sweep from standing variation. Given that we are considering sweeps that lead to substitutions, we know that the selected allele will start from a single mutation and proceed all the way to fixation. Until the beneficial allele reaches a sufficiently high frequency, i.e., while $2N_e s \cdot f^2 \ll 1$, its dynamics would be governed by genetic drift. The frequency at which selection will kick in can be approximated by $f \approx 1/\sqrt{2N_e s}$. The duration of the selected phase, $t_S$, can then be calculated using the diffusion approximation [39]. The interpretation of our estimates is then the same as for the case of sweeps from standing variation with these parameters. In turn, for beneficial alleles with dominance coefficients other than 0 or ½, the

1    mapping to the inferred semi-dominant selection coefficient is provided by the sojourn time for the

2    beneficial allele, again given by the diffusion approximation.

3    ***A mixture including different kinds of sweeps.*** In reality, we would expect a mixture of a

4    variety of sweeps to occur, raising the question about the best way to interpret our estimates. While

5    we discussed how various kinds of sweeps could "bias" our estimates, and in principle, these

6    "biases" could be considered together, we believe this would not be a very productive

7    interpretation. For example, it is not obvious whether, with soft sweeps in the mix, our estimate for

8    the fraction of substitutions driven by a given selection coefficient, $\alpha(s)$, is best interpreted as an

9    underestimate of the fraction of substitutions driven by such selection during some part of their

10   trajectory. Instead, since we are interested in the effects on diversity level, we consider our

11   estimate to be a summary of the effect of a continuous variety of partial sweeps, in which different

12   parts of the adaptive trajectory are driven by this selection coefficient—a classic sweep equivalent

13   that summarizes over a continuous set of possibilities. More generally, because the various kinds of

14   sweeps result in similar functional forms, we may interpret our estimates as designating a class of

15   mixtures that would result in similar expected diversity levels around substitutions.

16   In principle, the kind of derivations that we outlined could be used to write down equations for the

17   possible mixtures. However, if we were to use all the kinds of sweeps mentioned above then we

18   would end up with equations including numerous parameters, without, as of now, a way to solve for

19   them. Even when methods are developed to use other aspects of the data to learn about some of

20   these parameters, we will no doubt need some kind of coarse-graining scheme in order to think

21   about these mixtures. Both because there are more parameters than we are likely to be able to infer

22   with any certainty, and because, to begin with, we would have to have some coarse graining for the

23   continuous ranges of many of these parameters (e.g., selection and dominance coefficients and

24   initial frequencies for different sweep phases). Moreover, while we outlined how the main kinds of

25   sweeps considered to date would be recorded in our estimates, there is actually a continuous range

26   of possibilities. For example, there are hybrids of the kinds that we have considered, such as a

27   sweep that begins from standing variation but then proceeds through a selected phase divided into

28   several phases with different selection coefficients (in fact, such sweeps are expected under models

29   of polygenic adaptation [28,40]). Given these considerations, we do not try to dissect the different

30   mixtures of sweeps any further.

## E. Interpreting the inferences about background selection

***Imposing an upper bound on the mutation rate***. The unreasonably high estimates obtained for the deleterious mutation rate lead us to consider models in which we impose a biologically plausible upper bound. As an upper bound, we use an estimate for the total mutation rate in *D. melanogaster*, which we arrive at as follows. For point mutations, we use a rate of $3.5 \times 10^{-9}$ per bp per generation, which was estimated by Keightley et al. based on mutation accumulation experiments [41]. For indels, we use a rate of $1.6 \times 10^{-9}$ per bp per generation, which derives from estimates of the ratio of indel to point mutation rates ($8.4 \times 10^{-9}/5.8 \times 10^{-9}$) taken from Haag-Liautard et al. [42] and the aforementioned estimate for point mutations (while Haag-Liautard et al. also estimated the point mutation rate, they did so based on considerably less data). Additionally, we use a rate of transposable element (TE) insertions of $1.7 \times 10^{-9}$ per bp per generation, based on Nuzhdin & Mackay's [43] estimate of 0.2 TE insertions per genome per generation and assuming a uniform rate along the genome. Taken together these rates sum up to $\mu = 6.8 \times 10^{-9}$ mutations per bp per generation, which we use as an upper bound on the deleterious mutation rate. The rate of TE insertions might be somewhat higher [44] and the estimates based on mutation accumulation experiments might be noisy due to variation among lines [41,45]. However, moderate changes in the bound would not change the conclusions of our analysis.

We find that imposing an upper bound on the deleterious mutation rate introduces artifacts into the inference rather than making our estimates more plausible. The new estimates have a reduced genomic rate of deleterious mutations associated with strong selection ($t=10^{-1.5}$), and the bulk of these mutations are shifted from exons and UTRs to introns and intergenic regions (Table S4). Taken at face value, these new estimates do not make much sense because they imply a similar proportion of sites under strong purifying selection in exons and long introns, which clearly contradicts findings based on sequence conservation and diversity patterns [9-11]. Moreover, even if we assume that these estimates absorb the effects of other forms of linked selection, it would still make little sense for such effects to be similar (let alone greater) in long introns compared to exons and UTRs [9]. Instead, the new estimates appear to be an artifact (Fig S3). Because we cap the deleterious mutation rate, the model requires additional sites under strong purifying selection; and because strong selection ($t=10^{-1.5}$) affects diversity level in a non-localized fashion, assigning the strongly deleterious mutations to annotations other than exons and UTRs has a minor effect on the fit. These considerations suggest that the inference based on the unconstrained model better

1 reflects the effects of linked selection. If so, the unrealistically high estimates of the rate of

2 deleterious mutations around coding regions are likely absorbing other modes of linked selection,

3 which are also most likely to be acting on these regions.

4 ***Our uncertainty about $\pi_0$.*** We find that the constrained and unconstrained models yield very

5 similar fits, as measured by the log composite likelihood (where $\Delta CL$ is its increase compared to a

6 neutral model) and a variety of summaries (Table S4), while their estimates for the diversity levels

7 in the absence of linked selection, $\pi_0$, are substantially different. As noted, this difference

8 corresponds to different estimates for the deleterious mutation rate associated with strong

9 selection (Fig S3) and is possible because the data carries little direct information about $\pi_0$. As we

10 note in the manuscript, this introduces considerable uncertainty regarding the overall reduction in

11 diversity levels caused by linked selection (Table S4).

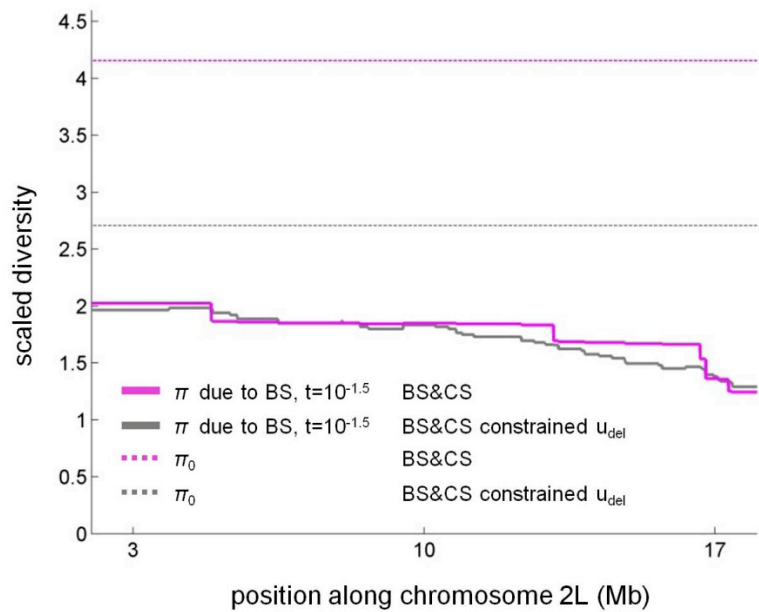

12 **Fig S3. The predicted effects of background selection due to strongly deleterious mutations**

13 **($t$=10^{-1.5}) in the constrained and unconstrained models**. Shown are the predicted scaled

14 diversity levels based on the two models, when considering only the effects of linked selection

15 caused by strongly deleterious mutations. See Table S4 for the parameter estimates for these

16 models.

| | Model | Background selection and classic sweeps | | | | Background selection and classic sweeps with constrained $u_{del}$ | | | |
|---|---|---|---|---|---|---|---|---|---|
| | $\pi_0 / \bar{\pi}$ | 4.4 | | | | 2.8 | | | |
| | $U_{del}$ (per diploid) | 1.60 | | | | 0.92 | | | |
| | Annotation | Exons | UTRs | Introns | Intergenic | Exons | UTRs | Introns | Intergenic |
| | $u_{del} / \mu$ | 437% | 603% | 19% | - | 90% | 88% | 87% | 44% |
| | $u(t=10^{-1.5}) / \mu$ | 377% | 577% | 19% | - | 18% | 67% | 85% | 44% |
| | $u(t=10^{-2.5}) / \mu$ | 2% | 2% | - | - | 2% | 2% | 1% | - |
| | $u(t=10^{-3.5}) / \mu$ | 56% | - | - | - | 69% | - | - | - |
| | $u(t=10^{-4.5}) / \mu$ | 2% | 23% | - | - | 1% | 20% | - | - |
| Parameters | $u(t=10^{-5.5}) / \mu$ | - | 2% | - | - | - | - | - | - |
| | $\sum_s \alpha(s) \cdot s$ | $3.2 \times 10^{-5}$ | $3.0 \times 10^{-6}$ | - | - | $3.0 \times 10^{-5}$ | $3.0 \times 10^{-6}$ | - | - |
| | $\alpha$ | 40% | 47% | - | - | 42% | 51% | - | - |
| | $\alpha(s=10^{-1.5})$ | - | - | - | - | - | - | - | - |
| | $\alpha(s=10^{-2.5})$ | 0.6% | - | - | - | 0.6% | - | - | - |
| | $\alpha(s=10^{-3.5})$ | 3.5% | - | - | - | 3.5% | - | - | - |
| | $\alpha(s=10^{-4.5})$ | - | 5.1% | - | - | - | 4.7% | - | - |
| | $\alpha(s=10^{-5.5})$ | 36.3% | 42.1% | - | - | 38.1% | 45.9% | - | - |

| | | Background selection and classic sweeps | Background selection and classic sweeps with constrained $u_{del}$ |
|---|---|---|---|
| | $\Delta CL$ | $3.9 \times 10^{-4}$ | $3.6 \times 10^{-4}$ |
| Diversity binned in local windows | $R^2$  1 Mb | 0.71 | 0.69 |
| | 100 kb | 0.44 | 0.43 |
| | 10 kb | 0.26 | 0.24 |
| | 1 kb | 0.20 | 0.19 |
| Diversity binned by distance from substitution | $R^2$  NS substitutions | 0.62 | 0.61 |
| | SYN substitutions | 0.66 | 0.69 |
| Diversity binned by predicted effect of linked selection | Spearman's $\rho$ | 0.913 | 0.905 |
| | Upper-to-lower tails observed diversity ratio | 5.3 | 5.4 |
| Diversity reduction measures | $k_{B\&S}$ $(1 - \bar{\pi} / \pi_0)$ | 73% | 59% |
| | $k_B$ | 67% | 49% |
| | $k_S$ | 41% | 32% |
| | $k_B/(k_B+k_S)$ | 62% | 61% |
| | $k_S/(k_B+k_S)$ | 38% | 39% |
| Coalescent rate measures | $r_B+r_S$ | 3.18 | 1.66 |
| | $r_B$ | 2.26 | 1.08 |
| | $r_S$ | 0.92 | 0.58 |
| | $r_B/(r_B+r_S)$ | 71% | 65% |
| | $r_S/(r_B+r_S)$ | 29% | 35% |

1  **Table S4. Parameter estimates, goodness-of-fit and other summaries for models**

2  **constraining $u_{del}$ or not.**

### F. Comparison with maps based on the Charlesworth approach

Pioneering work by Charlesworth [44,46] used estimates of the rates and distributions of selection coefficients of deleterious mutations and genetic maps, to demonstrate that background selection could account for the large-scale changes in diversity levels along chromosomes of *D. melanogaster*. More recently, this approach was extended by Comeron [47] to incorporate the spatial distributions of constrained genomic regions. This approach differs from ours in several ways, most notably in that estimates of selection parameters are not based on the effects of linked selection. Instead, estimates of the total rate of deleterious mutations come from mutation accumulation lines and estimates of the rate and distribution of selection coefficients at coding and non-coding annotations stem primarily from the direct effects of purifying selection on divergence and on the site frequency spectrum. Here we compare the predictions of this method with ours.

To this end, we introduce several modifications. At exonic regions, Charlesworth used a gamma distribution of selection coefficients truncated at $t=5\times10^{-6}$. We use a discretized approximation of this distribution on the grid $t=10^{-1.5}$, $10^{-2.5}$, $10^{-3.5}$, $10^{-4.5}$, $10^{-5.5}$ (integrating between mid points on the log-linear scale). For non-exonic regions, Charlesworth assumed that short segments switch between being under strong and weak selection, where the distribution of selection coefficients in each type of segment is also described by a gamma distribution truncated at $t=5\times10^{-6}$. We use a discretized approximation of the mixture of the two distributions, weighted by the relative length of segments; because the alternating segments are short, assuming a homogeneous mixture has a negligible effect on predicted diversity levels. The resulting selection parameters are shown in Table S5, alongside our estimates. Similar to Comeron, we use the spatial distribution of exonic and non-exonic regions along the genome. Finally, to determine the expected diversity level in the absence of background selection, $\pi_0$, we require that the mean diversity level predicted by the model matches the observed one.

Fig S4 shows a comparison between the predictions of the two methods and observed diversity levels. While the predictions along the chromosome appear similar, a quantitative comparison indicates that our method does better at all spatial scales (Fig S4A and Table S5). Both visual and quantitative comparison based on diversity patterns around substitutions (Fig S4B and Table S5) and on the stratification of diversity levels (Fig S4C and Table S5) confirms that our method performs better.

1    The fact that our method provides a better prediction of diversity levels is not surprising, given that
2    our inference relies on observed diversity levels. Our leave-one-out cross validation analysis of $R^2$
3    values on different spatial scales, however, suggests that over-fitting explains a negligible part of
4    the difference (Section C). Instead, we believe that two other factors are more important. First, as
5    we noted in the manuscript, our inferences for background selection likely absorb the effects of
6    other modes of linked selection and accounting for (some) of these effects lends better predictive
7    abilities. Second, Charlesworth's and Comeron's estimates for the distribution of selection
8    coefficients rely primarily on signatures of direct (in contrast to linked) purifying selection, which,
9    provided the population sample sizes currently available for *D. melanogaster*, only allows one to
10   probe a relatively narrow range of selection coefficients (cf. [48]). Using spatial diversity patterns
11   likely provides insight into a wider range of selection coefficients.

| Model | Background selection and classic sweeps | | | | Background selection alone | | | | Background selection based on Charlesworth | | | |
|---|---|---|---|---|---|---|---|---|---|---|---|---|
| $U_{del}$ (per diploid) | 1.60 | | | | 1.46 | | | | 0.56 | | | |
| Annotation | Exons | UTRs | Introns | Intergenic | Exons | UTRs | Introns | Intergenic | Exons | UTRs | Introns | Intergenic |
| **Parameters** — $u_{del} / \mu$ | 437% | 603% | 19% | - | 448% | 456% | 17% | - | 72% | 38% | 38% | 38% |
| $u(t{=}10^{-1.5}) / \mu$ | 377% | 577% | 19% | - | 369% | 453% | 17% | - | - | - | - | - |
| $u(t{=}10^{-2.5}) / \mu$ | 2% | 2% | - | - | - | - | - | - | 25% | 7% | 7% | 7% |
| $u(t{=}10^{-3.5}) / \mu$ | 56% | - | - | - | 77% | - | - | - | 27% | 8% | 8% | 8% |
| $u(t{=}10^{-4.5}) / \mu$ | 2% | 23% | - | - | 2% | 3% | - | - | 15% | 10% | 10% | 10% |
| $u(t{=}10^{-5.5}) / \mu$ | - | 2% | - | - | - | - | - | - | 5% | 12% | 12% | 12% |
| $\Delta CL$ | $3.9{\times}10^{-4}$ | | | | $2.8{\times}10^{-4}$ | | | | $-6.7{\times}10^{-5}$ | | | |
| **Diversity binned in local windows** — $R^2$ 1 Mb | 0.71 | | | | 0.76 | | | | 0.58 | | | |
| 100 kb | 0.44 | | | | 0.42 | | | | 0.19 | | | |
| 10 kb | 0.26 | | | | 0.23 | | | | 0.09 | | | |
| 1 kb | 0.20 | | | | 0.18 | | | | 0.08 | | | |
| **Diversity binned by distance from substitution** — $R^2$ NS substitutions | 0.62 | | | | 0.27 | | | | - | | | |
| SYN substitutions | 0.66 | | | | 0.53 | | | | 0.05 | | | |
| **Diversity binned by predicted effect of linked selection** — Spearman's $\rho$ | 0.913 | | | | 0.745 | | | | 0.773 | | | |
| Upper-to-lower tails observed diversity ratio | 5.3 | | | | 4.4 | | | | 3.5 | | | |
| **Diversity reduction measures** — $k_{B\&S}$ $(1-\bar{\pi}/\pi_0)$ | 73% | | | | 66% | | | | 38% | | | |
| $k_B$ | 67% | | | | 66% | | | | 38% | | | |
| $k_S$ | 41% | | | | - | | | | - | | | |
| $k_B/(k_B+k_S)$ | 62% | | | | 100% | | | | 100% | | | |
| $k_S/(k_B+k_S)$ | 38% | | | | - | | | | - | | | |
| **Coalescent rate measures** — $r_B+r_S$ | 3.18 | | | | 2.13 | | | | 5.23 | | | |
| $r_B$ | 2.26 | | | | 2.13 | | | | 5.23 | | | |
| $r_S$ | 0.92 | | | | - | | | | - | | | |
| $r_B/(r_B+r_S)$ | 71% | | | | 100% | | | | 100% | | | |
| $r_S/(r_B+r_S)$ | 29% | | | | - | | | | - | | | |

1    **Table S5. Comparison of the method of Charlesworth and ours.**

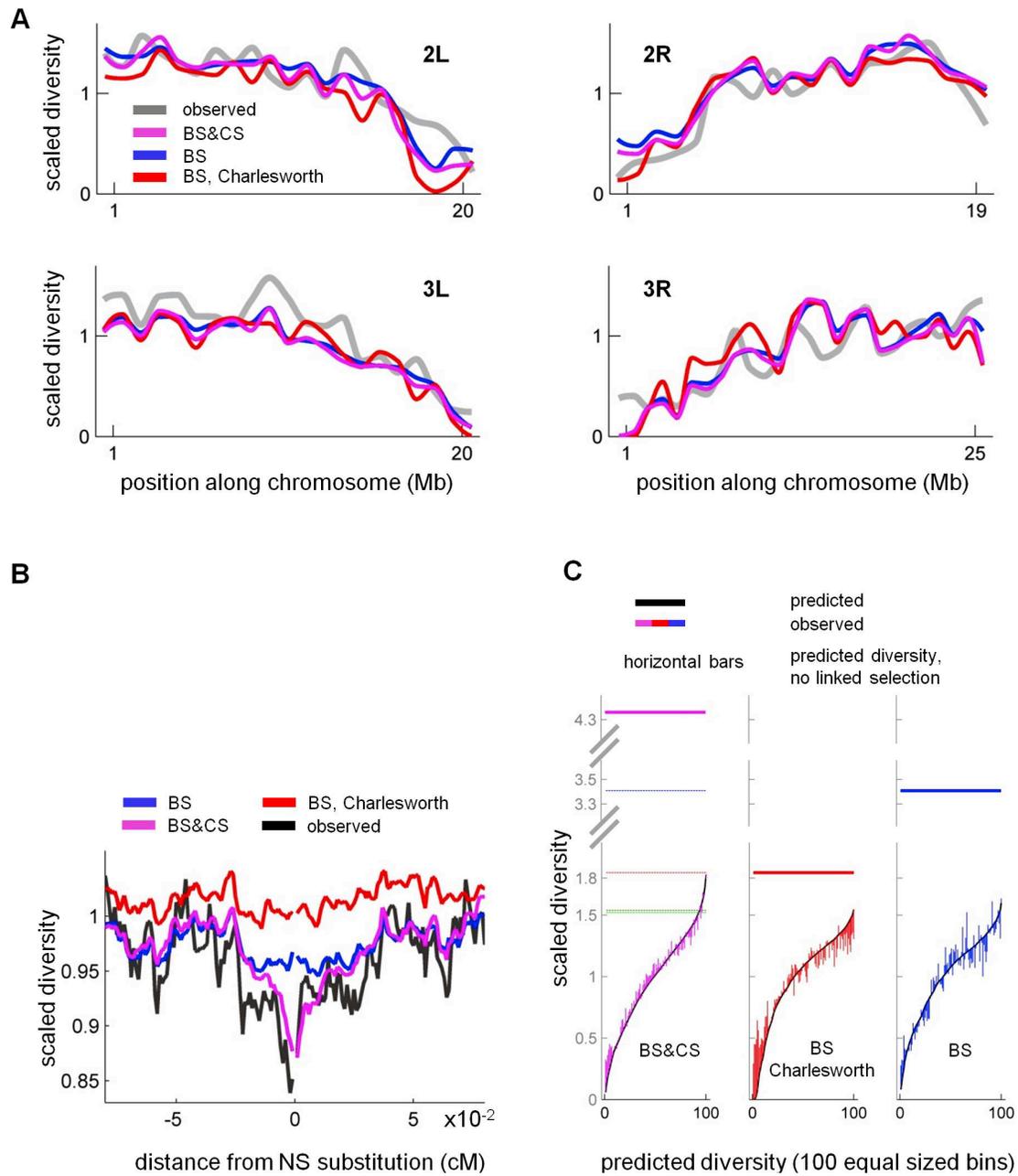

**Fig S4**. **Predicted scaled diversity levels based on the method of Charlesworth and ours.** (**A**) Diversity levels along the major autosomes. Other details are as in Fig 2A in the main text. (**B**) Diversity levels around non-synonymous substitutions. Other details are as in Fig 3B in the main text. (**C**) The impact of linked selection on diversity levels. Other details are as in Fig 6A in the main text. See Table S5 for the corresponding summaries.

### G. Comparison to other inference methods

To compare our results with those of other methods used to infer selection parameters based on signatures of linked selection (e.g., Fig 3 and 6 in the main text), we proceed as follows.

***Inferring sweep parameters based on the Sattath et al. method.*** We apply the Sattath et al. [21] method to estimate classic sweep parameters associated with amino acid substitutions. The method relies on finding the parameters that provide the best fit to the observed average scaled synonymous diversity as a function of distance from amino acid substitutions. We rely on the same synonymous diversity and divergence data used for the current method, including a recombination threshold of 0.75 cM/Mb or greater. We collate synonymous diversity levels and numbers of substitutions per codon as detailed in Section C, up to a distance of 0.11 cM from substitutions, using bins of $10^{-6}$ cM. We then apply the composite likelihood maximization described in Sattath et al. (2011) using a parametric distribution of selection coefficients consisting of three point masses (in addition to $s=0$), where in this case the values of the point masses are also parameters that are free to vary (i.e., there are six free parameters overall). In practice, only two of the point masses appear in the MLE (see "CS, Sattath et al." in Table S6).

| | Model | Background selection and classic sweeps | | | Classic sweeps | | | Classic sweeps based on Sattath et al. |
|---|---|---|---|---|---|---|---|---|
| | Annotation | Exons | UTRs | Introns | Exons | UTRs | Introns | Exons |
| **Parameters** | $\sum_s \alpha(s) \cdot s$ | $3.2 \times 10^{-5}$ | $3.0 \times 10^{-6}$ | . | $1.1 \times 10^{-4}$ | $3.6 \times 10^{-5}$ | . | $6.1 \times 10^{-5}$ |
| | $\alpha$ | 40% | 47% | . | 40% | 51% | . | 20% |
| | $\alpha(s=10^{-1.5})$ | . | . | . | 0.3% | 0.1% | . | |
| | $\alpha(s=10^{-2.5})$ | 0.6% | . | . | 0.4% | . | . | 1.5% $\quad s=10^{-2.4}$ |
| | $\alpha(s=10^{-3.5})$ | 3.5% | . | . | 3.7% | . | . | |
| | $\alpha(s=10^{-4.5})$ | . | 5.1% | . | . | 4.4% | . | |
| | $\alpha(s=10^{-5.5})$ | 36.3% | 42.1% | . | 35.7% | 46.1% | . | 18.8% $\quad s=10^{-5.2}$ |
| | $\Delta CL$ | $3.9 \times 10^{-4}$ | | | $2.4 \times 10^{-4}$ | | | |
| **Diversity binned in local windows** | $R^2$  1 Mb | 0.71 | | | 0.67 | | | |
| | 100 kb | 0.44 | | | 0.39 | | | |
| | 10 kb | 0.26 | | | 0.21 | | | |
| | 1 kb | 0.20 | | | 0.16 | | | |
| **Diversity binned by distance from substitution** | $R^2$ NS substitutions | 0.62 | | | 0.51 | | | 0.56 |
| | SYN substitutions | 0.66 | | | 0.49 | | | 0.65 |
| **Diversity binned by predicted effect of linked selection** | Spearman's $\rho$ | 0.913 | | | 0.890 | | | |
| | Upper-to-lower tails observed diversity ratio | 5.3 | | | 5.0 | | | |
| **Diversity reduction measures** | $k_{B\&S}$ $(1 - \bar{\pi} / \pi_0)$ | 73% | | | 43% | | | 21% |
| | $k_B$ | 67% | | | . | | | . |
| | $k_S$ | 41% | | | 43% | | | 21% |
| | $k_B/(k_B+k_S)$ | 62% | | | . | | | . |
| | $k_S/(k_B+k_S)$ | 38% | | | 100% | | | 100% |
| **Coalescent rate measures** | $r_B+r_S$ | 3.18 | | | 0.87 | | | 0.29 |
| | $r_B$ | 2.26 | | | . | | | . |
| | $r_S$ | 0.92 | | | 0.87 | | | 0.29 |
| | $r_B/(r_B+r_S)$ | 71% | | | . | | | . |
| | $r_S/(r_B+r_S)$ | 29% | | | 100% | | | 100% |

**Table S6. Comparison with the Sattath et al. (2011) inference method.**

1   ***An error in Sattath et al. (2011)***. In the course of this analysis, we found an error in Sattath et
2   al. (2011) that caused a three-fold increase in estimates of the fraction of substitutions associated
3   with non-zero selection coefficients. While this does not change the qualitative conclusions in
4   Sattath et al. (2011), it does imply that all estimated fractions in their Table S5 should be divided by
5   3. For example, for a distribution of fitness effects consisting of two point masses, the correct
6   estimates are that 4.8% of the amino acid substitutions were driven by classic selective sweeps,
7   1.1% of them with a selection coefficient of $s$=5.1×10$^{-3}$ and the remaining 3.7% with $s$=1.1×10$^{-4}$.

8   ***Inferring selection parameters using the Wiehe, Kim and Stephan method***. Wiehe and
9   Stephan (1993) [49] introduced a method to infer compound classic sweep parameters from the
10  observed relationship between levels of heterozygosity and recombination rates. Kim and Stephan
11  (2000) [50] extended the underlying model to include both classic sweeps and background
12  selection. Macpherson et al. [51] and Andolfatto [52] also extended the method, this time to infer
13  compound classic sweep parameters from the relationship between levels of heterozygosity and
14  rates of non-synonymous substitutions.

15  The model for expected heterozygosity underlying these studies can be summarized by

16
$$\pi(c,d_n) = \pi_0 \frac{c \cdot Exp(-u_{del}/c)}{c + I \cdot d_n \cdot (\sum_s \alpha(s) \cdot s) \cdot Exp(-u_{del}/c)},$$

17  where $c$ and $d_n$ are the rates of recombination and non-synonymous substitutions per bp per
18  generation measured in a specified window, $\pi_0$ is the expected heterozygosity in the absence of
19  linked selection, $I \sim 0.075$ is a constant, and the compound selection parameters are: $u_{del}$, the total
20  rate of deleterious mutations per bp per generation and $\sum_s \alpha(s) \cdot s$, where $\alpha(s)$ is the fraction of
21  beneficial non-synonymous substitutions with selection coefficient $s$. To infer the compound
22  parameter for background selection alone, the second term in the denominator is set to 0
23  (corresponding to no sweeps) and $u_{del}$ is inferred using least squares to fit the remaining functional
24  form with the observed relationship between $\pi$ and $c$. To infer the compound parameter for classic
25  sweeps alone, the exponential is set to 1 (corresponding to $u_{del}$=0); for the inference based on the
26  relationship between $\pi$ and $c$, $d_n$ is set to its genomic average, and for the inference based on the
27  relationship between $\pi$ and $d_n$, $c$ is set to its genomic average. In applying these methods to our
28  data, we partition codons into 100 bins with equal amounts of data based on $d_n$, measured in a window of 0.03 cM around each codon, or based on the local estimate of the recombination rate (corresponding to a spatial resolution of ~100 kb [13]).

Estimates of the compound selection parameters based on these inferences are shown in Table S7, alongside our own. As we note in the Results, the Wiehe, Kim and Stephan methods underestimate the compound parameters and thus the effects of linked selection because, by ignoring some genomic features affecting the strength of linked selection, they suffer from the equivalent of regression toward the mean (also see Fig 6 in the main text). This effect is best seen by comparing estimates for a single mode of linked selection, where all methods face the problem of absorbing other modes of linked selection. This underestimation effect is also apparent in the comparison of the Sattath et al. estimates and ours under a model of classic sweeps alone, as well as in the comparison of our inferences based on the joint model for background selection and classic sweeps with models of one or the other (Fig 6A in the main text).

| Model | $\sum_s \alpha(s) \cdot s$ | $U_{del}$ per diploid |
|---|---|---|
| Background selection and classic sweeps | $3.5 \times 10^{-5}$ | 1.60 |
| Background selection alone | . | 1.46 |
| Classic sweeps alone | $1.5 \times 10^{-4}$ | . |
| Classic sweeps based on Sattath et al. | $6.1 \times 10^{-5}$ | . |
| Background selection based on Charlesworth | . | 0.56 |
| Background selection based on Kim & Stephan by recombination rate | . | 0.74 |
| Classic sweeps based on Wiehe & Stephan by recombination rate | $4.2 \times 10^{-5}$ | . |
| Classic sweeps based on Macpherson et al. by NS divergence | $3.8 \times 10^{-5}$ | . |

**Table S7. Estimates of compound selection parameters based on different inference methods.** The compound selection parameters estimated are: $\sum_s \alpha(s) \cdot s$ for non-synonymous substitutions and $U_{del}$, the total number of deleterious mutations on autosomes per genome per generation.

### H. Sensitivity to the recombination rate threshold and to codon usage bias

***The recombination rate threshold***. To examine the sensitivity of our estimates to the choice of recombination rate threshold, we inferred the parameters of the combined model excluding sites with sex-averaged recombination rates <0.375 cM/Mb and <1.125 cM/Mb (instead of <0.75 cM/Mb), again excluding the distal 5% of chromosome arms (Table S8). We find that the goodness-of-fit statistics (evaluated across data from all sites with recombination rate >0.75 cM/Mb) and most parameter estimates are robust to the change of threshold. The one notable exception is that with the lower threshold, we no longer infer ~40% of substitutions at UTRs to have been driven by sweeps associated with very weak selection (with $s=10^{-5.5}$). As a result, our total estimate of the fraction of beneficial substitutions at UTRs drops dramatically from ~47% to ~4%.

Further analysis suggests that our inference about a large proportion of weak sweeps at UTRs is reliable. Direct evidence comes from collated plots of the average diversity levels around substitutions in UTRs (Fig S5A). Our predictions based on a threshold of 0.375 cM/Mb or on a threshold of 0.75 but excluding the weakly selected sweeps overestimate the observed diversity levels near substitutions, as expected if weaker sweeps are indeed present. In contrast, when we use the predictions based on the higher thresholds of 0.75 cM/Mb or 1.25 cM/Mb, both of which include a large proportion of weakly selected sweeps, we explain the trough close to substitutions much better.

| | Model | Background selection and classic sweeps | | | Background selection and classic sweeps | | | Background selection and classic sweeps | | |
|---|---|---|---|---|---|---|---|---|---|---|
| | Recombination threshold | 1.125 cM/Mb | | | 0.75 cM/Mb | | | 0.375 cM/Mb | | |
| | Annotation | Exons | UTRs | Introns | Exons | UTRs | Introns | Exons | UTRs | Introns |
| Parameters | BS parameters | | | | | | | | | |
| | $u(t=10^{-1.5})/\mu$ | 321% | 710% | 24% | 377% | 577% | 19% | 403% | 671% | 3% |
| | $u(t=10^{-2.5})/\mu$ | . | . | . | 2% | 2% | . | 3% | 18% | . |
| | $u(t=10^{-3.5})/\mu$ | 63% | . | . | 56% | . | . | 56% | . | . |
| | $u(t=10^{-4.5})/\mu$ | . | 12% | . | 2% | 23% | . | . | 11% | . |
| | $u(t=10^{-5.5})/\mu$ | . | . | . | . | 2% | . | . | . | . |
| | CS parameters | | | | | | | | | |
| | $\alpha(s=10^{-1.5})$ | . | . | . | . | . | . | . | . | . |
| | $\alpha(s=10^{-2.5})$ | 0.7% | . | . | 0.6% | . | . | 0.4% | . | . |
| | $\alpha(s=10^{-3.5})$ | 4.0% | . | . | 3.5% | . | . | 4.1% | . | . |
| | $\alpha(s=10^{-4.5})$ | . | 6.4% | . | . | 5.1% | . | . | 3.7% | . |
| | $\alpha(s=10^{-5.5})$ | 44.2% | 34.2% | 0.2% | 36.3% | 42.1% | . | 31.9% | 0.1% | . |
| | $\Delta CL$ | 4.08×10⁻⁴ | | | 4.11×10⁻⁴ | | | 4.06×10⁻⁴ | | |
| Diversity binned in local windows | $R^2$  1 Mb | 0.63 | | | 0.62 | | | 0.63 | | |
| | 100 kb | 0.30 | | | 0.30 | | | 0.30 | | |
| | 10 kb | 0.18 | | | 0.18 | | | 0.18 | | |
| | 1 kb | 0.14 | | | 0.14 | | | 0.14 | | |
| Diversity binned by distance from substitution | $R^2$ NS substitutions | 0.61 | | | 0.58 | | | 0.56 | | |
| | SYN substitutions | 0.65 | | | 0.57 | | | 0.56 | | |
| Diversity binned by predicted effect of linked selection | Spearman's $\rho$ | 0.796 | | | 0.795 | | | 0.794 | | |
| | Upper-to-lower tails observed diversity ratio | 3.9 | | | 4.0 | | | 4.3 | | |
| Diversity reduction measures | $k_{B\&S}\,(1-\bar{\pi}/\pi_0)$ | 73% | | | 73% | | | 72% | | |
| | $k_B$ | 66% | | | 67% | | | 68% | | |
| | $k_S$ | 43% | | | 41% | | | 34% | | |
| | $k_B/(k_B+k_S)$ | 61% | | | 62% | | | 67% | | |
| | $k_S/(k_B+k_S)$ | 39% | | | 38% | | | 33% | | |
| Coalescent rate measures | $r_B+r_S$ | 3.2 | | | 3.2 | | | 3.0 | | |
| | $r_B$ | 2.2 | | | 2.3 | | | 2.3 | | |
| | $r_S$ | 1.0 | | | 0.9 | | | 0.6 | | |
| | $r_B/(r_B+r_S)$ | 69% | | | 71% | | | 78% | | |
| | $r_S/(r_B+r_S)$ | 31% | | | 29% | | | 22% | | |

1  **Table S8. Sensitivity of our inference to the choice of recombination rate threshold.**

2  Goodness-of-fit measures are evaluated for data from regions with recombination rate

3  > 0.75 cM/Mb.

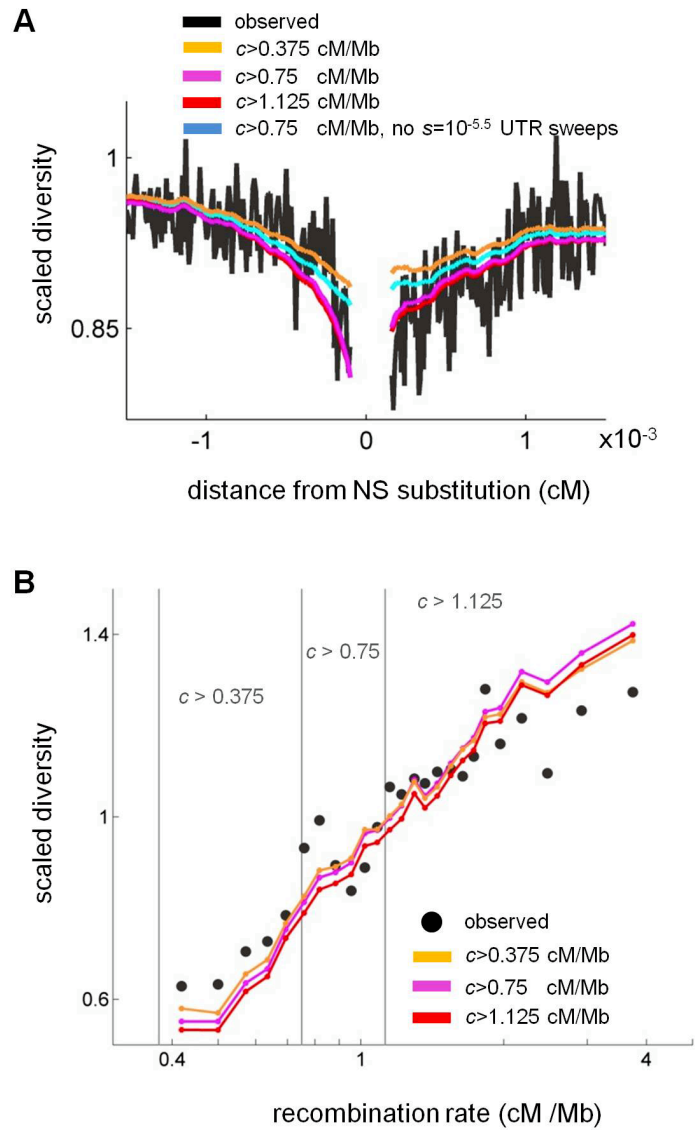

**Fig S5. Estimates associated with weak sweeps are more reliable with the higher recombination threshold.**

Moreover, we expect that including regions with lower recombination rates will lead to underestimates of the fraction of substitutions caused by sweeps, particularly for weak sweeps at UTRs, for the following reasons:

(1) Interference among beneficial mutations, which we do not model, is greater in regions with lower recombination [53-63], and will likely have a greater effect on the number of fixations caused by weakly beneficial mutations (because when strongly and weakly beneficial alleles interfere, the stronger one is more likely to prevail; also see [64]). Also, selection is less effective in regions with lower recombination, so both weakly selected beneficial and deleterious mutations will be less effectively selected and may even be rendered effectively neutral. If these speculations are true, then we would expect there to be fewer weakly selected sweeps in regions with lower recombination. Indirect evidence for fewer selected mutations causing the reduction in diversity in regions with lower recombination rates can be seen by comparing observed and predicted levels as a function of recombination rate (Fig S5B). Notably, we see that our predictions tend to overestimate the reduction in diversity levels in regions with lower recombination rates and to overestimate them in regions with higher rates.

(2) Rather than reflecting an average over the range of recombination rates used, our estimates are disproportionally affected by data from regions with lower recombination rates—simply because each sweep in these regions affects many more codons than in higher recombination regions. Thus, by including lower recombination regions, we downwardly bias our estimates of the fraction of beneficial substitutions, and most strongly those under weak selection.

(3) This bias should be the stronger for weakly selected substitutions in UTRs than in exons, given that we measure diversity levels in exons, and therefore have proportionately fewer synonymous sites at short genetic distances from substitutions in UTRs in regions of high recombination than we do for exons. Consistent with this interpretation, when we decrease the recombination rate threshold below 0.1 cM/Mb, thus increasing the influence of low recombination regions on our estimates even further, estimates of the fraction of weakly beneficial amino acid substitutions begin to decrease.

Given these considerations, we conclude that our inference about a substantial number of weakly selected substitutions at UTRs to be correct. We choose to work with a threshold of 0.75 cM/Mb, because our results appear to be robust to increasing it to 1.25cM/Mb. Thus, compromising between minimizing the biases associated with regions with low recombination and maximizing our data lead us to this threshold. We note finally that even if weakly selected sweep are less common in low recombination regions, these regions still comprise a fairly small portion of the genome, such that our estimates based on our higher threshold are more reflective of the genomic rates.

   ***Codon usage bias***. While we use all synonymous changes as our proxy for neutral diversity, weak
2   selection for preferred codon usage is known to exist in *D. melanogaster* ([5,65,66] (but see
3   [67,68]). To examine whether such selection affects our results, we apply our inference to subsets
4   of synonymous changes that are thought to be subject to considerably less selection. Specifically, we
5   restrict the analysis to (i) preferred codons and preferred to preferred changes (P2P|P), as defined
6   by Vicario et al. (method A) [69]; or to (ii) unpreferred codons and unpreferred to unpreferred
7   changes (U2U|U). This restriction comes at the cost of a considerable reduction in the amount of
8   data: for (i), this restricts us to 61% of codons and 6% of synonymous changes, and for (ii), to 39%
9   of codons and 10% of synonymous changes. To control for variation in mutation rates, we use
10   estimates of synonymous divergence between *D. simulans* and *D. yakuba* in windows of 1780 kb
11   (not correcting for multiple hits in this case) using the same class of codons and changes as we do
12   for polymorphism.

13   Table S9 shows the parameter estimates and summaries for the two subsets of codons alongside
14   those for the entire dataset. To gauge the uncertainty in our estimates due to the dramatic
15   reduction in sample size ($\sim$10-20-fold), we divide each dataset into two subsets with half the data
16   (set A and set B in Table S9, see Section C for details, but without controlling for the correlation
17   between diversity and recombination) and infer the parameters for each half separately. This
18   analysis suggests that our estimates of individual selection parameters based on restricted (but not
19   the full) datasets are associated with considerable uncertainty, so that drawing conclusions about
20   the effects of weak selection based on these estimates would be misleading. Instead, we limit
21   ourselves to qualitative patterns and summaries that appear to be fairly robust despite the smaller
22   amount of data.

23   These patterns and summaries suggest that our inference is fairly insensitive to weak selection on
24   synonymous codon usage. Notably, both background selection and classic sweeps substantially
25   reduce average diversity levels (together between 49-76%), with a larger (2$\sim$3-fold) reduction due
26   to background selection compared to that of classic sweeps. The effects of classic sweeps are mainly
27   due to three classes of substitutions and selection effects, whose presence appear to be robust but
28   whose exact size varies: a minority of non-synonymous substitutions (0.6-3.7%) with large effects
29   ($s=10^{-2.5}$-$10^{-3.5}$) and a majority ($\sim$30-50%) with small effect sizes ($s=10^{-5.5}$), as well as many
30   substitutions at UTRs ($\sim$17-90%) with small effects ($s=10^{-4.5}$-$10^{-5.5}$). In turn, the effects of
31   background selection are mainly due to strongly selected mutations ($t=10^{-1.5}$), with a smaller contribution from intermediate-effect mutations ($t=10^{-2.5}$-$10^{-4.5}$) in exons. In conclusion, we see no clear indication that weak selection biases our inferences, though we lack power to detect a small effect.

| | All | | | P2P\|P | | | U2U\|U | | |
|---|---|---|---|---|---|---|---|---|---|
| | **All** | | | **P2P\|P** | | | **U2U\|U** | | |
| $k_{B\&S}(1-\bar{\pi}/\pi_0)$ | 76% | | | 49% | | | 72% | | |
| $k_B/(k_B+k_S)$ | 63% | | | 66% | | | 53% | | |
| $k_S/(k_B+k_S)$ | 37% | | | 34% | | | 47% | | |
| **Annotation** | Exons | UTRs | Introns | Exons | UTRs | Introns | Exons | UTRs | Introns |
| **Full** | | | | | | | | | |
| $u(t=10^{-1.5})/\mu$ | 451.0% | 612.0% | 11.0% | 2.0% | 152.0% | 50.0% | 390.0% | 540.0% | . |
| $u(t=10^{-2.5})/\mu$ | . | . | . | 6.0% | 81.0% | . | . | 2.0% | . |
| $u(t=10^{-3.5})/\mu$ | 61.0% | . | . | 14.0% | . | . | 65.0% | . | . |
| $u(t=10^{-4.5})/\mu$ | 8.0% | 10.0% | . | 9.0% | 1.0% | . | . | 36.0% | . |
| $u(t=10^{-5.5})/\mu$ | . | . | . | . | 17.0% | 1.0% | . | . | . |
| $\alpha(s=10^{-1.5})$ | . | . | . | . | . | . | . | . | . |
| $\alpha(s=10^{-2.5})$ | 0.8% | . | . | 0.5% | . | . | 0.6% | 0.3% | . |
| $\alpha(s=10^{-3.5})$ | 2.9% | . | . | 1.4% | . | . | . | . | . |
| $\alpha(s=10^{-4.5})$ | . | 0.9% | . | . | 16.6% | . | . | 5.1% | 2.6% |
| $\alpha(s=10^{-5.5})$ | 32.9% | 46.4% | . | 31.8% | . | 58.1% | 50.7% | 87.5% | . |
| $k_{B\&S}(1-\bar{\pi}/\pi_0)$ | 77% | | | 54% | | | 75% | | |
| $k_B/(k_B+k_S)$ | 66% | | | 75% | | | 70% | | |
| $k_S/(k_B+k_S)$ | 34% | | | 25% | | | 30% | | |
| **Annotation** | Exons | UTRs | Introns | Exons | UTRs | Introns | Exons | UTRs | Introns |
| **Set A** | | | | | | | | | |
| $u(t=10^{-1.5})/\mu$ | 482.0% | 583.0% | . | . | 3.0% | 4.0% | 320.0% | 795.0% | 8.0% |
| $u(t=10^{-2.5})/\mu$ | . | . | . | 144.0% | 2.0% | . | 4.0% | 3.0% | . |
| $u(t=10^{-3.5})/\mu$ | 86.0% | . | . | 8.0% | . | . | 120.0% | 3.0% | . |
| $u(t=10^{-4.5})/\mu$ | 3.0% | 15.0% | . | . | 58.0% | . | 2.0% | 2.0% | . |
| $u(t=10^{-5.5})/\mu$ | . | . | . | . | 2.0% | . | . | 2.0% | . |
| $\alpha(s=10^{-1.5})$ | . | . | . | . | . | . | . | . | . |
| $\alpha(s=10^{-2.5})$ | 0.5% | . | . | 0.2% | . | . | . | 0.4% | . |
| $\alpha(s=10^{-3.5})$ | 2.9% | . | . | 3.0% | . | . | . | . | . |
| $\alpha(s=10^{-4.5})$ | . | . | . | . | 3.1% | . | . | . | 0.9% |
| $\alpha(s=10^{-5.5})$ | 27.6% | 30.4% | . | 26.6% | . | 33.8% | 47.6% | 62.4% | . |
| $k_{B\&S}(1-\bar{\pi}/\pi_0)$ | 74% | | | 46% | | | 71% | | |
| $k_B/(k_B+k_S)$ | 60% | | | 59% | | | 58% | | |
| $k_S/(k_B+k_S)$ | 40% | | | 41% | | | 42% | | |
| **Annotation** | Exons | UTRs | Introns | Exons | UTRs | Introns | Exons | UTRs | Introns |
| **Set B** | | | | | | | | | |
| $u(t=10^{-1.5})/\mu$ | 253.0% | 540.0% | 101.0% | 2.0% | 299.0% | 56.0% | 351.0% | 655.0% | . |
| $u(t=10^{-2.5})/\mu$ | . | . | . | 1.0% | 3.0% | . | . | 2.0% | . |
| $u(t=10^{-3.5})/\mu$ | 43.0% | . | . | . | . | . | 17.0% | 18.0% | . |
| $u(t=10^{-4.5})/\mu$ | 13.0% | . | . | 14.0% | . | . | 1.0% | 45.0% | . |
| $u(t=10^{-5.5})/\mu$ | . | . | . | . | 17.0% | 3.0% | . | 6.0% | . |
| $\alpha(s=10^{-1.5})$ | . | . | . | . | . | . | . | . | . |
| $\alpha(s=10^{-2.5})$ | 1.0% | . | . | 0.5% | . | . | 1.2% | 0.1% | . |
| $\alpha(s=10^{-3.5})$ | 3.0% | . | . | 0.2% | . | 0.1% | . | . | . |
| $\alpha(s=10^{-4.5})$ | . | 6.5% | . | 5.2% | 30.9% | . | . | 13.2% | 1.6% |
| $\alpha(s=10^{-5.5})$ | 39.5% | 37.5% | . | 27.6% | 15.7% | 99.5% | 61.7% | 113.6% | 17.8% |

1  **Table S9. Sensitivity of our inference to selection for synonymous codon usage.** All summaries
2  and estimates correspond to inferences based on the combined model with background selection
3  and classic sweeps. Estimates associated with intergenic regions are negligible and are therefore
4  omitted from the table.

### *I. Inference based on additional models*

In the Results, we focus on three models. The main one considers the combined effects of background selection and classic selective sweeps at each of four functional annotations, with a grid of selection coefficients for each annotation and mode of selection consisting of 5 point masses, i.e., $t$ and $s = 10^{-5.5}$, $10^{-4.5}$, $10^{-3.5}$, $10^{-2.5}$ and $10^{-1.5}$. To investigate the support for including both modes of selection in our inferences, we also report the results for models with background selection alone and classic sweeps alone. To complete our analysis of the main model, here, we consider two additional variants of it. First, we consider how our inference is affected by using a finer grid of selection coefficients, where instead of 5 point masses, we use 11 with $t$ and $s = 10^{-6}$, $10^{-5.5}$, $10^{-5}$, $10^{-4.5}$, $10^{-4}$, $10^{-3.5}$, $10^{-3}$, $10^{-2.5}$, $10^{-2}$, $10^{-1.5}$ and $10^{-1}$. Second, we consider how our inference is affected by focusing only on the annotations whose effects appear to dominate, namely on exons and UTRs alone. The parameter estimates and goodness-of-fit statistics for these models are shown in Table S10.

We find that both variants of the model have little effect on our inferences. Using a finer grid of selection coefficients offers little improvement to the quality of our predictions and has only minor effects on our parameter estimates, leading to our choice to work primarily with the simpler model with only 5 grid points. We also find that including long introns and intergenic regions has little effect. This finding does not imply little selection in these annotations (see [9,11,70]) as we use synonymous changes to measure diversity levels and thus have less power to make reliable inferences about selection in more distant annotations.

| Model | Background selection and classic sweeps | | | Background selection and classic sweeps 11 point masses grids | | | BS and CS Selection at exons and UTRs only | |
|---|---|---|---|---|---|---|---|---|
| $k_{B\&S}(1-\bar\pi/\pi_0)$ | 73% | | | 86% | | | 75% | |
| Annotation | Exons | UTRs | Introns | Exons | UTRs | Introns | Exons | UTRs |
| **Parameters** — BS parameters | | | | | | | | |
| $u(t=10^{-1})\ /\ \mu$ | | | | 679% | 298% | 10% | | |
| $u(t=10^{-1.5})\ /\ \mu$ | 377% | 577% | 19% | 193% | 536% | . | 441% | 665% |
| $u(t=10^{-2})\ /\ \mu$ | | | | 1% | 3% | . | | |
| $u(t=10^{-2.5})\ /\ \mu$ | 2% | 2% | | . | 2% | . | . | . |
| $u(t=10^{-3})\ /\ \mu$ | | | | . | 1% | . | | |
| $u(t=10^{-3.5})\ /\ \mu$ | 56% | . | . | 11% | . | . | 58% | . |
| $u(t=10^{-4})\ /\ \mu$ | | | | 41% | 1% | . | | |
| $u(t=10^{-4.5})\ /\ \mu$ | 2% | 23% | | 1% | . | . | . | 28% |
| $u(t=10^{-5})\ /\ \mu$ | | | | . | . | . | | |
| $u(t=10^{-5.5})\ /\ \mu$ | . | 2% | | . | . | . | . | . |
| $u(t=10^{-6})\ /\ \mu$ | | | | . | . | . | | |
| CS parameters | | | | | | | | |
| $\alpha(s=10^{-1})$ | | | | 0.1% | . | . | | |
| $\alpha(s=10^{-1.5})$ | . | . | | . | . | . | . | . |
| $\alpha(s=10^{-2})$ | | | | | | | | |
| $\alpha(s=10^{-2.5})$ | 0.6% | . | | 0.6% | . | . | 0.6% | . |
| $\alpha(s=10^{-3})$ | | | | . | 0.4% | . | | . |
| $\alpha(s=10^{-3.5})$ | 3.5% | 0.0% | | 1.5% | . | . | 3.6% | . |
| $\alpha(s=10^{-4})$ | | | | 3.9% | . | . | | |
| $\alpha(s=10^{-4.5})$ | 0.0% | 5.1% | | . | 0.2% | . | . | 4.6% |
| $\alpha(s=10^{-5})$ | | | | 0.4% | 24.0% | . | | |
| $\alpha(s=10^{-5.5})$ | 36.3% | 42.1% | | 4.8% | 1.3% | . | 37.1% | 44.0% |
| $\alpha(s=10^{-6})$ | | | | 40.4% | . | | | |
| $\Delta CL$ | $3.9\times10^{-4}$ | | | $4.0\times10^{-4}$ | | | $3.9\times10^{-4}$ | |
| **Diversity binned in local windows** — $R^2$ 1 Mb | 0.71 | | | 0.69 | | | 0.71 | |
| 100 kb | 0.44 | | | 0.43 | | | 0.44 | |
| 10 kb | 0.26 | | | 0.26 | | | 0.26 | |
| 1 kb | 0.20 | | | 0.20 | | | 0.20 | |
| **Diversity binned by distance from substitution** — $R^2$ NS substitutions | 0.62 | | | 0.64 | | | 0.63 | |
| SYN substitutions | 0.66 | | | 0.65 | | | 0.68 | |
| **Diversity binned by predicted effect of linked selection** — Spearman's $\rho$ | 0.913 | | | 0.913 | | | 0.910 | |
| Upper-to-lower tails observed diversity ratio | 5.3 | | | 5.1 | | | 5.3 | |
| **Diversity reduction measures** — $k_{B\&S}(1-\bar\pi/\pi_0)$ | 73% | | | 86% | | | 75% | |
| $k_B$ | 67% | | | 81% | | | 69% | |
| $k_S$ | 41% | | | 69% | | | 43% | |
| $k_B/(k_B+k_S)$ | 62% | | | 54% | | | 62% | |
| $k_S/(k_B+k_S)$ | 38% | | | 46% | | | 38% | |
| **Coalescent rate measures** — $r_B+r_S$ | 3.18 | | | 7.59 | | | 3.51 | |
| $r_B$ | 2.26 | | | 4.72 | | | 2.52 | |
| $r_S$ | 0.92 | | | 2.87 | | | 0.98 | |
| $r_B/(r_B+r_S)$ | 71% | | | 62% | | | 72% | |
| $r_S/(r_B+r_S)$ | 29% | | | 38% | | | 28% | |

Table S10. Comparison to models with a finer grid of selection coefficients and fewer functional annotations.

1   *J. Additional figures and tables*

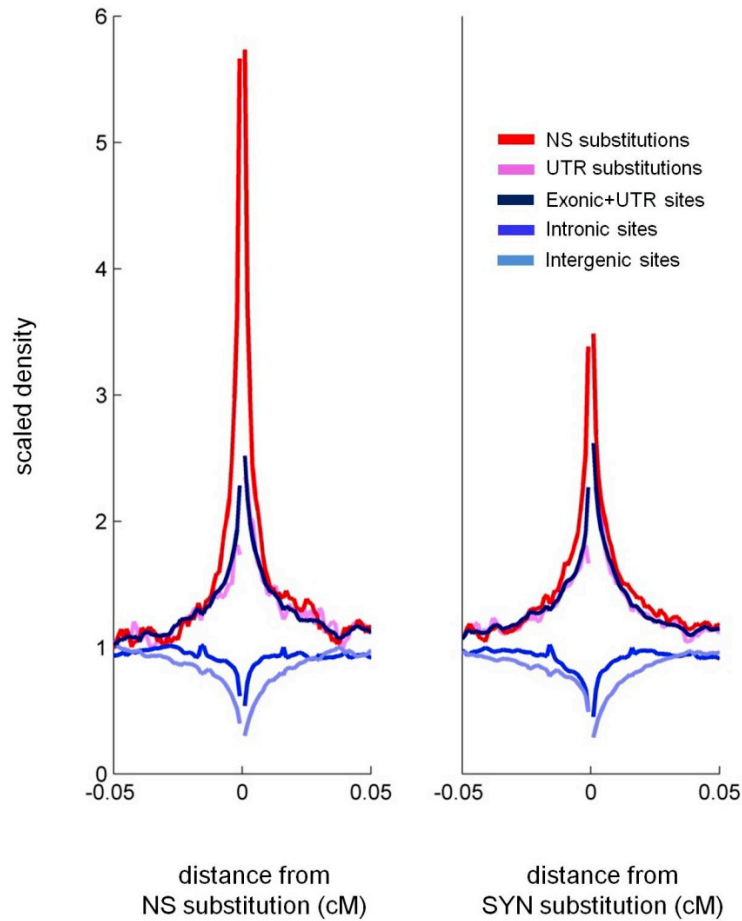

2   **Fig S6**. **Densities of functional annotations around synonymous and non-synonymous**

3   **substitutions (complementary to Fig 4 in the main text).** Shown are the densities of all

4   annotations for which we inferred non-negligible selection parameters: non-synonymous

5   (NS) substitutions (red), UTR substitutions (pink), exonic and UTR sites (black), intronic

6   sites (dark blue), and intergenic sites (light blue). Densities are plotted relative to their

7   level at distance 0.1 cM away from the focal substitutions.

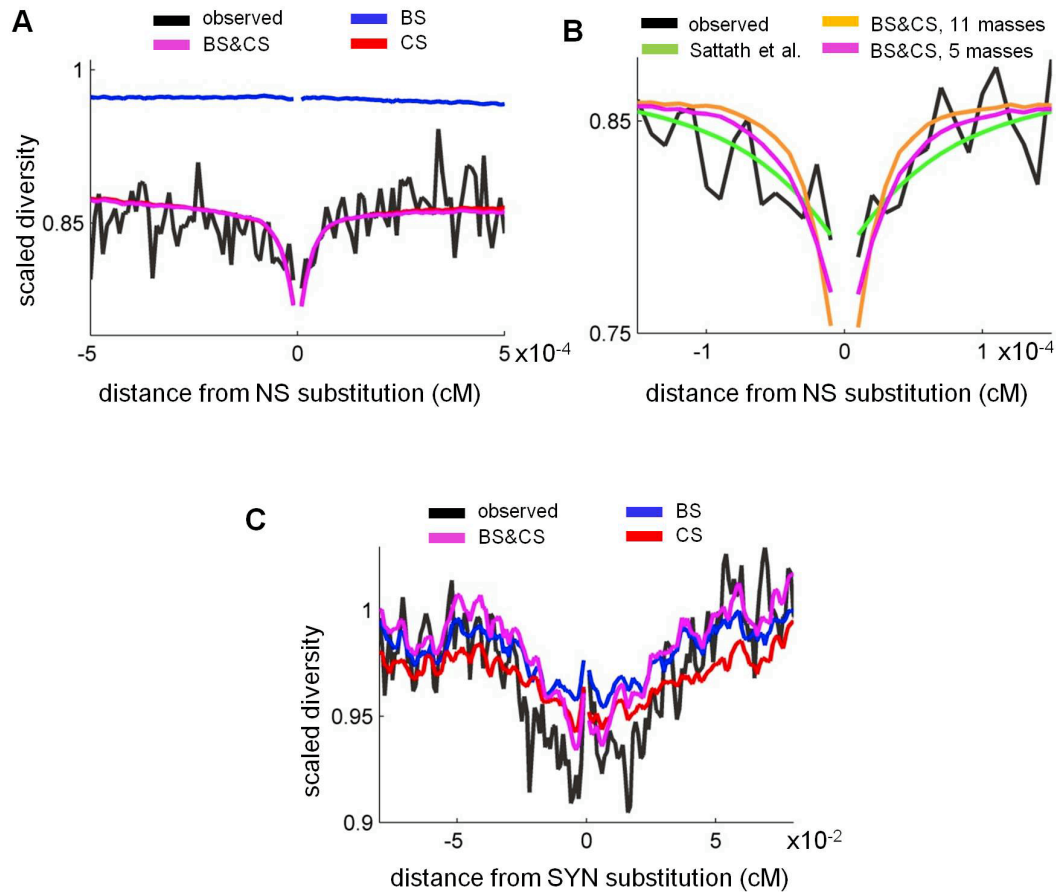

1 **Fig S7**. **Observed and predicted scaled diversity levels around exonic substitutions**
2 **(complementary to Fig 3B and 5A in the main text).** (**A**) A close up on Fig 5A in the
3 main text near non-synonymous substitutions (<5×10⁻⁴ cM). (**B**) A close up on Fig 3B in the
4 main text near non-synonymous substitutions (<1.5×10⁻⁴ cM). Also shown are the
5 predictions based on the model with a finer grid of selection coefficients (11 points). (**C**)
6 The equivalent of Fig 5A in the main text around synonymous (rather than non-
7 synonymous) substitutions.

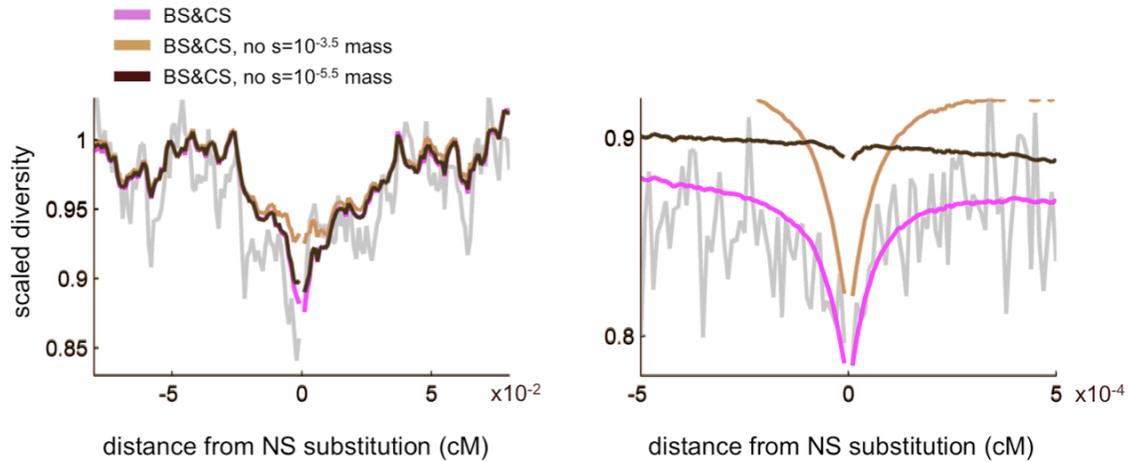

1  **Fig S8**. **The contribution of the two inferred modes of sweeps to diversity levels**
2  **around non-synonymous substitutions.** To isolate the contribution of each mode, we
3  remove it from our predictions. Removing sweeps with $s=10^{-3.5}$ (brown, left) removes a
4  trough in diversity levels on the scale of $10^{-2}$ cM that is apparent in the full model (pink,
5  left) and in the data (gray, left). In turn, removing sweeps with $s=10^{-5.5}$ (black, right)
6  removes a trough on the scale of $10^{-4}$ cM apparent in both the full model (pink, right) and
7  data (gray, right). In the models where a mode of sweep was excluded, we scaled the
8  predicted diversity levels in the absence of linked selection such that the predicted
9  genome-wide average will coincide with the full model and data.

**Table S11. Detailed parameter estimates for all models.**

**A. Classic sweeps parameters**. For introns we always find $\alpha$=0 and have therefore omitted these estimates from the table.

| Model | Background selection and classic sweeps | | Background selection and classic sweeps 11 masses | | Background selection and classic sweeps constrained | | Classic sweeps | | Classic sweeps Sattath et al. |
|---|---|---|---|---|---|---|---|---|---|
| $k_{B\&S}(1-\bar{\pi}/\pi_0)$ | 73% | | 86% | | 59% | | 43% | | 35% |
| $\sum_s \alpha(s)\cdot s$ | 3.5×10⁻⁵ | | 1.2×10⁻⁴ | | 3.3×10⁻⁵ | | 1.5×10⁻⁴ | | 6.1×10⁻⁵ |
| Annotation | Exons | UTRs | Exons | UTRs | Exons | UTRs | Exons | UTRs | Exons |
| $\sum_s \alpha(s)\cdot s$ | 3.2×10⁻⁵ | 3.0×10⁻⁶ | 1.1×10⁻⁴ | 6.3×10⁻⁶ | 3.0×10⁻⁵ | 3.0×10⁻⁶ | 1.1×10⁻⁴ | 3.6×10⁻⁵ | 6.1×10⁻⁵ |
| $\alpha$ | 40% | 47% | 52% | 26% | 42% | 51% | 40% | 51% | 20% |
| $\alpha(s=10^{-1.0})$ | | | 0.1% | . | | | | | |
| $\alpha(s=10^{-1.5})$ | . | . | . | . | . | . | 0.3% | 0.1% | |
| $\alpha(s=10^{-2.0})$ | | | . | . | | | | | 1.5% (s=4.1E-03) |
| $\alpha(s=10^{-2.5})$ | 0.6% | . | 0.6% | . | 0.6% | . | 0.4% | . | |
| $\alpha(s=10^{-3.0})$ | | | . | 0.4% | | | | | |
| $\alpha(s=10^{-3.5})$ | 3.5% | . | 1.5% | . | 3.5% | . | 3.7% | . | |
| $\alpha(s=10^{-4.0})$ | | | 3.9% | . | | | | | |
| $\alpha(s=10^{-4.5})$ | . | 5.1% | . | 0.2% | . | 4.7% | . | 4.4% | |
| $\alpha(s=10^{-5.0})$ | | | 0.4% | 24.0% | | | | | 18.8% (s=5.8E-06) |
| $\alpha(s=10^{-5.5})$ | 36.3% | 42.1% | 4.8% | 1.3% | 38.1% | 45.9% | 35.7% | 46.1% | |
| $\alpha(s=10^{-6.0})$ | | | 40.4% | . | | | | | |

## B. Background selection parameters.

| Model | Background selection and classic sweeps | | | | Background selection | | | | Background selection and classic sweeps 11 masses | | | |
|---|---|---|---|---|---|---|---|---|---|---|---|---|
| $k_{B\&S}(1-\bar{\pi}/\pi_0)$ | 73% | | | | 65% | | | | 86% | | | |
| $U_{del}$ (per diploid) | 1.60 | | | | 1.46 | | | | 2.90 | | | |
| Annotation | Exons (20%) | UTRs (5%) | Introns (40%) | Intergenic (35%) | Exons | UTRs | Introns | Intergenic | Exons | UTRs | Introns | Intergenic |
| $u_{del}/\mu$ | 437% | 603% | 19% | . | 448% | 456% | 17% | . | 926% | 842% | 10% | 1% |
| $u(t=10^{-1.0})/\mu$ | | | | | | | | | 679% | 298% | 10% | 1% |
| $u(t=10^{-1.5})/\mu$ | 377% | 577% | 19% | . | 369% | 453% | 17% | . | 193% | 536% | . | . |
| $u(t=10^{-2.0})/\mu$ | | | | | | | | | 1% | 3% | . | . |
| $u(t=10^{-2.5})/\mu$ | 2% | 2% | . | . | . | . | . | . | . | 2% | . | . |
| $u(t=10^{-3.0})/\mu$ | | | | | | | | | . | 1% | . | . |
| $u(t=10^{-3.5})/\mu$ | 56% | . | . | . | 77% | . | . | . | 11% | . | . | . |
| $u(t=10^{-4.0})/\mu$ | | | | | | | | | 41% | 1% | . | . |
| $u(t=10^{-4.5})/\mu$ | 2% | 23% | . | . | 2% | 3% | . | . | 1% | . | . | . |
| $u(t=10^{-5.0})/\mu$ | | | | | | | | | . | . | . | . |
| $u(t=10^{-5.5})/\mu$ | . | 2% | . | . | . | . | . | . | . | . | . | . |
| $u(t=10^{-6.0})/\mu$ | | | | | | | | | . | . | . | . |

| Model | Background selection and classic sweeps constrained | | | | Background selection constrained | | | | Background selection Charlesworth | | | |
|---|---|---|---|---|---|---|---|---|---|---|---|---|
| $k_{B\&S}(1-\bar{\pi}/\pi_0)$ | 59% | | | | 49% | | | | 37% | | | |
| $U_{del}$ (per diploid) | 0.92 | | | | 0.90 | | | | 0.56 | | | |
| Annotation | Exons | UTRs | Introns | Intergenic | Exons | UTRs | Introns | Intergenic | Exons | UTRs | Introns | Intergenic |
| $u_{del}/\mu$ | 90% | 88% | 87% | 44% | 90% | 88% | 85% | 38% | 72% | 38% | 38% | 38% |
| $u(t=10^{-1.0})/\mu$ | | | | | | | | | | | | |
| $u(t=10^{-1.5})/\mu$ | 18% | 67% | 85% | 44% | 1% | 84% | 85% | 38% | . | . | . | . |
| $u(t=10^{-2.0})/\mu$ | | | | | | | | | | | | |
| $u(t=10^{-2.5})/\mu$ | 2% | 2% | 1% | . | 2% | 3% | . | . | 25% | 7% | 7% | 7% |
| $u(t=10^{-3.0})/\mu$ | | | | | | | | | | | | |
| $u(t=10^{-3.5})/\mu$ | 69% | . | . | . | 85% | . | . | . | 27% | 8% | 8% | 8% |
| $u(t=10^{-4.0})/\mu$ | | | | | | | | | | | | |
| $u(t=10^{-4.5})/\mu$ | 1% | 20% | . | . | 2% | 2% | . | . | 15% | 10% | 10% | 10% |
| $u(t=10^{-5.0})/\mu$ | | | | | | | | | | | | |
| $u(t=10^{-5.5})/\mu$ | . | . | . | . | . | . | . | . | 5% | 12% | 12% | 12% |
| $u(t=10^{-6.0})/\mu$ | | | | | | | | | | | | |

**Table S12. Goodness-of-fit and other summaries for all models.**

| | Model | BS & CS | BS & CS 11 masses | BS & CS constrained | BS & CS excluding UTR sweeps | BS | BS constrained | CS | BS Charlesworth | CS Sattath et al. | BS Kim & Stephan by $c$ | CS Wiehe & Stephan by $c$ | CS Wiehe & Stephan by $D_n$ |
|---|---|---|---|---|---|---|---|---|---|---|---|---|---|
| | $\Delta CL$ | $3.9\times10^{-4}$ | $4.0\times10^{-4}$ | $3.6\times10^{-4}$ | $3.8\times10^{-4}$ | $2.8\times10^{-4}$ | $2.5\times10^{-4}$ | $2.4\times10^{-4}$ | $-6.7\times10^{-5}$ | | | | |
| **Diversity binned in local windows** | $R^2$  1 Mb | 0.71 | 0.69 | 0.69 | 0.72 | 0.76 | 0.72 | 0.67 | 0.58 | | | | |
| | 100 kb | 0.44 | 0.43 | 0.43 | 0.45 | 0.42 | 0.41 | 0.39 | 0.19 | | | | |
| | 10 kb | 0.26 | 0.26 | 0.24 | 0.27 | 0.23 | 0.22 | 0.21 | 0.09 | | | | |
| | 1 kb | 0.20 | 0.20 | 0.19 | 0.20 | 0.18 | 0.16 | 0.16 | 0.08 | | | | |
| **Diversity binned by distance from substitution** | $R^2$ NS substitutions | 0.62 | 0.64 | 0.61 | 0.64 | 0.27 | 0.24 | 0.51 | 0 | 0.56 | | | |
| | $R^2$ SYN substitutions | 0.66 | 0.65 | 0.69 | 0.68 | 0.53 | 0.53 | 0.49 | 0.05 | 0.65 | | | |
| **Diversity binned by predicted linked selection effect** | Spearman's $\rho$ | 0.913 | 0.913 | 0.905 | 0.906 | 0.745 | 0.737 | 0.89 | 0.773 | 0.869 | 0.732 | 0.732 | 0.807 |
| | Upper tail diversity reduction | 60% | 81% | 36% | 62% | 57% | 32% | 18% | 28% | -13%(*) | 9% | 11% | -5%(*) |
| | Upper-to-lower tails observed diversity ratio | 5.3 | 5.1 | 5.4 | 5.4 | 4.4 | 4.3 | 5.0 | 3.5 | 4.5 | 3.2 | 3.2 | 4.1 |
| **Diversity reduction measures** | $\pi_0/\bar{\pi}$ | 4.4 | 9.0 | 2.8 | 4.7 | 3.4 | 2.3 | 2.0 | 1.8 | 1.5 | 1.5 | 1.5 | 1.6 |
| | $k_{B\&S}\,(1-\bar{\pi}/\pi_0)$ | 73% | 86% | 59% | 74% | 65% | 49% | 43% | 37% | | | | |
| | $k_B$ | 67% | 81% | 49% | 69% | 66% | 49% | 0% | 38% | | | | |
| | $k_S$ | 41% | 69% | 32% | 37% | 0% | 0% | 43% | 0% | | | | |
| | $k_B/(k_B+k_S)$ | 62% | 54% | 61% | 65% | 100% | 100% | 0% | 100% | | | | |
| | $k_S/(k_B+k_S)$ | 38% | 46% | 39% | 35% | 0% | 0% | 100% | 0% | | | | |
| **Coalescent rate measures** | $r_B+r_S$ | 3.18 | 7.59 | 1.66 | 3.28 | 2.13 | 1.07 | 0.87 | 5.23 | | | | |
| | $r_B$ | 2.26 | 4.72 | 1.08 | 2.51 | 2.13 | 1.07 | 0 | 5.23 | | | | |
| | $r_S$ | 0.92 | 2.87 | 0.58 | 0.77 | 0 | 0 | 0.87 | 0 | | | | |
| | $r_B/(r_B+r_S)$ | 71% | 62% | 65% | 77% | 100% | 100% | 0% | 100% | | | | |
| | $r_S/(r_B+r_S)$ | 29% | 38% | 35% | 23% | 0% | 0% | 100% | 0% | | | | |

(*) The negative value reflects the fact that the observed diversity level is higher than the level predicted in the absence of linked selection.

# References

1. Mackay TF, Richards S, Stone EA, Barbadilla A, Ayroles JF, Zhu D, Casillas S, Han Y, Magwire MM, Cridland JM, Richardson MF, Anholt RR, Barron M, Bess C, Blankenburg KP, Carbone MA, Castellano D, Chaboub L, Duncan L, Harris Z, Javaid M, Jayaseelan JC, Jhangiani SN, Jordan KW, Lara F, Lawrence F, Lee SL, Librado P, Linheiro RS, Lyman RF, Mackey AJ, Munidasa M, Muzny DM, Nazareth L, Newsham I, Perales L, Pu LL, Qu C, Ramia M, Reid JG, Rollmann SM, Rozas J, Saada N, Turlapati L, Worley KC, Wu YQ, Yamamoto A, Zhu Y, Bergman CM, Thornton KR, Mittelman D, Gibbs RA (2012) The Drosophila melanogaster Genetic Reference Panel. Nature 482: 173-178.

2. Hu TT, Eisen MB, Thornton KR, Andolfatto P (2013) A second-generation assembly of the Drosophila simulans genome provides new insights into patterns of lineage-specific divergence. Genome Res 23: 89-98.

3. Adams MD, Celniker SE, Holt RA, Evans CA, Gocayne JD, Amanatides PG, Scherer SE, Li PW, Hoskins RA, Galle RF, George RA, Lewis SE, Richards S, Ashburner M, Henderson SN, Sutton GG, Wortman JR, Yandell MD, Zhang Q, Chen LX, Brandon RC, Rogers YH, Blazej RG, Champe M, Pfeiffer BD, Wan KH, Doyle C, Baxter EG, Helt G, Nelson CR, Gabor GL, Abril JF, Agbayani A, An HJ, Andrews-Pfannkoch C, Baldwin D, Ballew RM, Basu A, Baxendale J, Bayraktaroglu L, Beasley EM, Beeson KY, Benos PV, Berman BP, Bhandari D, Bolshakov S, Borkova D, Botchan MR, Bouck J, Brokstein P, Brottier P, Burtis KC, Busam DA, Butler H, Cadieu E, Center A, Chandra I, Cherry JM, Cawley S, Dahlke C, Davenport LB, Davies P, de Pablos B, Delcher A, Deng Z, Mays AD, Dew I, Dietz SM, Dodson K, Doup LE, Downes M, Dugan-Rocha S, Dunkov BC, Dunn P, Durbin KJ, Evangelista CC, Ferraz C, Ferriera S, Fleischmann W, Fosler C, Gabrielian AE, Garg NS, Gelbart WM, Glasser K, Glodek A, Gong F, Gorrell JH, Gu Z, Guan P, Harris M, Harris NL, Harvey D, Heiman TJ, Hernandez JR, Houck J, Hostin D, Houston KA, Howland TJ, Wei MH, Ibegwam C, Jalali M, Kalush F, Karpen GH, Ke Z, Kennison JA, Ketchum KA, Kimmel BE, Kodira CD, Kraft C, Kravitz S, Kulp D, Lai Z, Lasko P, Lei Y, Levitsky AA, Li J, Li Z, Liang Y, Lin X, Liu X, Mattei B, McIntosh TC, McLeod MP, McPherson D, Merkulov G, Milshina NV, Mobarry C, Morris J, Moshrefi A, Mount SM, Moy M, Murphy B, Murphy L, Muzny DM, Nelson DL, Nelson DR, Nelson KA, Nixon K, Nusskern DR, Pacleb JM, Palazzolo M, Pittman GS, Pan S, Pollard J, Puri V, Reese MG, Reinert K, Remington K, Saunders RD, Scheeler F, Shen H, Shue BC, Siden-Kiamos I, Simpson M, Skupski MP, Smith T, Spier E,

Spradling AC, Stapleton M, Strong R, Sun E, Svirskas R, Tector C, Turner R, Venter E, Wang AH, Wang X, Wang ZY, Wassarman DA, Weinstock GM, Weissenbach J, Williams SM, WoodageT, Worley KC, Wu D, Yang S, Yao QA, Ye J, Yeh RF, Zaveri JS, Zhan M, Zhang G, Zhao Q, Zheng L, Zheng XH, Zhong FN, Zhong W, Zhou X, Zhu S, Zhu X, Smith HO, Gibbs RA, Myers EW, Rubin GM, Venter JC (2000) The genome sequence of Drosophila melanogaster. Science 287: 2185-2195.

4. McDonald JH, Kreitman M (1991) Adaptive protein evolution at the Adh locus in Drosophila. Nature 351: 652-654.

5. Akashi H (1995) Inferring Weak Selection from Patterns of Polymorphism and Divergence at Silent Sites in Drosophila DNA. Genetics 139: 1067-1076.

6. Andolfatto P, Przeworski M (2000) A genome-wide departure from the standard neutral model in natural populations of Drosophila. Genetics 156: 257-268.

7. Smith NG, Eyre-Walker A (2002) Adaptive protein evolution in Drosophila. Nature 415: 1022-1024.

8. Zeng K, Charlesworth B (2010) Studying Patterns of Recent Evolution at Synonymous Sites and Intronic Sites in Drosophila melanogaster. J Mol Evol 70: 116-128.

9. Andolfatto P (2005) Adaptive evolution of non-coding DNA in Drosophila. Nature 437: 1149-1152.

10. Halligan DL, Keightley PD (2006) Ubiquitous selective constraints in the Drosophila genome revealed by a genome-wide interspecies comparison. Genome Res 16: 875-884.

11. Casillas S, Barbadilla A, Bergman CM (2007) Purifying selection maintains highly conserved noncoding sequences in Drosophila. Mol Biol Evol 24: 2222-2234.

12. St Pierre SE, Ponting L, Stefancsik R, McQuilton P, FlyBase C (2014) FlyBase 102--advanced approaches to interrogating FlyBase. Nucleic acids research 42: D780-788.

13. Comeron JM, Ratnappan R, Bailin S (2012) The Many Landscapes of Recombination in Drosophila melanogaster. PLoS Genet 8: e1002905.

14. Chan AH, Jenkins PA, Song YS (2012) Genome-Wide Fine-Scale Recombination Rate Variation in Drosophila melanogaster. PLoS Genet 8: e1003090.

15. Parsch J, Novozhilov S, Saminadin-Peter SS, Wong KM, Andolfatto P (2010) On the utility of short intron sequences as a reference for the detection of positive and negative selection in Drosophila. Mol Biol Evol 27: 1226-1234.

16. Clemente F, Vogl C (2012) Unconstrained evolution in short introns? - an analysis of genome-wide polymorphism and divergence data from Drosophila. Journal of evolutionary biology 25: 1975-1990.

17. Stapleton M, Carlson J, Brokstein P, Yu C, Champe M, George R, Guarin H, Kronmiller B, Pacleb J, Park S, Wan K, Rubin GM, Celniker SE (2002) A Drosophila full-length cDNA resource. Genome biology 3: RESEARCH0080.

18. Stone EA (2012) Joint genotyping on the fly: identifying variation among a sequenced panel of inbred lines. Genome Res 22: 966-974.

19. Cridland JM, Macdonald SJ, Long AD, Thornton KR (2013) Abundance and distribution of transposable elements in two Drosophila QTL mapping resources. Mol Biol Evol 30: 2311-2327.

20. Yang Z (1997) PAML: a program package for phylogenetic analysis by maximum likelihood. Comput Appl Biosci 13: 555-556.

21. Sattath S, Elyashiv E, Kolodny O, Rinott Y, Sella G (2011) Pervasive Adaptive Protein Evolution Apparent in Diversity Patterns around Amino Acid Substitutions in Drosophila simulans. PLoS Genet 7: e1001302.

22. McVicker G, Gordon D, Davis C, Green P (2009) Widespread Genomic Signatures of Natural Selection in Hominid Evolution. PLoS Genet 5: e1000471.

23. Nordborg M, Charlesworth B, Charlesworth D (1996) The effect of recombination on background selection. Genet Res 67: 159-174.

24. Kaplan NL, Hudson RR, Langley CH (1989) The "hitchhiking effect" revisited. Genetics 123: 887-899.

25. The MathWorks I, Natick, Massachusetts, United States MATLAB and Optimization Toolbox Release 2013b.

26. Arlot S, Celisse A (2010) A survey of cross-validation procedures for model selection. Stat Surv 4: 40-79.

27. Coop G, Ralph P (2012) Patterns of Neutral Diversity Under General Models of Selective Sweeps. Genetics 192: 205-U438.

28. Chevin LM, Hospital F (2008) Selective sweep at a quantitative trait locus in the presence of background genetic variation. Genetics 180: 1645-1660.

29. Ralph P, Coop G (2010) Parallel adaptation: one or many waves of advance of an advantageous allele? Genetics 186: 647-668.

30. Pennings PS, Hermisson J (2006) Soft sweeps II--molecular population genetics of adaptation from recurrent mutation or migration. Mol Biol Evol 23: 1076-1084.

31. Pennings PS, Hermisson J (2006) Soft sweeps III: the signature of positive selection from recurrent mutation. PLoS Genet 2: e186.

32. Innan H, Kim Y (2004) Pattern of polymorphism after strong artificial selection in a domestication event. Proc Natl Acad Sci U S A 101: 10667-10672.

33. Przeworski M, Coop G, Wall JD (2005) The signature of positive selection on standing genetic variation. Evolution Int J Org Evolution 59: 2312-2323.

34. Hermisson J, Pennings PS (2005) Soft sweeps: molecular population genetics of adaptation from standing genetic variation. Genetics 169: 2335-2352.

35. Peter BM, Huerta-Sanchez E, Nielsen R (2012) Distinguishing between Selective Sweeps from Standing Variation and from a De Novo Mutation. PLoS Genet 8: e1003011.

36. Berg JJ, Coop G (2015) A Coalescent Model for a Sweep of a Unique Standing Variant. Genetics 201: 707-725.

37. Teshima KM, Przeworski M (2006) Directional positive selection on an allele of arbitrary dominance. Genetics 172: 713-718.

38. Ewing G, Hermisson J, Pfaffelhuber P, Rudolf J (2011) Selective sweeps for recessive alleles and for other modes of dominance. Journal of mathematical biology 63: 399-431.

39. Ewens WJ (2004) Mathematical population genetics. New York: Springer. v. 1.

40. de Vladar HP, Barton N (2014) Stability and Response of Polygenic Traits to Stabilizing Selection and Mutation. Genetics 197: 749-767.

41. Keightley PD, Trivedi U, Thomson M, Oliver F, Kumar S, Blaxter ML (2009) Analysis of the genome sequences of three Drosophila melanogaster spontaneous mutation accumulation lines. Genome Res 19: 1195-1201.

42. Haag-Liautard C, Dorris M, Maside X, Macaskill S, Halligan DL, Houle D, Charlesworth B, Keightley PD (2007) Direct estimation of per nucleotide and genomic deleterious mutation rates in Drosophila. Nature 445: 82-85.

43. Nuzhdin SV, Mackay TF (1995) The genomic rate of transposable element movement in Drosophila melanogaster. Mol Biol Evol 12: 180-181.

44. Charlesworth B (1996) Background selection and patterns of genetic diversity in Drosophila melanogaster. Genet Res 68: 131-149.

45. Houle D, Nuzhdin SV (2004) Mutation accumulation and the effect of copia insertions in Drosophila melanogaster. Genet Res 83: 7-18.

46. Charlesworth B (2012) The Role of Background Selection in Shaping Patterns of Molecular Evolution and Variation: Evidence from Variability on the Drosophila X Chromosome. Genetics 191: 233-246.

47. Comeron JM (2014) Background Selection as Baseline for Nucleotide Variation across the Drosophila Genome. PLoS Genet 10: e1004434.

48. Keightley PD, Eyre-Walker A (2007) Joint inference of the distribution of fitness effects of deleterious mutations and population demography based on nucleotide polymorphism frequencies. Genetics 177: 2251-2261.

49. Wiehe TH, Stephan W (1993) Analysis of a genetic hitchhiking model, and its application to DNA polymorphism data from Drosophila melanogaster. Mol Biol Evol 10: 842-854.

50. Kim Y, Stephan W (2000) Joint effects of genetic hitchhiking and background selection on neutral variation. Genetics 155: 1415-1427.

51. Macpherson JM, Sella G, Davis JC, Petrov DA (2007) Genomewide spatial correspondence between nonsynonymous divergence and neutral polymorphism reveals extensive adaptation in Drosophila. Genetics 177: 2083-2099.

52. Andolfatto P (2007) Hitchhiking effects of recurrent beneficial amino acid substitutions in the Drosophila melanogaster genome. Genome Res 17: 1755-1762.

53. Kliman RM, Hey J (1993) Reduced natural selection associated with low recombination in Drosophila melanogaster. Mol Biol Evol 10: 1239-1258.

54. Comeron JM, Kreitman M, Aguade M (1999) Natural selection on synonymous sites is correlated with gene length and recombination in Drosophila. Genetics 151: 239-249.

55. Betancourt AJ, Presgraves DC (2002) Linkage limits the power of natural selection in Drosophila. Proc Natl Acad Sci U S A 99: 13616-13620.

56. Hey J, Kliman RM (2002) Interactions between natural selection, recombination and gene density in the genes of Drosophila. Genetics 160: 595-608.

57. Marais G, Domazet-Loso T, Tautz D, Charlesworth B (2004) Correlated evolution of synonymous and nonsynonymous sites in Drosophila. J Mol Evol 59: 771-779.

58. Presgraves DC (2005) Recombination enhances protein adaptation in Drosophila melanogaster. Current biology : CB 15: 1651-1656.

59. Zhang Z, Parsch J (2005) Positive correlation between evolutionary rate and recombination rate in Drosophila genes with male-biased expression. Mol Biol Evol 22: 1945-1947.

60. Haddrill PR, Halligan DL, Tomaras D, Charlesworth B (2007) Reduced efficacy of selection in regions of the Drosophila genome that lack crossing over. Genome biology 8: R18.

61. Comeron JM, Williford A, Kliman RM (2008) The Hill-Robertson effect: evolutionary consequences of weak selection and linkage in finite populations. Heredity 100: 19-31.

62. Larracuente AM, Sackton TB, Greenberg AJ, Wong A, Singh ND, Sturgill D, Zhang Y, Oliver B, Clark AG (2008) Evolution of protein-coding genes in Drosophila. Trends in genetics : TIG 24: 114-123.

63. Betancourt AJ, Welch JJ, Charlesworth B (2009) Reduced effectiveness of selection caused by a lack of recombination. Current biology : CB 19: 655-660.

64. Weissman DB, Barton NH (2012) Limits to the rate of adaptive substitution in sexual populations. PLoS Genet 8: e1002740.

65. Shields DC, Sharp PM, Higgins DG, Wright F (1988) "Silent" sites in Drosophila genes are not neutral: evidence of selection among synonymous codons. Mol Biol Evol 5: 704-716.

66. Moriyama EN, Hartl DL (1993) Codon usage bias and base composition of nuclear genes in Drosophila. Genetics 134: 847-858.

67. Nielsen R, Bauer DuMont VL, Hubisz MJ, Aquadro CF (2007) Maximum likelihood estimation of ancestral codon usage bias parameters in Drosophila. Mol Biol Evol 24: 228-235.

68. Andolfatto P, Wong KM, Bachtrog D (2011) Effective population size and the efficacy of selection on the X chromosomes of two closely related Drosophila species. Genome biology and evolution 3: 114-128.

69. Vicario S, Moriyama EN, Powell JR (2007) Codon usage in twelve species of Drosophila. BMC evolutionary biology 7: 226.

70. Haddrill PR, Bachtrog D, Andolfatto P (2008) Positive and negative selection on noncoding DNA in Drosophila simulans. Mol Biol Evol 25: 1825-1834.